\documentclass[10pt]{amsart2000}
\usepackage{latexsym,amssymb,amsmath2000,epic,
graphics,
}
\newcommand{\arr}{\longrightarrow}
\newcommand{\ar}{\rightarrow}

\newcommand{\mx}[1]{\mbox{{#1}}}

\newcommand{\ch}{\mx{ch}}
\newcommand{\Sym}{{\tt S}}

\DeclareMathOperator{\End}{End}
\DeclareMathOperator{\IM}{Im}
\DeclareMathOperator{\Ker}{Ker}
\DeclareMathOperator{\Coker}{Coker}

\DeclareMathOperator{\Ind}{Ind}
\DeclareMathOperator{\rank}{rank}

\def\gr{{\mathrm{gr}\,}}
\def\Gr{{\mathrm{Gr}\,}}
\def\Im{{\mathrm{Im}\,}}

\newcommand{\dst}{\mbox{{\sf d}}}

\newcommand{\dprt}{\mbox{$\partial$}}
\newcommand{\dpind}[1]{\mbox{${\partial}_{#1}$}}
\newcommand{\dhind}[1]{\mbox{$\hat{\partial}_{#1}$}}

\newcommand{\lsp}[1]{\langle {#1} \rangle}

\newcommand{\lag}{L}
\newcommand{\gind}[1]{{\tt g}_{#1}}

\newcommand{\DIV}{\mathop{\rm div}}

\def\CC{{\mathbb{C}}}

 \def\MM{{\bold M}}

\def\RR{{\mathbb{R}}}

\def\ZZ{{\mathbb{Z}}}

\newcommand{\al}{\alpha}

\newcommand{\De}{\Delta}
\newcommand{\tdn}{\triangledown}
\newcommand{\ep}{\varepsilon}
\newcommand{\ze}{\zeta}

\newcommand{\ka}{\kappa}
\newcommand{\la}{\lambda}
\newcommand{\La}{\Lambda}
\newcommand{\ro}{\rho}
\newcommand{\si}{\sigma}
\newcommand{\ta}{\tau}
\newcommand{\ph}{\varphi}

\newcommand{\tht}{\theta}

\newcommand{\Om}{\Omega}

\def\Bw{{\mathcal B}}

\def\Hw{{\mathcal H}}

\def\Pw{{\mathcal P}}

\def\Ww{{\mathcal W}}

\newcommand{\fg}{\mbox{{\tt g}}}

\newcommand{\fk}{\mathfrak{k}}

\newcommand{\ad}{\mathop{\rm ad \, }}

\newcommand{\st}[1]{
\ensuremath{^{\scriptstyle \textrm{#1}}}}
\renewcommand{\tilde}{\widetilde}
\renewcommand{\theequation}%
  {\arabic{section}.\arabic{equation}}
\makeatletter
\renewcommand\section%
 {\@startsection {section}{1}{\z@}%
  {-3.5ex \@plus -1ex \@minus -.2ex}%
    {2.3ex \@plus.2ex}%
       {\normalfont\large\bfseries}}
\@addtoreset{equation}{section}
\makeatother

\newtheorem{Proposition}{Proposition}[section]
\newcommand{\bPr}{\begin{Proposition}}
\newcommand{\ePr}{\end{Proposition}}

\newtheorem{Theorem}[Proposition]{Theorem}
\newcommand{\bTh}{\begin{Theorem}}
\newcommand{\eTh}{\end{Theorem}}

\newtheorem{Lemma}[Proposition]{Lemma}
\newcommand{\bLe}{\begin{Lemma}}
\newcommand{\eLe}{\end{Lemma}}

\newtheorem{Definition}[Proposition]{Definition}
\newcommand{\bDe}{\begin{Definition}}
\newcommand{\eDe}{\end{Definition}}

\newtheorem{Corollary}[Proposition]{Corollary}
\newcommand{\bCo}{\begin{Corollary}}
\newcommand{\eCo}{\end{Corollary}}

\newtheorem{Conjecture}[Proposition]{Conjecture}
\newcommand{\bCj}{\begin{Conjecture}}
\newcommand{\eCj}{\end{Conjecture}}

\theoremstyle{remark}
\newtheorem{remark}[Proposition]{Remark}

\newcommand{\bRe}{\begin{remark}}
\newcommand{\eRe}{\end{remark}}

\newcommand{\bEq}{\begin{equation}}
\newcommand{\eEq}{\end{equation}}

\newcommand{\bEa}{\begin{eqnarray}}
\newcommand{\eEa}{\end{eqnarray}}

\newcommand{\bEaz}{\begin{eqnarray*}}
\newcommand{\eEaz}{\end{eqnarray*}}

\newcommand{\bAr}{\begin{array}}
\newcommand{\eAr}{\end{array}}

\newcommand{\bN}{\begin{enumerate}}
\newcommand{\eN}{\end{enumerate}}

\newcommand{\bD}{\begin{description}}
\newcommand{\eD}{\end{description}}

\newcommand{\prf}{{\sl Proof.}}

\newcommand{\pLe}{{\sl Proof of Lemma}}

\newcommand{\epf}{$\Box$}

\newcommand{\alphaparenlist}{%
  \renewcommand{\theenumi}{\alph{enumi}}%
  \renewcommand{\labelenumi}{(\theenumi)}%
}
\newcommand{\Alphaparenlist}{%
  \renewcommand{\theenumi}{\Alph{enumi}}%
  \renewcommand{\labelenumi}{(\theenumi)}%
}

\newcommand{\arabicparenlist}{%
  \renewcommand{\theenumi}{\arabic{enumi}}%
  \renewcommand{\labelenumi}{(\theenumi)}%
}
\newcommand{\romanparenlist}{%
  \renewcommand{\theenumi}{\roman{enumi}}%
  \renewcommand{\labelenumi}{(\theenumi)}%
}
\begin{document}

%\begin{quote}
%{\raggedleft \mbox{{\small {\it Preliminary version 2.2 }}}\\}
%\end{quote}

\title[Representations of $E(3,6)$~II:~~Four series of degenerate modules.]
{Representations of the exceptional
Lie superalgebra
$E(3,6)$~II:~~Four series of degenerate modules.
}

%\subjclass{28A78,32H20,42B25.}
%\keywords{   }
%%%%%%%%\date{{\small \today}}
%\date{{\small September 5, 2000}}
\author{Victor G. Kac${}^*$ and Alexei Rudakov${}^\dagger$}
\thanks{${}^*$~Supported in part by NSF grant
    DMS-9970007. \\
\hspace*{1.5em}${}^\dagger$~Research was
partially conducted by Alexei Rudakov
for the Clay Mathematics Institute.
%financed by {\it European Commission}
%(TMR 1998-2001 Network {\it Harmonic Analysis}).
}

\begin{abstract}
Four $\ZZ_+$-bigraded complexes with the action of the exceptional
infinite-dimensional Lie superalgebra $E(3,6)$ are
constructed. We show that all the images and cokernels and all but three
kernels of the
differentials are irreducible $E(3,6)$-modules. This is based on the
list of singular vectors and the calculation of homology of these complexes.
As a result, we obtain an explicit construction of all degenerate irreducible
$E(3,6)$-modules and compute their characters and sizes. Since the group
of symmetries of the Standard Model $SU(3) \times SU(2) \times U(1)$
(divided by a central subgroup of order six) is a maximal compact
subgroup of the group of automorphisms of $E(3,6)$, our results
may have applications to particle physics.

\end{abstract}
\maketitle

%s0
\section{Introduction.}

It has been established by A.~Rudakov \cite{R} some 25~years ago
that all degenerate irreducible continuous modules over the Lie
algebra $W_n$ of all formal vector fields in $n$ indeterminates
occur as kernels and as cokernels of the differential of the
$\ZZ_+$-graded de~Rham complex $\Omega_n$ of all formal
differential forms in $n$ indeterminates.

The main objective of the present paper is to obtain a similar
result for the exceptional infinite-dimensional Lie superalgebra
$E(3,6)$ from the list of simple linearly compact Lie
superalgebras classified by V.~Kac \cite{K}.  It turned
out that the situation is much more interesting:  we have
constructed four $\ZZ_+$-bigraded complexes with the action of
$E(3,6)$ and certain connecting homomorphisms between these
complexes.  All images and cokernels and all but three kernels of
differentials turn out to be irreducible, and all degenerate irreducible
$E(3,6)$-modules occur among them.  The
failure of irreducibility is connected to non-triviality of
homology of these complexes, which we compute as well.
At the end of the paper we compute the characters and sizes
of all degenerate irreducible $E(3,6)$-modules and speculate
on their relation to the Standard Model.

In our two other papers \cite{KR1} and \cite{KR2} on the subject
we show that all locally finite with respect to any non-trivial
open subalgebra irreducible $E(3,6)$-modules,
that do not occur as subquotients in our four
complexes, are non-degenerate (and therefore induced).

%s1
\section{A reminder on $E(3,6)$ and induced modules.}
\label{sec:1}

Recall that we view $E(3,6)$ as a subalgebra of the exceptional
Lie superalgebra $E(5,10)$.  The latter is constructed as follows
(see \cite{CK} for details).  The even part, $E(5,10)_{\bar{0}}$,
is isomorphic to the Lie algebra $S_5$ of divergenceless formal
vector fields in the indeterminates $x_1 , \ldots , x_5$, while the odd
part, $E(5,10)_{\bar{1}}$, is the space $d\Omega^1_5$ of closed
($\equiv$ exact) $2$-forms, and the remaining brackets are
defined by:
\begin{displaymath}
  [X,w] = L_X w, \,\, [w,w'] =w \wedge w', \,\,
  X \in S_5 ,\,\, w,w' \in d\Omega^1_5 \, .
\end{displaymath}
In the second formula a closed $4$-form $w \wedge w'$ is
identified with the vector field whose contraction with $dx_1
\wedge \cdots \wedge dx_5$ produces this $4$-form.  As in
\cite{KR1}, we use the notation:
\begin{displaymath}
  d_{jk} = dx_j \wedge dx_k, \quad \partial_i = \partial /
  \partial x_i \, .
\end{displaymath}
Elements from $E(5,10)_{\bar{0}}$ are of the form
\begin{displaymath}
  \sum_i a_i \partial_i , \hbox{ where }
  a_i \in \CC [[x_1 , \ldots ,x_5]],
  \sum_i \partial_i a_i =0 \, ,
\end{displaymath}
and elements from $E(5,10)_{\bar {1}}$ are of the form
\begin{displaymath}
  w=\sum_{j,k} b_{jk} d_{jk} \hbox{ where }
  b_{jk} \in \CC [[x_1 ,\ldots ,x_5]], \,\, d w=0 \, .
\end{displaymath}
In particular, the commutator of two odd elements can be
computed using bilinearity and the rule $(a,b \in \CC [[x_1,
\ldots ,x_5]])$
\begin{displaymath}
  [ad_{jk}, bd_{\ell m}] = \epsilon_{ijk\ell m}ab \, \partial_i
  \, ,
\end{displaymath}
where $\epsilon_{ijk\ell m}$ is the sign of the permutation
$\scriptstyle{(ijk\ell m)}$ if all indices are distinct and $0$ otherwise.

The Lie superalgebra $E(5,10)$ carries a unique consistent
irreducible $\ZZ$-grading defined by
\begin{displaymath}
  \deg x_i = 2 =-\deg \partial_i , \,\, \deg dx_i = -\tfrac{1}{2} \, .
\end{displaymath}

In order to define $E(3,6)$ as a subalgebra of $E(5,10)$, let
$z_+ = x_4$, $z_- =x_5$, $\partial_+ = \partial_4$,
$\partial_-=\partial_5$ and define the secondary $\ZZ$-grading
by:
\begin{displaymath}
  \deg x_i =0 =\deg \partial_i, \,\,
  \deg z_{\pm} =1=-\deg \partial_{\pm}, \,\,
  \deg d=-\tfrac{1}{2} \, .
\end{displaymath}
Then $E(3,6)$ is the $0$\st{th} piece of the secondary grading.
The consistent $\ZZ$-grading of $E(5,10)$ induces the consistent
$\ZZ$-grading $E(3,6)=\lag =\oplus_{j \geq -2} \fg_j$, where
\begin{displaymath}
  \fg_{-2} = \langle \partial_i |\, i=1,2,3 \rangle ,\,\,
  \fg_{-1} = \langle d^+_i:= d_{i4},\,
  d^-_i := d_{i5} |\, i=1,2,3 \rangle \, .
\end{displaymath}
Furthermore, $\fg_0 \simeq s \ell (3) \oplus s \ell (2) \oplus
g\ell (1)$, where
\begin{eqnarray}
  \label{eq:1.1}
  s \ell (3) &=& \langle h_1 = x_1 \partial_1 -x_2 \partial_2 ,\,
       h_2 =x_2 \partial_2 -x_3 \partial_3 ,\,
       e_1 =x_1 \partial_2,\\
\nonumber
  &&\,\,\,e_2 = x_2 \partial_3 ,\, e_{12} = x_1 \partial_3 ,\,
        f_1 =x_2 \partial_1 ,\, f_2=x_3\partial_2 ,\,
        f_{12} =x_3\partial_1 \rangle \, , \\
  \label{eq:1.2}
   s\ell (2) &=& \langle h_3 = z_+ \partial_+ -z_- \partial_- ,\,\,
        e_3 = z_+ \partial_- , \,f_3 =z_-\partial_+ \rangle \, ,\\
  \label{eq:1.3}
   g\ell (1)&=&  \langle Y =\mbox{$\frac{2}{3} \sum x_i\partial_i$}
        - (z_+\partial_+ + z_- \partial_-) \,\rangle \, .
\end{eqnarray}
The eigenspace decomposition of $\ad (3Y)$ coincides with the
consistent $\ZZ$-grading of $E(3,6)$.

Below we often mark the elements $\dpind i \in \gind {-2}$ by a hat,
writing $\dhind i$ whenever we need to distinguish  them from
$\dpind i$ in another role.

As has been mentioned in [CK] and [KR1],
the even part $E(3,6)_{\bar{0}}$ of $E(3,6)$ contains a subalgebra $W$
isomorphic to $W_3$ with the isomorphism given by the formula
\begin{equation}
\label{eq:Wsub}
D \longmapsto D - \tfrac{1}{2}\DIV{\! D} \,
(z_+\partial_+ + z_- \partial_-) \, .
\end{equation}

We fix the Cartan subalgebra $\Hw = \langle h_1,h_2,h_3,Y
\rangle$ and the Borel subalgebra\\
 $\Bw=\Hw + \langle e_i \,\,
{\scriptstyle (i=1,2,3)},\,
e_{12} \rangle$ of $\fg_0$.  Then $f_0 := d^+_1$ is the highest
weight vector of the (irreducible) $\fg_0$-module $\fg_{-1}$,
\begin{displaymath}
  e'_0 := x_3 d^-_3  \hbox{  and  }  e_0 :=x_3 d^-_2 -
     x_2 d^-_3 + 2x_5d_{23}
\end{displaymath}
are the lowest weight vectors of the $\fg_0$-module $\fg_1$, and
one has:
\begin{eqnarray}
  \label{eq:1.4}
  [e'_0 , f_0] &=& f_2 \, , \\
  \label{eq:1.5}
  [e_0 ,f_0] &=&
\mbox{$\frac{2}{3} h_1 + \frac{1}{3} h_2 -h_3-Y$}
=:h_0 \, , \\
\noalign{
\hbox{so that } }
h_0 &=& -x_2 \partial_2 -x_3 \partial_3 +2z_-\partial_- \, .
  \label{eq:1.6}
\end{eqnarray}
%
%\noalign{\nonumber\hbox{so that}}\\
%
%  \label{eq:1.6}
%  h_0 &=& -x_2 \partial_2 -x_3 \partial_3 +2z_-\partial_- \, .
%\end{eqnarray}
%
The following relations are also important to keep in mind:
\begin{eqnarray}
  \label{eq:1.7}
  [\,e'_0 , d^+_1] &=& f_2 , \,\,
      [e'_0 ,d^+_2]  =-f_{12}, \,\,
      [e'_0 ,d^+_3] =0 , \,\,
  [e'_0 , d^-_i] = 0 \, , \\
  \label{eq:1.8}
  [\,d^{\pm}_i , d^{\pm}_j] &=& 0 ,\,\,\qquad
  [d^+_i ,d^-_j] + [d^+_j , d^-_i] = 0 \, .
\end{eqnarray}
Recall that $\fg_0$ along with the elements $f_0$, $e_0$, $e'_0$
generate the Lie superalgebra $E(3,6)$.

Sometimes we use the following shorthand notation for the
elements of the universal enveloping algebra of $E(3,6)$:
\begin{displaymath}
  d^{\pm}_{ijk} = d^{\pm}_i d^{\pm}_j d^{\pm}_k \, .
\end{displaymath}

We have the triangular decomposition:
\begin{displaymath}
  L:=E(3,6) = L_- + \fg_0 + L_+ \, ,
\end{displaymath}
where $L_- =\oplus_{j<0} \fg_j , \, L_+ =\prod_{j>0} \fg_j$.
Given a $\fg_0$-module $V$, we extend it to $L_0 := \fg_0 +L_+$
by letting $L_+$ act trivially and consider the induced
$L$-module
\begin{equation}
\label{eq:1.10}
  M(V) = U(L) \otimes_{U(L_0)}V \simeq U(L_-) \otimes_{\CC} V \, ,
\end{equation}
the latter being a $\fg_0$-isomorphism.
If $V$ is a finite-dimensional irreducible $\fg_0$-module, the $L$-module
$M(V)$ is called a \emph{generalized Verma module}. Any quotient of a
generalized
Verma module is called a highest weight module.

Recall that all finite-dimensional irreducible $\fg_0$-modules are
modules with highest weight $\lambda$ such that
\begin{displaymath}
  p=\lambda (h_1), q=\lambda (h_2) , \, r=\lambda (h_3) \in
  \ZZ_+ , \,\,\, y=\lambda (Y) \in \CC \, .
\end{displaymath}
Such a module is denoted by $F (p,q ; r;y)$ and the corresponding
generalized Verma
$L$-module is denoted by $M(p,q;r;y)$. A highest weight vector
of the $\fg_0$-module $F(p,q;r;y)$ is called a highest weight vector
of any corresponding highest weight $L$-module.

Note that
the hypercharge operator $Y$ acts diagonally on $M (p,q ;r;y)$
with eigenvalues from the set $y-\frac{1}{3} \ZZ_+$,
its eigenspaces are $\fg_0$-invariant and finite-dimensional, and the
\mx{$y$-eigenspace} is isomorphic to the $\fg_0$-module $F(p,q;r;y)$.

A highest weight vector $s$ of an irreducible finite-dimensional
$\fg_0$-submodule of an \mx{$L$-module} is called \emph{singular}
if
\begin{equation}
\label{eq:defsing}
  e'_0 s=0, \quad e_0s=0  \,\quad (equivalently: \,L_+ s = 0\,).
\end{equation}
For example, the highest weight vector of
$F(p,q;r,y) \subset M(p,q;r;y)$
is a singular vector, called a \emph{trivial singular
vector}.

Recall that the $L$-module $M(p,q;r;y)$ has a unique
irreducible quotient denoted by $I(p,q;r;y)$.  The irreducible
$L$-module $I(p,q;r;y)$ is called \emph{non-degenerate} if
it coincides with $M (p,q;r;y)$, the module
$I(p,q;r;y)$ and the module $M(p,q;r;y)$ are called \emph{ degenerate }
if $M(p,q;r;y) \neq I(p,q;r;y)$.

It follows that $M(p,q,;r;y)$ is a degenerate $L$-module
iff it has a non-trivial singular vector.

The main result of \cite{KR1} is the following.

\begin{Theorem}
  \label{th:1.1}
If $pq \neq 0$, then the $E(3,6)$-module $I(p,q;r;y)$ is non-degenerate.
\end{Theorem}

The corollary of the main result of \cite{KR2} is the following.

\begin{Theorem}
  \label{th:1.2}
The following list consists of all degenerate $E(3,6)$-modules
$I(p,q;r;y)$\break $(p,q,r \in \ZZ_+)$:

\begin{list}{type}{}
\item {A}: \,\,  $I (p,0;r;y_A)$, where $y_A =\frac{2}{3} p-r$,

\item {B}: \,\, $I (p,0;r;y_B)$, where $y_B = \frac{2}{3} p+r+2$,

\item {C}: \,\, $I(0,q;r,y_C)$, where $y_C=-\frac{2}{3} q-r-2$,

\item {D}: \,\, $I(0,q;r,y_D)$, where $y_D=-\frac{2}{3} q+r$.
\end{list}
\end{Theorem}

In the present paper we construct four complexes of degenerate
generalized Verma modules, which allows us
to construct the four series of degenerate modules given by
Theorem~\ref{th:1.2}.

%%%%%%%%%%%%%%%%%%%%%%%%%%%%%%%%end section 1

%s2

\section{The four bigraded complexes.}
\label{sec:2}

Given a $\fg_0$-module $V$ and a number $a \in \CC$, we can
define a new $\fg_0$-module, denoted by $V_{[a]}$ with the same
action of $s\ell (3) \oplus s\ell (2)$ but the changed action of
$Y$ by adding to it  $aI_V$.  Consider the following
$\fg_0$-modules:
\begin{eqnarray*}
  V_A &=& \CC [x_1,x_2,x_3,z_+,z_-] \, , \\
  V_B &=& \CC [x_1,x_2,x_3,\partial_+,\partial_-]_{[2]} \, , \\
  V_C &=& \CC [\partial_1,\partial_2,\partial_3,z_+,z_-]_{[-2]}
  \, , \\
  V_D &=& \CC [\partial_1,\partial_2,\partial_3,\partial_+,\partial_-] \, ,
\end{eqnarray*}
with the action of $\fg_0$ given by
(\ref{eq:1.1})--(\ref{eq:1.3}).  The decomposition into a sum of
irreducible $\fg_0$-modules is given by the bigrading
$(X=A,B,C,D)$
\begin{displaymath}
  V_X = \bigoplus_{m,n \in \ZZ} V^{m,n}_X \, ,
\end{displaymath}
where
\begin{displaymath}
  V^{m,n}_X = \{ f \in V_X | \,
\mbox{$\sum\nolimits_i x_i \partial_i f  = mf$},\,\,
\mbox{$\sum\nolimits_{\epsilon} z_{\epsilon} \partial_{\epsilon} f=nf$}
\} \, .
\end{displaymath}

Let $M_X = M(V_X)$ be the corresponding induced $L$-module.
The bigrading of $V_X$ gives rise to a bigrading of $M_X$ by
generalized Verma modules:
\begin{equation}
  \label{eq:2.1}
  M_X = \bigoplus_{m,n \in \ZZ} M (V^{m;n}_X) \, .
\end{equation}
Note that we have the following isomorphisms of $\fg_0$-modules
(see~(\ref{eq:1.1})):
\begin{eqnarray}
  \label{eq:2.2}
  V^{p,r}_A & \simeq & F (p,0;r;y_A) \, , \\
\nonumber
  V^{p,-r}_B & \simeq & F (p,0;r;y_B) \, , \\
\nonumber
  V^{-q,r}_C & \simeq & F(0,q;r;y_C) \, , \\
\nonumber
  V^{-q,-r}_D & \simeq & F(0,q;r;y_D) \, .
\end{eqnarray}
Consequently, the bigrading (\ref{eq:2.1}) takes the form:
\begin{eqnarray*}
  M_X &=& \oplus_{p,r \in \ZZ_+} M(p,0;r;y_X),
  \hbox{ for } X=A \hbox{ or } B \, ,\\
  M_X &=& \oplus_{q,r \in \ZZ_+} M(0,q;r,y_X),
  \hbox{ for } X=C \hbox{ or } D \, .
\end{eqnarray*}

We let the algebra $U(L_-) \otimes \End V_X$ act on $M_X =U(L_-)
\otimes V_X$ by the formula:
\begin{equation}
  \label{eq:2.3}
  (u \otimes \varphi) (u'\otimes v)
  = u'u \otimes \varphi (v), \,\,
  u,u' \in U (L_-), \,\, \varphi \in \End V_X, \,\, v \in V_X \, .
\end{equation}
We shall often drop the $\otimes$ sign, e.g.~we shall write
$\partial_i$ in place of $1 \otimes \partial_i$, etc.

Introduce the following operators on $M_X$ (acting
by~(\ref{eq:2.3})):
\begin{eqnarray}
  \label{eq:2.3n}
  \Delta^{\pm} &=& \sum^3_{i=1} d^{\pm}_i \otimes \partial_i \, , \\
  \delta_i \,\,&=& d^+_i \otimes \partial_+ +d^-_i \otimes  \partial_-,
    \,\, i = 1,2,3 \, .
\end{eqnarray}
Here $\partial_i$ acts as $d/dx_i$ (resp. as multiplication by
$\partial_i$) in the cases $X=A,B$ (resp. $X=C,D$) and similarly
for $\partial_{\pm}$.

\begin{Lemma}
  \label{lem:2.1}
\alphaparenlist
  \begin{enumerate}
  \item %%a
    The action of $s\ell (3)$ (resp. $s\ell (2)$) $\subset \fg_0
    \subset E(3,6)$ on the $E(3,6)$-module $M_X$ commutes with
    the operators $\Delta^{\pm}$ and $\partial_{\pm}$
    (resp.~$\delta_i$ and $\partial_i$).

\item %%b
  One has the following commutation relations for $Y$:
  \begin{displaymath}
    [Y,\partial_{\pm}] = \partial_{\pm},         \,\,
    [Y,\partial_i]     = -\tfrac{2}{3} \partial_i ,   \,\,
    [Y,d^{\pm}_i]      = -\tfrac{1}{3} d^{\pm}_i ,\,\,
    [Y,\Delta^{\pm}]   = -\Delta^{\pm} , \,\,
    [Y,\delta_i]       =  \tfrac{2}{3}\delta_i \, .
  \end{displaymath}

\item %%c
  \,$(\Delta^{\pm})^2 =0 , \,\,\,
  \Delta^+\Delta^- + \Delta^-\Delta^+ =0 , \,\,\,
  \delta_i \delta_j + \delta_j \delta_i =0$.
  \end{enumerate}
\end{Lemma}

\begin{proof}
  The proof of (a) and (b) is straightforward, (c)~follows from
  (\ref{eq:1.8}).
\end{proof}

Now we introduce our basic operator
\begin{displaymath}
  \triangledown := \Delta^+ \partial_+ + \Delta^- \partial_-
%  \equiv
=\delta_1\partial_1 + \delta_2 \partial_2 +
  \delta_3 \partial_3  \, .
\end{displaymath}

\begin{Proposition}
  \label{prop:2.2}
\alphaparenlist
\begin{enumerate}
\item %%a
  $\triangledown^2 =0$.

\item %%b
  The operator $\triangledown$ commutes with the action of
  $E(3,6)$ on $M_X$.
\end{enumerate}
\end{Proposition}

\begin{proof}
  (a) is immediate by Lemma~\ref{lem:2.1}c.  In order to
  prove~(b), notice that the action (\ref{eq:2.3}) commutes with
  the left multiplication action of $L_-$ on $M_X$, hence, in
  particular, $\triangledown$ commutes with $L_-$.  Furthermore,
  it follows from Lemma~\ref{lem:2.1}a (resp.~\ref{lem:2.1}b)
  that $\triangledown$ commutes with the action of $s\ell (3)$
  and $s\ell (2)$ (resp.~$Y$), hence $\tdn$ commutes with $\gind 0$.

Now the proof that the operator $\triangledown$ commutes with $L$
is based on the following lemma.

\begin{Lemma}
  \label{lem:2.3}

Let $L=\prod_j \fg_j$ be a $\ZZ$-graded Lie superalgebra and let
$M$ be an $L$-module induced from an $L_{0}$-module $V$
such that $\fg_j |_V =0$ for $j >0$.  Let $\triangledown$ be an
operator
%from $U(L_-) \otimes \End V$
acting on $M$
%by (\ref{eq:2.3})
that commutes with $L_-$.
Suppose that $\triangledown$ commutes with
$\fg_0$ and
\begin{equation}
  \label{eq:2.4}
  g \triangledown (v) = 0\,\, \hbox{ for all }\,
  g \in L_+, \,\, v \in 1 \otimes V \, .
\end{equation}
Then $\triangledown$ commutes with $L$.
\end{Lemma}

\begin{proof}
It is sufficient to show that $\tdn$ commutes with $L_+$.
  For $g \in L_+$, $u \in L_-$ we can write:
  \begin{displaymath}
    gu = \sum_{n \geq 0} u_ng_n , \hbox{ where }
    u_n \in U(L_-), \,\,    g_n \in \fg_n \, .
  \end{displaymath}
Then we have:
\begin{eqnarray*}
  g \triangledown (u \otimes v) &=& gu\triangledown (v)
  =\sum_{n \geq 0} u_ng_n \triangledown (v)\\
  &=& u_0g_0 \triangledown (v) = u_0 \triangledown g_0 (v)
      = \sum_{n \geq 0} u_n \triangledown g_n (v)
      = \triangledown g (u \otimes v) \, .
\end{eqnarray*}
\end{proof}

We shall establish that the conditions of Lemma~\ref{lem:2.3} are
valid in our situation.
Since $\{ g \in L_+ |\, g \triangledown (v) =0$ for all $v \in 1
\otimes V_X \}$ is a $\fg_0$-invariant subalgebra of $L_+$ (because
$\triangledown$ commutes with $\fg_0$), in order to establish
(\ref{eq:2.4}), it suffices to check only the following:
\begin{eqnarray}
  \label{eq:2.5}
  e'_0 \triangledown (v) =0 \, , \, v \in 1 \otimes V_X \, ,\\
\label{eq:2.6}
e_0 \triangledown (v) =0 \, , \, v \in 1 \otimes V_X \, .
\end{eqnarray}
(Recall that $\fg_1$ generates $L_+$ and $e_0,e'_0$ generate the
$\fg_0$-module $\fg_1$.)  Using (\ref{eq:1.7}), we
get~(\ref{eq:2.5}):
\begin{eqnarray*}
  e'_0 \triangledown (v)
  =  \sum_{\epsilon =+,-}
       (f_{2} \partial_1 \partial_{\epsilon} v
       - f_{12} \partial_2 \partial_{\epsilon}v)
  = ((x_3 \partial_2) \partial_1 - (x_3\partial_1) \partial_2)
      \sum_{\epsilon} \partial_{\epsilon} v=0 \, .
\end{eqnarray*}

Furthermore, we have:
\begin{equation}
  \label{eq:2.7}
  e_0 \triangledown (v) = %-
h_0 (\partial_1\partial_+v)
  + f_1 (\partial_2 \partial_+ v)+
  f_{12} (\partial_3\partial_+ v) -
  2f_3 (\partial_1\partial_- v)\, .
\end{equation}

In the $X=A$ case, $h_0$ acts on $V_X$ by (\ref{eq:1.6}) (where $
x_4=z_+ ,\, x_5 =z_-$), $f_1 =x_2\partial_1$, $f_{12}
=x_3\partial_1$ and $f_3 = z_- \partial_+$, hence (\ref{eq:2.7})
becomes:
\begin{displaymath}
  e_0 \triangledown (v) = (h_0 + x_2\partial_2 +
     x_3\partial_3 -2z_-\partial_-) \partial_1\partial_+ v=0 \, .
\end{displaymath}

In the $X=B$ case, $h_0$ acts on $V_X$ by ``(\ref{eq:1.6}) minus
$2I$'' due to the twist for $Y$, which is necessary to include in
order to compensate for the additional term $2 \partial_1
\partial_+ v$ occurring from the last term on the right of
(\ref{eq:2.7}).

Similar calculations show that (\ref{eq:2.6})
 holds in the $X=C$ and $D$ cases as well.
\end{proof}

\begin{remark}
  \label{rem:2.4}
\alphaparenlist
\begin{enumerate}

\item %%a
  Since $\triangledown$ commutes with the representation of
  $E(3,6)$ in $M_X$, the non-zero image under $\triangledown$ of a singular
  vector is a singular vector.

\item %%b
  Let $V$ be a direct sum of finite-dimensional irreducible
  $\gind 0$-modules, $M(V)$ be the corresponding
  induced $E(3,6)$-module, and let
  $\triangledown$ be an operator on $M(V)$ defined via
  (\ref{eq:2.3}), such that $\triangledown v$ is a singular
  vector for each of the highest weight vectors $v$ of $\fg_0$ in
  $V$.
  Suppose that
  $\triangledown $ commutes with $Y$ and all $f_i$,
  $i=1,2,3$, then $\triangledown$ commutes with $E(3,6)$.  This
  follows from Lemma~\ref{lem:2.3}.
\item %%c
  Suppose that for the same module $M(V)$ there is a linear
  map $\ph:M(V)\ar N$ to another $\lag$-module $N$. If
  $\ph$ commutes with the action of $\lag$, then the non-zero
  image of a singular vector is a singular vector. If $\ph$
  commutes with the action of $L_-$, \break the image
  $\ph(v)$ is a singular
  vector for each of the highest weight vectors $v$ of $\fg_0$ in
  $V$, \break $\ph$ commutes with $\gind 0$, then $\ph$ is
  a morphism of $\lag$-modules. The arguments are just the same as
  for (a), (b).
\item %%d
  Lemma~\ref{lem:2.3} and the above remark (c) could be generalized
  to the case when $M$ is a highest weight $\lag$-module, or a sum of
  such modules.
\end{enumerate}
\end{remark}

Proposition~\ref{prop:2.2} implies
immediately the following corollary about
singular vectors.

\begin{Corollary}
  \label{cor:2.5}
\alphaparenlist
\begin{enumerate}
\item %%a
The $E(3,6)$-module $M (p,0;r;y_A)$ has a non-trivial singular
vector
\begin{displaymath}
  \triangledown (x^{p+1}_1 z^{r+1}_+) =
  (p+1)(r+1)d^+_1 x^p_1 z^r_+
\end{displaymath}
for all $p,r \in \ZZ_+$.

\item  %%b
  The $E(3,6)$-module $M (p,0;r;y_B)$ has a non-trivial singular
  vector
  \begin{displaymath}
    \triangledown (x^{p+1}_1 \partial^{r-1}_-)
    = (p+1) \delta_1 x^p_1 \partial^{r-1}_-
  \end{displaymath}
for all $p,r \in \ZZ_+$, $r>0$.

\item %%c
  The $E(3,6)$-module $M (0,q;r;y_C)$ has a nontrivial singular
  vector
  \begin{displaymath}
    \triangledown (\partial^{q-1}_3 z^{r+1}_+)=
    (r+1) \Delta^+ \partial^{q-1}_3 z^r_+
  \end{displaymath}
for all $q,r \in \ZZ_+$, $q >0$.

\item  %%d
  The $E(3,6)$-module $M (0,q;r;y_D)$ has a nontrivial singular
  vector
  \begin{displaymath}
    \triangledown (\partial^{q-1}_3 \partial^{r-1}_-)
  \end{displaymath}
for all $q,r \in \ZZ_+$, $\,q,r >0$.
\end{enumerate}
\end{Corollary}

In order to construct more singular vectors, consider smaller
$\fg_0$-modules $V_{X'} \subset V_X$ and the corresponding
induced $\lag$-modules $M_{X'} \subset M_X$:
\[
\begin{array}{lclcl}
  V_{A'} &=& \CC [x_1,x_2,x_3] &\hbox{ and }&
  M_{A'} = M(V_{A'}) \, ,\\
   V_{B'} &=& \CC [x_1,x_2,x_3]_{[2]} &\hbox{ and }&
  M_{B'} = M(V_{B'}) \, ,\\
  V_{C'} &=& \CC [\partial_1,\partial_2,\partial_3]_{[-2]} &\hbox{ and }&
  M_{C'} = M(V_{C'}) \, ,\\
  V_{D'} &=& \CC [\partial_1,\partial_2,\partial_3] &\hbox{ and }&
  M_{D'} = M(V_{D'}) \, .
\end{array}
\]
Let $\tau_1 : V_{A'} \underrightarrow{\sim} V_{B'}$ and $\tau_2:
V_{C'} \underrightarrow{\sim} V_{D'}$ be the identity
isomorphisms.  Then we have:
\begin{equation}
  \label{eq:2.8}
  [Y,\tau_i] =2\tau_i \, .
\end{equation}
Let
\begin{eqnarray*}
  \triangledown_2 &=& \Delta^+\Delta^-\tau_1 :
  M_{A'} \to M_{B'} \, , \\
  \triangledown_2 &=&  \Delta^+\Delta^-\tau_2 :
  M_{C'} \to M_{D'}
\end{eqnarray*}
act via (\ref{eq:2.3}).

\begin{Proposition}
  \label{prop:2.6}
\alphaparenlist
\begin{enumerate}
\item %%a
  $\triangledown \triangledown_2 = \triangledown_2
  \triangledown =0$.

\item %%b
  $\triangledown_2$ is an $E(3,6)$-module isomorphism:
$M_{X'} \underrightarrow{\sim} M_{Y'}$, where $X'=A'$
  (resp. $C'$) and $Y'=B'$ (resp.,~$D'$).
\end{enumerate}
\end{Proposition}

\begin{proof}
  (a) is clear from Lemma~\ref{lem:2.1}c.  Next, $L_-$ commutes
  with $\triangledown_2$ by definition, $s\ell (3)$ commutes due
  to Lemma~\ref{lem:2.1},  $s\ell (2)$ commutes for trivial
  reasons and $Y$ commutes due to Lemma~\ref{lem:2.1}b and
  (\ref{eq:2.8}).

As before, due to Lemma~\ref{lem:2.3}, in order to establish (b),
it suffices to show that
\begin{displaymath}
  e'_0 \triangledown_2 v=0 \,\hbox{ and }\,
  e_0 \triangledown_2 v=0 \,\,\,
  \hbox{ for } \,v \in 1 \otimes V_{X'} \, .
\end{displaymath}
Let $X'=A'$.  We have:
\begin{eqnarray*}
  e'_0 \triangledown_2 v
  &=&  e'_0(- \Delta^-\Delta^+ \tau_1) (v) \\
  &=& \Delta^- (-f_2 \partial_1 + f_{12} \partial_2) \tau_1 (v)
 = \Delta^- (-(x_3\partial_2) \partial_1 +
       (x_3\partial_1)\partial_2) \tau_1 (v) =0 \, .
\end{eqnarray*}
Next, we have:
\begin{eqnarray*}
  e_0 \triangledown_2 v
  &=& ((2 f_3 \partial_1)  \Delta^{+} +
      \Delta^{-} (h_0 \partial_1 + f_1 \partial_2
      + f_{12} \partial_3 )) \tau_i (v) \\
  &=& \Delta^- ((2+ h_0) \partial_1 + (x_2\partial_1)\partial_2
    + (x_3 \partial_1)\partial_3) \tau_i (v), \\
\end{eqnarray*}
because $f_3 \tau_i (v)=0$, since $f_3$ annihilates $B'$ and $D'$.
Notice that
for $v=x^{a_1}_1 x^{a_2}_2 x^{a_3}_3$:
\[
(x_j \partial_1)\partial_j \tau_1 (v)=
    (x_j \partial_j)\partial_1 \tau_1 (v), \,\,\,j=2,3,
\]
but for
$v=\partial^{a_1}_1 \partial^{a_1}_2 \partial^{a_1}_3$:
\[
(x_j \partial_1)\partial_j \tau_2 (v)=
    (x_j \partial_j\,-1)\partial_1 \tau_2 (v)
,  \,\,\,j=2,3.
\]
At the same time
$h_0|_{B'}=-x_2 \partial_2-x_3 \partial_3+2z_-\partial_- -2$ and
$h_0|_{D'}=-x_2 \partial_2-x_3 \partial_3+2z_-\partial_-$.
Therefore
\begin{eqnarray*}
  e_0 \triangledown_2 v
  &=& \Delta^- ((2+ h_0) \partial_1 + (x_2\partial_1)\partial_2
    + (x_3 \partial_1)\partial_3) \tau_i (v), \\
  &=& \Delta^- (2z_-\partial_-)
\tau_i (v)=0.
\end{eqnarray*}
\vspace{-6ex}

\end{proof}

Finally, in a similar fashion, consider the $\fg_0$-modules
$V_{X''} \subset V_X$ and the corresponding induced
$\lag$-modules $M_{X''} \subset M_X$:
\[
\begin{array}{lclcl}
  V_{A''} &=& \CC [z_+,z_-] &\hbox{ and }&
              M_{A''} = M(V_{A''}) \, ,\\
  V_{B''} &=& \CC [\partial_+ ,\partial_-]_{[2]} &\hbox{ and }&
              M_{B''} = M(V_{B''}) \, , \\
  V_{C''} &=& \CC [z_+,z_-]_{[-2]} &\hbox{ and }&
              M_{C''} = M(V_{C''}) \, ,\\
  V_{D''} &=& \CC [\partial_+ ,\partial_-] &\hbox{ and }&
              M_{D''} = M(V_{D''}) \, .
\end{array}
\]
We let $\rho_1 : V_{A''} \underrightarrow{\sim} V_{C''}$ and
$\rho_2 : V_{B''}\underrightarrow{\sim} V_{D''}$ be the identity
isomorphisms, so that
\begin{equation}
  \label{eq:2.9}
  [Y,\rho_i] = -2\rho_i \, .
\end{equation}
Let
\begin{eqnarray*}
  \triangledown_3 &=& \delta_1\delta_2\delta_3 \rho_1 :
     M_{A''} \to M_{C''} \, , \\
  \triangledown_3 &=&  \delta_1\delta_2\delta_3 \rho_2:
     M_{B''} \to M_{D''}
\end{eqnarray*}
act via (\ref{eq:2.3}).

\begin{Proposition}
  \label{prop:2.7}
\alphaparenlist
\begin{enumerate}
\item %%a
  $\triangledown \triangledown_3 = \triangledown_3 \triangledown
  =0$.

\item %%b
  $\triangledown_3$ is an $E(3,6)$-module isomorphism $M_{X''}
  \to M_{Y''}$, where $X''=A''$ (resp.~$B''$) and $Y'' = C''$
  (resp.~$D''$).
\end{enumerate}
\end{Proposition}

\begin{proof}
  As in the proof of Proposition~\ref{prop:2.6}, due to
  Lemma~\ref{lem:2.1}, (a)~is clear and (b)~reduces to checking
  the relations
\begin{displaymath}
  e'_0 \triangledown_3 v =0 \,\hbox{ and } \,
  e_0 \triangledown_3 v=0 \,\,\,\hbox{ for }\, v \in 1 \otimes V_X \, .
\end{displaymath}
  We have
\begin{displaymath}
  e'_0 \triangledown_3 v = (f_2 \partial_+
  \delta_2 \delta_3 + \delta_1 f_{12} \partial_+ \delta_3) \rho_1 v =
  \partial_+ \delta^2_3 \rho_1 v =0\,\,
\end{displaymath}
 if
  $X'' = A'' $  because $f_2\rho_1 v =f_{12}\rho_1 v =0$,
and similarly for $X''=B''$.

Next, we have for $v \in V_{X''}$:
\begin{eqnarray*}
  [e_0,\delta_1\delta_2\delta_3] v &=&
    ((h_0 \partial_+ -2f_3 \partial_-) \delta_2 \delta_3 - \delta_1
      f_1 \partial_+\delta_3   +\delta_1\delta_2 f_{12} \partial_+) v\\
    &=& (h_0 \partial_+ -2f_3 \partial_-)
        \delta_2\delta_3 (v) \,  .
\end{eqnarray*}
Since $[h_0 , \delta_i]=-\delta_i$ for $i=1,2$ and
$[f_3,\delta_i]=0$, we obtain for $v \in V_{X''}$:
\begin{displaymath}
  [e_0 , \delta_1\delta_2\delta_3]v =
    \delta_2\delta_3 ((h_0-2)\partial_+ v
    - 2(z_- \partial_+) \partial_- v)
\, .
\end{displaymath}
Again
$(z_- \partial_+) \partial_- v=(z_- \partial_-) \partial_+ v$
for $v \in C''$
and
$(z_- \partial_+) \partial_- v=(z_- \partial_-\, -1) \partial_+ v$
for $v \in D''$.

In the case $X'' =A''$
we have:
\begin{displaymath}
 e_0\delta_1\delta_2\delta_3 \rho_1 v
  = \delta_2 \delta_3 (h_0- 2z_- \partial_+-2)\partial_- \rho_1v=0\,.
\end{displaymath}
In the case $X''=B''$,
we similarly deduce:
\begin{displaymath}
 e_0\delta_1\delta_2\delta_3 \rho_2 v
  = \delta_2 \delta_3 (h_0- 2z_- \partial_+)\partial_- \rho_2v=0\,.
\end{displaymath}%
\vspace{-6ex}

\end{proof}

Propositions~\ref{prop:2.6} and \ref{prop:2.7} lead to the next
corollary about singular vectors.

\begin{Corollary}
  \label{cor:2.8}
\alphaparenlist
\begin{enumerate}
\item %%a
  The $E(3,6)$-module $M(p,0;0;y_B)$ has a non-trivial singular
  vector
  \begin{displaymath}
    \triangledown_2 (x_1^{p+2}) = (p+2) (p+1)
    d^+_1 d^-_1 x^p_1
  \end{displaymath}
for all $p \in \ZZ_+$.

\item %%b
  The $E(3,6)$-module $M(0,q;0;y_D)$ has a non-trivial singular
  vector
  \begin{displaymath}
    \triangledown_2 (\partial_3^{q-2}) =
    \Delta^+\Delta^- \partial_3^{q-2}
  \end{displaymath}
for all $q \in \ZZ$, $q \geq 2$.

\item %%c
  The $E(3,6)$-module $M(0,0;r;y_C)$ has a non-trivial singular
  vector
  \begin{displaymath}
    \triangledown_3 (z_+^{r+3}) =(r+3)(r+2)(r+1)
    d^+_{123} z^r_+
  \end{displaymath}
for all $r \in \ZZ_+$.

\item %%d
  The $E(3,6)$-module $M(0,0;r;y_D)$ has a non-trivial singular
  vector
  \begin{displaymath}
    \triangledown_3 (\partial_-^{r-3})
  \end{displaymath}
for all $r \in \ZZ_+$, $r \geq 3$.
\end{enumerate}
\end{Corollary}

There are a few more singular vectors. One is the vector
\begin{equation}
  \label{eq:2.10}
  w_1=d^+_{123} \Delta^- 1
\end{equation}
of the $E(3,6)$-module $M(0,1;0;y_D)$.  It is straightforward to
show by checking that $e'_0 w_1=0$ and $e_iw_1=0$ for $i=0, \ldots ,3$.
This singular vector generates a non-zero homomorphism of
$E(3,6)$-modules $\triangledown'_4: M(0,0;2;y_A) \to M(0,1;0;y_D)$
such that for the highest weight vector $z^2_+$ of $M(0,0;2;y_A)$
we have
\begin{equation}
  \label{eq:2.11}
  \triangledown'_4 (z^2_+) =d^+_{123} \Delta^- 1 \, .
\end{equation}
We shall be looking for $\triangledown'_4$ in the form:
\begin{equation}
  \label{eq:2.12}
  \triangledown'_4 = a \Delta^- \partial^2_+
    + b \Delta^- \partial_+ \partial_-
    + c \Delta^- \partial^2 _-\, , \quad a=d^+_{1}d^+_{2}d^+_{3}
%d^+_{123}
\,,
\end{equation}
where $b,c \in U(L_-)$.  Here all three operators $\partial^2_+$,
$\partial_+ \partial_-$ and $\partial^2_-$ map $V^{0;2}_A$, which
is the space of quadratic polynomials in $z_{\pm}$, to $\CC$,
hence operator (\ref{eq:2.12}) may be viewed as a linear map from
$M(0,0;2;y_A)$ to $M(0,1;0;y_D)$.  It is clear that, applied to the
highest weight vector $\tfrac{1}{2} z^2_+$ of $M(0,0;2;y_A)$ the
operator (\ref{eq:2.12}) gives the singular vector $w$.  Hence,
due to Remark~\ref{rem:2.4}b, it suffices to check that
$\triangledown'_4$ commutes with $f_1,f_2,f_3$.  Commuting with
$f_3$ gives the relations:  $[f_3,a] =b$, $[f_3,b]=2c$,
which gives:
\begin{equation}
  \label{eq:2.13}
  b= d^-_1 d^+_2 d^+_3 + d^+_1 d^-_2 d^+_3
       + d^+_1 d^+_2 d^-_3 \, ,\,\,\,
 c= d^-_1 d^-_2 d^+_3 + d^-_1 d^+_2 d^-_3
       + d^+_1 d^-_2 d^-_3 \, .
\end{equation}
With these $b$ and $c$, the operator $\triangledown'_4$ given by
(\ref{eq:2.12}) clearly commutes with $s\ell (3)$.
Thus, $\triangledown'_4 : M(0,0;2;-2) \to M(0,1;0;-\tfrac{2}{3})$
defined by
(\ref{eq:2.12}) and (\ref{eq:2.13}) is a non-zero homomorphism of
$E(3,6)$-modules.

Another singular vector is the vector
\begin{equation}
  \label{eq:2.10n}
  w_2=d^-_{1}(a \partial^2_+
    + b  \partial_+ \partial_-
    + c \partial^2 _-)
\end{equation}
where $a,b,c$ are the same as in (\ref{eq:2.12}), (\ref{eq:2.13}),
and  $w_2 \in M(0,0;2;y_D)$. Similarly the checking is
straightforward although somewhat tedious. Again
this singular vector generates a non-zero homomorphism of
$E(3,6)$-modules $\triangledown''_4: M(1,0;0;y_A) \to M(0,0;2;y_D)$
such that for the highest weight vector $x_1$ of $M(1,0;0;y_A)$
we have $\triangledown''_4 x_1=w_2$. This allows us to find
that
\begin{equation}
  \label{eq:2.12n}
\bAr{rl}
  \triangledown''_4 = &
 \quad \! d^-_1( a \partial^2_+
               + b \partial_+ \partial_-
               + c \partial^2 _-)\dpind{1} \\
&
      + d^-_2( a \partial^2_+
             + b \partial_+ \partial_-
             + c \partial^2 _-)\dpind{2} \\
&
      + d^-_3( a \partial^2_+
             + b \partial_+ \partial_-
             + c \partial^2 _-)\dpind{3} \,.
\eAr
\end{equation}
Clearly $\triangledown''_4 $ commutes with  $f_1,f_2,f_3$ and
Remark~\ref{rem:2.4}b shows that $\triangledown''_4 $ determines
the morphism of
$E(3,6)$-modules $\triangledown''_4: M(1,0;0;\tfrac{2}{3}) \to M(0,0;2;2)$.

%*%last-sing-vector
The most complicated singular vector is $w_3\in M(0,0;1;y_D)$, given by
\begin{equation}
  \label{eq:2.10nn}
\bAr{l}
w_3=
\left(
\dst^{-}_{123}\dst^{+}_{123}
+(\dst^{-}_{13}\dst^{+}_{12}-\dst^{-}_{12}\dst^{+}_{13})\dhind1 +
(\dst^{-}_{21}\dst^{+}_{23}-\dst^{-}_{23}\dst^{+}_{21})\dhind2 +
\right.\\
\qquad\,\,\,
\,+
(\dst^{-}_{32}\dst^{+}_{31}-\dst^{-}_{31}\dst^{+}_{32})\dhind3 -
\dst^{-}_{1}\dst^{+}_{1}\hat{\partial}_1^2 -
\dst^{-}_{2}\dst^{+}_{2}\hat{\partial}_2^2-
\dst^{-}_{3}\dst^{+}_{3}\hat{\partial}_3^2-
\\
\qquad\,\,\,
\,
-(\dst^{-}_{1}\dst^{+}_{2}+\dst^{-}_{2}\dst^{+}_{1})\dhind1\dhind2 -
(\dst^{-}_{1}\dst^{+}_{3}+\dst^{-}_{3}\dst^{+}_{1})\dhind1\dhind3 -
\left.(\dst^{-}_{2}\dst^{+}_{3}+\dst^{-}_{3}\dst^{+}_{2})\dhind2\dhind3
\right)\dpind{-}
+
\\
\qquad\qquad\qquad
+\left(
\dst^{-}_{1}\dst^{+}_{123}\dhind1 +
\dst^{-}_{2}\dst^{+}_{123}\dhind2 +
\dst^{-}_{3}\dst^{+}_{123}\dhind3
\right)\dpind{+}=\\
\quad
=
%\left(\,
\dst^{-}_{123} a \dpind{-}
+\left(
\dst^{-}_{1} \,\dhind1+
\dst^{-}_{2} \,\dhind2+
\dst^{-}_{3} \,\dhind3\right) (b\dpind{-}+a\dpind{+})\,.
%+\left(
%\dst^{-}_{1}a\,\dhind1 +
%\dst^{-}_{2}a\,\dhind2 +
%\dst^{-}_{3}a\,\dhind3
%\right)\dpind{+}\,.
\eAr
\end{equation}
This singular vector generates a homomorphism of $E(3,6)$-modules
\[
\tdn_{\!6}: M(0,0;1;y_A) \ar M(0,0;1;y_D)
\]
such that
$\tdn_{\!6} \,z_+=w_3$. In (\ref{eq:2.10nn}) we keep the
notations from (\ref{eq:2.12}), (\ref{eq:2.13}) and
denote by $ \dhind i$ elements $\dpind i \in \gind{-2}$
($i=1,2,3$).

It is straightforward to check that
$\tdn w_3=0$, hence
\begin{equation}
\label{eq:tdn6}
\tdn\cdot\tdn_{\!6}=0\,.
\end{equation}
In Proposition~\ref{prop:5.25} of Section 5
we will write $\tdn_6$ explicitely
and show
that $\tdn_{\!6}\cdot\tdn=0$ as well.
%*%last-s-v

%xx
\begin{remark}
\label{rem:2.9}
  The operators
$\triangledown$, $\triangledown_2$, $\triangledown_3$,
$\triangledown'_4$,  $\triangledown''_4$ and $\tdn_{\!6}$
   have degree $1$,  $2$,  $3$,  $4$, $4$  and $6$ respectively,
with respect to the
  $\ZZ$-gradation of $U(L_-)$ induced by the consistent gradation
  of $E(3,6)$.
\end{remark}

The maps $\triangledown$, $\triangledown_2$, $\triangledown_3$,
$\triangledown'_4$ , $\triangledown''_4$ and $\tdn_6$
are illustrated by Figure~1. The black nods in quadrants $A$, $B$,
$C$, $D$
represent
generalized Verma modules
$M(p,0;r;y_X)$, $X=A,B$ or $M(0,q;r;y_X)$, $X=C,D$.
The plain arrows
represent $\triangledown$, the dotted arrows represent
$\triangledown_2$, the interrupted arrows represent
$\triangledown_3$ and the bold arrows represent
$\triangledown'_4$, $\triangledown''_4$ and $\tdn_6$.\\

\begin{figure}[htbp]
  \begin{center}
    \leavevmode
  \setlength{\unitlength}{0.25in}
%%%%%%%%%%%%%%%%%%%%%%%%%%%%%%%%%%%%%%%%%%%%%%%   corner labels
%\begin{picture}(20,19)
\begin{picture}(21,19)

\put(18.7,17.5){A}
\put(18.7,-1){B}

\put(1,17.5){C}
\put(1,-1){D}

%%%%%%%%%%%%%%%%%%%%%%%%%%%%%%%%%%%%%%%%%%%%%%%   vertical lines
\put(8.5,17.5){\line(0,-1){7} }%5 C-vertical line
%\put(10,17.5){\line(0,-1){18.5} }%6 full vertical line
\put(11.5,17.5){\line(0,-1){8.5} }%6 A-vertical line
\put(8.4,18){r}
\put(11.4,18){r}

\put(10,-.5){\line(0,1){9.5} }% D-vertical line
\put(9.9,-1){r}

\put(13,-.5){\line(0,1){8} }% B-vertical line
%\put(11.5,-1){\line(0,1){8.5} }%7 partial vertical line
\put(12.9,-1){r}

%%%%%%%%%%%%%%%%%%%%%%%%%%%%%%%%%%%%%%%%%%%%%%%   horizontal lines
\put(13,7.5){\line(1,0){5} }%5 B-horizontal line
\put(18.5,7.4){p}

\put(1.8,9){\line(1,0){8} }%6 full horizontal line
\put(11.5,9){\line(1,0){6.5} }%6 full horizontal line
\put(1,8.9){q}
\put(18.5,8.9){p}

\put(1.8,10.5){\line(1,0){6.6} }%7 C-horizontal line
\put(1,10.4){q}

%%%%%%%%%%%%%%%%%%%%%%%%%%%%%%%%%%%%    dots set up in vertical rows
%\multiput(1,0)(0,1.5){12}{\circle*{.25} }%
\multiput(2.5,0)(0,1.5){12}{\circle*{.25} }%1
\multiput(4,0)(0,1.5){12}{\circle*{.25} }%2
\multiput(5.5,0)(0,1.5){12}{\circle*{.25} }%3
\multiput(7,0)(0,1.5){12}{\circle*{.25} }%4
\multiput(8.5,0)(0,1.5){12}{\circle*{.25} }%5
\multiput(10,0)(0,1.5){7}{\circle*{.25} }%6
%\multiput(10,0)(0,1.5){12}{\circle*{.25} }%
\multiput(11.5,9)(0,1.5){6}{\circle*{.25} }%7
\multiput(13,0)(0,1.5){12}{\circle*{.25} }%8
\multiput(14.5,0)(0,1.5){12}{\circle*{.25} }%9
\multiput(16,0)(0,1.5){12}{\circle*{.25} }%10
\multiput(17.5,0)(0,1.5){12}{\circle*{.25} }%11

%%%%%%%%%%%%%%%%%%%%%%%%%%%%%%%%%%%%    arrows

%%%\put(4,16.5){\vector(-1,-1){1.35}}
%\multiput(4,16.5)(1.5,0){10}{\vector(-1,-1){1.35}}%%1
%\multiput(4,15)(1.5,0){10}{\vector(-1,-1){1.35}}%%2
%\multiput(4,13.5)(1.5,0){10}{\vector(-1,-1){1.35}}%%3
%\multiput(4,12)(1.5,0){10}{\vector(-1,-1){1.35}}%%4
%\multiput(4,10.5)(1.5,0){10}{\vector(-1,-1){1.35}}%%5
%\multiput(4,9)(1.5,0){10}{\vector(-1,-1){1.35}}%%6
%\multiput(4,7.5)(1.5,0){10}{\vector(-1,-1){1.35}}%%7
%\multiput(4,6)(1.5,0){10}{\vector(-1,-1){1.35}}%%8
%\multiput(4,4.5)(1.5,0){10}{\vector(-1,-1){1.35}}%%9
%\multiput(4,3)(1.5,0){10}{\vector(-1,-1){1.35}}%%10
%\multiput(4,1.5)(1.5,0){10}{\vector(-1,-1){1.35}}%%11

%%%%%%%%%%%%%%%%%%%%%%%%%%%%%%%%%%%%%%%%%%   endings

\multiput(1.95,-.55)(0,1.5){7}{\vector(+1,+1){.65}}%1v-ends
\multiput(1.95,11.45)(0,1.5){4}{\vector(+1,+1){.65}}%1v-ends

\multiput(3.45,-.55)(1.5,0){5}{\vector(+1,+1){.65}}%1h-ends
\multiput(13.95,-.55)(1.5,0){3}{\vector(+1,+1){.65}}%1h-ends

\multiput(18.2,.7)(0,1.5){12}{\vector(-1,-1){.55}}%v-ends
\multiput(3.2,17.2)(1.5,0){5}{\vector(-1,-1){.55}}%h-ends
\multiput(12.2,17.2)(1.5,0){5}{\vector(-1,-1){.55}}%h-ends

%%%%%%%%%%%%%%%%%%%%%%%%%%%%%%%%%%%%%%%%%%   dashed lines

\dashline[-10]{.2}(10.3,16.8)(8.5,15)
       \put(8.8,15.35){\vector(-1,-1){.25}}

\multiput(4,16.5)(1.5,0){4}{\vector(-1,-1){1.35}}%%1
\dashline[-10]{.2}(11.5,16.5)(8.5,13.5)
       \put(8.8,13.85){\vector(-1,-1){.25}}
\multiput(13,16.5)(1.5,0){4}{\vector(-1,-1){1.35}}%%1

\multiput(4,15)(1.5,0){4}{\vector(-1,-1){1.35}}%%2
\dashline[-10]{.2}(11.5,15)(8.5,12)%%2
       \put(8.8,12.35){\vector(-1,-1){.25}}
\multiput(13,15)(1.5,0){4}{\vector(-1,-1){1.35}}%%2

\multiput(4,13.5)(1.5,0){4}{\vector(-1,-1){1.35}}%%3
\dashline[-10]{.2}(11.5,13.5)(8.5,10.5)%%3
       \put(8.8,10.85){\vector(-1,-1){.25}}
\multiput(13,13.5)(1.5,0){4}{\vector(-1,-1){1.35}}%%3

\multiput(4,12)(1.5,0){4}{\vector(-1,-1){1.35}}%%4

\thicklines
\put(11.5,12){\vector(-1,-1){2.81}}
\put(13,9){\vector(-1,-1){2.81}}

%*!*?
\put(11.49,10.49){\vector(-1,-2){1.44}}
\put(11.51,10.51){\line(-1,-2){1.41}}
\put(11.47,10.47){\line(-1,-2){1.41}}

\thinlines
\multiput(13,12)(1.5,0){4}{\vector(-1,-1){1.35}}%%4

\dashline[-10]{.1}(4,10.5)(2.5,9)%%5
       \put(7.33,9.37){\vector(-1,-1){.25}}
\dashline[-10]{.1}(5.5,10.5)(4,9)%%5
       \put(5.8,9.37){\vector(-1,-1){.25}}
\dashline[-10]{.1}(7,10.5)(5.5,9)%%5
       \put(4.3,9.37){\vector(-1,-1){.25}}
\dashline[-10]{.1}(8.5,10.5)(7,9)%%5
       \put(2.9,9.35){\vector(-1,-1){.25}}

\multiput(13,10.5)(1.5,0){4}{\vector(-1,-1){1.35}}%%5

\multiput(4,9)(1.5,0){5}{\vector(-1,-1){1.35}}%%6
%\dashline[-10]{.1}(11.5,9)(10,7.5)%%6
%\dashline[-10]{.1}(13,9)(11.5,7.5)%%6
\dashline[-10]{.1}(14.5,9)(13,7.5)%%6
\dashline[-10]{.1}(16,9)(14.5,7.5)%%6
\dashline[-10]{.1}(17.5,9)(16,7.5)
\dashline[-10]{.1}(2.25,10.25)(2,10)

%%%%%%%%%%%%%%%%%%%%%%%%%%%%%%%%arrow heads for dashed lines 7
%     \put(11.9,7.9){\vector(-1,-1){.25}}
      \put(13.4,7.9){\vector(-1,-1){.25}}
      \put(14.9,7.9){\vector(-1,-1){.25}}
      \put(16.4,7.9){\vector(-1,-1){.25}}
%     \put(17.9,7.9){\vector(-1,-1){.25}}
%new
\multiput(4,7.5)(1.5,0){5}{\vector(-1,-1){1.35}}%%7
\dashline[-10]{.2}(13,7.5)(10,4.5)%%7
      \put(10.4,4.9){\vector(-1,-1){.25}}
\multiput(14.5,7.5)(1.5,0){3}{\vector(-1,-1){1.35}}%%7

\multiput(4,6)(1.5,0){5}{\vector(-1,-1){1.35}}%%8
\dashline[-10]{.2}(13,6)(10,3)%%8
      \put(10.4,3.4){\vector(-1,-1){.25}}
\multiput(14.5,6)(1.5,0){3}{\vector(-1,-1){1.35}}%%8

\multiput(4,4.5)(1.5,0){5}{\vector(-1,-1){1.35}}%%9
\dashline[-10]{.2}(13,4.5)(10,1.5)%%9
      \put(10.4,1.9){\vector(-1,-1){.25}}
\multiput(14.5,4.5)(1.5,0){3}{\vector(-1,-1){1.35}}%%9

\multiput(4,3)(1.5,0){5}{\vector(-1,-1){1.35}}%%10
\dashline[-10]{.2}(13,3)(10,0)%%10
     \put(10.4,.4){\vector(-1,-1){.25}}
\multiput(14.5,3)(1.5,0){3}{\vector(-1,-1){1.35}}%%10

\multiput(4,1.5)(1.5,0){3}{\vector(-1,-1){1.35}}%%11

\put(8.5,1.5){\vector(-1,-1){1.35}}%%5
\put(10,1.5){\vector(-1,-1){1.35}}%%5

\dashline[-10]{.2}(13,1.5)(11,-.5)%%end
\dashline[-10]{.2}(13,0)(12.5,-.5)%%end

\multiput(14.5,1.5)(1.5,0){2}{\vector(-1,-1){1.35}}%%11

\put(16,1.5){\vector(-1,-1){1.35}}%%5

\put(17.5,1.5){\vector(-1,-1){1.35}}%%5

\put(17.5,1.5){\vector(-1,-1){1.35}}%%5

\end{picture}
\vspace{3ex}
    \caption{}
    \label{fig:1}
  \end{center}
\end{figure}

There is still one more singular vector---the vector
\begin{equation}
  \label{eq:2.14}
  w_4=d^+_1 (\Delta^+ \partial_+ + \Delta^- \partial_-)
\end{equation}
of the $E(3,6)$-module $M(0,1; 1; y_D)$.  It occurs as the image
of a non-trivial singular vector of $M (0,0; 0; 0)$ under the map
$\triangledown : M(0,0;0;0) \to M (0,1;1;y_D)$ (cf.~Corollary~2.5a
for $p=r=0$ and 2.5d for $q=r=1$).  The reason for its existence
is the fact that the composition of the following maps is
non-zero:
\begin{displaymath}
  M (1,0;1;y_A) \overset{\triangledown}{\longrightarrow}
  M (0,0;0;0) \overset{\triangledown}{\longrightarrow}
  M (0,1;1;y_D) \, .
\end{displaymath}

The main result of \cite{KR2} is the following theorem.
\begin{Theorem}
  \label{th:2.10}
A complete list of non-trivial singular vectors (up to a constant
factor) of the $E (3,6)$-modules $M(p,q;r;y)$ is given by
Corollary~\ref{cor:2.5},
Corollary~\ref{cor:2.8}, (\ref{eq:2.10}), (\ref{eq:2.10n}), (\ref{eq:2.10nn})
and~(\ref{eq:2.14}).
\end{Theorem}
\bCo
\label{cor:2.11}
All degenerate modules $M(p,q;r;y)$, except for $M(0,1;1;\tfrac{2}{3})$
(of type D),
have a unique (up to a constant factor) non-trivial singular vector.
The module $M(0,1;1;\tfrac{2}{3})$ has two such vectors.
\eCo

%%%%%%%%%%%%%%%%%%%%%%%%%%%%%%%%%% end of sec 2 %%%
%s3
\section{Homology of complexes $(M_X,\triangledown).$}
\label{sec:3}

Recall that the complexes
\begin{displaymath}
  M_X =U (L_-) \otimes V_X
\end{displaymath}
are $\ZZ$-bigraded:
\begin{equation}
  \label{eq:3.1}
  M_X = \oplus_{m,n \in \ZZ} M^{m,n}_X , \,\,\,\,\,
  M^{m,n}_X = U(L_-) \otimes V^{m,n}_X \, ,
\end{equation}
where $V^{m,n}_X =0$ if $X=A$, $m,n<0$;
if $X=B$, $m<0$, $n>0$;
if $X=C$, $m>0$, $n<0$; and
if $X=D$, $m>0$, $n>0$, and $V^{m,n}_X$ are
irreducible $\fg_0$-modules described by (\ref{eq:2.2})
otherwise.

Note that
\begin{equation}
  \label{eq:3.2}
  \triangledown : M^{m,n}_X \to M^{m-1,n-1}_X \, .
\end{equation}
Due to (\ref{eq:3.2}) the $\ZZ$-bigrading (\ref{eq:3.1}) of $M_X$
induces a $\ZZ$-bigrading on its homology:
\begin{equation}
  \label{eq:3.3}
  H(M_X) = \oplus_{m,n} H^{m,n} (M_X) \, .
\end{equation}

Note also that $M_X$ is a free $\Sym(\fg_{-2})$-module:
$M_X\simeq\Sym (\fg_{-2})\otimes (\Lambda (\fg_{-1}) \otimes V_X)$ and that
$\triangledown$ commutes with $\Sym(\fg_{-2})$.  Hence $H(M_X)$ and
each $H^{m,n} (M_X)$ are $\Sym(\fg_{-2})$-modules as well.

The canonical filtration of $U(L_-)$:
\begin{displaymath}
  \CC =F_{0} U(L_-) \subset ... \subset F_i U(L_-)
  = L_- F_{i-1} U(L_-) + F_{i-1} U(L_-) \subset \ldots
\end{displaymath}
induces a filtration of $M_X$ by letting
\begin{equation}
\label{eq:3.3n}
  F_i M_X = F_i U(L_-) \otimes V_X.
\end{equation}
Moreover $\triangledown (F_iM_X) \subset F_{i+1}M_X$, so that $M_X$
becomes a filtered complex with the differential $\tdn$ and the bigrading
(\ref{eq:3.1}) and the filtration is bounded below.
% The same properties hold for
% $M_{X^0}$ with the naturally induced filtration.

As we discuss in Appendix, one can use a spectral sequence to study
the homology for such a complex and the spectral sequence converges
when the filtration is bounded below.
This applied to $(M_X, \triangledown)$ produces a sequence of
complexes $\{(E^i,\triangledown^i)\}$, such that $E^{i+1}$ is the
homology of ${(E^i, \triangledown^i)}$, $\lim_{i \to \infty} E^i
=\Gr (H(M_X))$, and $E^0 = H(\Gr M_X)$.

%ww
\begin{remark}
\label{rem:W}
In the subalgebra $W\simeq W_3$
described in (\ref{eq:Wsub})  consider
the standard filtration of $W_3$ :
  $\,W=L^W_{-1}\supset L^W_0 \supset \ldots\,$,
then
\[ L^W_j \cdot F_i M_X \subset F_{i-j}\,M_X\,.
\]
Therefore the action of $W$ on $M_X$ descends to
the action of $W$ on $\Gr M_X$, because  $W\simeq \Gr W$.
The actions commute with $\tdn$,
and thus $W$ acts also on the spectral sequence and homologies.
\end{remark}

To get hold on $H(\Gr M_X)$ we notice that,
by the PBW theorem, the associated graded algebra $\Gr\, U(L_-)
\simeq \Sym (\fg_{-2}) \otimes \Lambda (\fg_{-1})$ (tensor product
of associative algebras).  Then we get the associated graded complex
\begin{displaymath}
  \Gr M_X = \Gr U(L_-) \otimes V_X
\simeq (\Sym (\fg_{-2}) \otimes \Lambda (\fg_{-1})) \otimes V_X\,.
\end{displaymath}
Clearly $L^W_1$ annihilates $\Lambda (\fg_{-1}) \otimes V_X$ and
the above isomorpism can be interpreted as the following
isomorphism of $W$-modules
\begin{equation}
\label{eq:Wisom}
\Gr M_X \simeq
\Sym (\fg_{-2}) \otimes (\Lambda (\fg_{-1}) \otimes V_X)\simeq
\Ind^W_{L_0^W}(\Lambda (\fg_{-1}) \otimes V_X)\,.
\end{equation}

The differential of the complex $\Gr M_X$ we denote again by
$\triangledown$,
because it is given by the same formula as for $M_X$, except
that the multiplication by $d^{\pm}_i$ is to be taken in $\Gr U(L_-)$
instead of the multiplication in $U(L_-)$.

It follows that $G_X=\Lambda (\fg_{-1}) \otimes V_X$ is the subcomplex
of the complex $(\Gr M_X, \triangledown)$, and that the latter is
obtained from the former by extending coefficients from $\CC$ to
$\Sym=\Sym(\fg_{-2})$. Homologies $H^{m,n} (G_X)$, that
are computed
with the differential $\triangledown $ restricted from $\Gr M_X$,
are also annihilated
by $L^W_1$,
 so we have an isomorphism of $W$-modules (and
$\fg_0$-modules):
\begin{equation}
  \label{eq:3.4}
  H^{m,n} (\Gr M_X) \simeq \Sym
      \otimes_{\CC} H^{m,n} (G_X)
     \simeq
             \Ind^W_{L_0^W}(H^{m,n} (G_X) )\,,
\end{equation}

Thus, (\ref{eq:3.4}) and the theory of spectral sequence give us
the following result:

\begin{Proposition}
  \label{prop:3.1}
 $\,H^{m,n}(G_X)=0\,
\Longrightarrow
 \,H^{m,n}(\Gr M_X)=0\,
\Longrightarrow\,
 H^{m,n}(M_X)=0$\,, \\ and \quad
$\rank_{\Sym} H^{m,n} (M_X)
\leq \rank_{\Sym} H^{m,n} (\Gr M_X) =
\dim_{\CC} H^{m,n} (G_X)$.
\end{Proposition}

Figure 1 and Propositions 2.6, 2.7 make it reasonable to
consider complexes not restricted to the quadrants of the figure.
In the following we make the first step in this direction
(see Figure~\ref{fig:2} several pages below).

Let $M_{AB}$ be the module $M_{A}\oplus M_{B}$ with the bigrading
and the filtration induced from the summands
provided with the differential
$
%\tilde
{\triangledown}$ that coincides with $\triangledown_2$
on $M_{A'}\subset M_{A}\oplus M_{B}$ and with $\triangledown$
on the bigraded components of $M_{A}\oplus M_{B}$ that do not
belong to $M_{A'}$.

Let us define  $M_{CD}$ and also
$G_{AB},\,G_{CD}$
in the same way as the sum of the spaces
with the differential constructed from $\triangledown_2$ and $\triangledown$.

Clearly  $M_{AB}$ and $M_{CD}$ are filtered modules with the differential
and the bigrading introduced above,
and we can use the spectral sequence to study their homology.
Moreover
the isomorphism (\ref{eq:3.4}) holds
   for $ X=AB,\,CD$
%generalizes to
%
%\begin{equation}
%  \label{eq:3.5}
%  H^{m,n} (\Gr M_X) \simeq \Sym
%      \otimes_{\CC} H^{m,n} (G_X) \text{ for } X=AB,\,CD  \, ,
%\end{equation}
%
and Proposition~\ref{prop:3.1} remains valid for $X=AB,\,CD$. This
and the calculations of the homology $H^{m,n}(G_{X})$ made in the next section
provide us with the following proposition.

\begin{Proposition}
  \label{prop:3.2}
 $ H^{m,n}(M_{AB})=H^{m,n}(G_{\!AB})=0$ when
$m\geq2$ or $m=1,\,n\neq 0,1,2$,\\
 $ H^{m,n}(M_{CD})=H^{m,n}(G_{CD})=0$ when $m\leq-2$ or $m=-1,\,n\neq 0,-1,-2$.
\end{Proposition}

%%%%%%%%%%%%%%%%%%%%%%%%% end of sec 3 %%%%
%s4

\section{Homology of $G_X.$}
\label{sec:4}

Let us notice that
along with natural inclusions $V_{X'} \subset V_X$
there are natural projections
$V_{X} \ar V_{X'}$ defined by  substituting
%zeros
 $z_{\pm}$ and $\dprt_{\pm}$
by zero.
One has the corresponding projections
$G_{X} \ar G_{X'}$.

We will consider the compositions
$G_{X} \ar G_{X'} \ar G_{Y'}$
where the projection is the first map and $\triangledown_2 $ is
the second one, as well as compositions
$G_{X'} \ar G_{Y'} \ar G_Y$
where the first map is $\triangledown_2 $ and the second is
the natural inclusion. These compositions
are  morphisms of $E(3,6)$-modules,
and we allow ourselves
to denote them by $\triangledown_2 $ as well.

Let
\begin{equation}
\label{eq:AB0}
\begin{array}{ll}
   G_{A^o}=\Ker(\triangledown_2 :  G_{A} \ar   G_{B'})\,,&\,\,\,
   G_{B^o}= \Coker(\triangledown_2 :  G_{A'} \ar   G_{B})\,, \\
\end{array}
\end{equation}
\begin{equation}
\label{eq:CD0}
\begin{array}{ll}
   G_{C^o}=\Ker(\triangledown_2 :  G_{C} \ar   G_{D'})\,,&\,\,\,
   G_{D^o}= \Coker(\triangledown_2 :  G_{C'} \ar   G_{D})\,.
\end{array}
\end{equation}

It follows from Proposition~\ref{prop:2.6}a that the differential $\triangledown$
is defined for $G_{X^o}$. It is clear from the above definitions that
\begin{equation}
\label{eq:H-AB}
H^{m,n}(G_{\!AB})=\left\{
\begin{array}{ll}
H^{m,n}(G_{A^o}) &\text{ for } n> 0,\\
H^{m,n}(G_{B^o}) &\text{ for } n <0,\\
H^{m,0}(G_{A^o})\oplus H^{m,0}(G_{B^o})  &\text{ for } n= 0;
\end{array}\right.
\end{equation}
\begin{equation}
\label{eq:H-CD}
H^{m,n}(G_{CD})=\left\{
\begin{array}{ll}
H^{m,n}(G_{C^o}) &\text{ for } n> 0,\\
H^{m,n}(G_{D^o}) &\text{ for } n <0,\\
H^{m,0}(G_{C^o})\oplus H^{m,0}(G_{D^o})  &\text{ for } n= 0.
\end{array}\right.
\end{equation}
This means that the computation of the homologies $H^{m,n}(G_{\!AB})$,
$H^{m,n}(G_{CD})$ reduces to finding $H^{m,n}(G_{X^o})$.

In the computations we need to go from $G_X$ to $G_{X^o}$ and the
following lemma is helpful.
\begin{Lemma}
\label{lem:4.1}
Let $(M,d)$ be a differential complex
\[
M_0 \leftarrow M_1 \leftarrow M_2 \leftarrow \cdots,
\]
let $(N,d)$ be another differential complex of the same type and let
$\al:M\ar N$ be a morphism of complexes.
%\alphaparenlist
  \begin{enumerate}
  \item %%a
Suppose $M^o=\Ker \al$ and $N$ is concentrated at 0
(i.e., $N_i=0$ for $i\neq 0$).
Then $H_n(M)\simeq H_n(M^o)$ for $n\neq 0$ and
there is an
exact sequence:
\[ 0\ar H_0(M^o) \ar H_0(M) \ar H_0(N)  \,.\]
  \item %%b
Suppose $N^o=\Coker \al$ and $M$ is concentrated at 0
%$M_i=0$ for $i\neq 0$
. Then
$H_n(N)\simeq H_n(N^o)$ for $n\neq 0$ and
there is
an exact sequence:
\[ H_n(M) \ar H_n(N) \ar H_n(N^o) \ar 0 \,.\]
  \end{enumerate}
\end{Lemma}
The statement follows immediately from definitions.

We can apply this lemma to connect the
$\triangledown$-homology of $G_{X^o}$ and $G_X$ since one can check,
using the definitions (\ref{eq:AB0}), (\ref{eq:CD0}),
that the conditions of the lemma are valid. \\

In the rest of the section we also consider the following
$\ZZ$-bigrading of
modules $G_X$. Let for $X=A,B,C,D$
\begin{equation}
\label{eq:4.5}
\begin{array}{l}
(V_X)_{[p,q]}=\{f\in V_X \,| \,(z_+\dprt_+)f=pf,\, (z_-\dprt_-)f=qf\,\}\,,
\text{ and }\\
(G_X)_{[p,q]}=\La( \gind{-1})\otimes (V_X)_{[p,q]},
\end{array}
\end{equation}
It is important to mention that $(G_X^{m,n})_{[p,q]}\neq 0$ only if
$p+q=n$.

We hope that the notations allow one to distinguish what
grading is refered to.
The new bigrading naturally descends to $G_{X^o}$.

The definitions of $\De^{\pm}$, Lemma~\ref{lem:2.1}c and the formula
\begin{displaymath}
  \triangledown = \Delta^+ \partial_+ + \Delta^- \partial_-
\end{displaymath}
allow us to conclude that the bigraded modules $G_X$,
$G_{X^o}$ provided with differentials
$\dst'=\Delta^+ \partial_+$, $\dst''=\Delta^- \partial_-$
become  bicomplexes and
their homologies with respect to $\triangledown$ (that we are interested in) are
nothing but
the total homoligies of the bicomplexes. So the classical theory of
spectral sequences of a bicomplex is relevant here
([ML, Chapter XI, section 6]) and
the following lemma contains
well-known statements about two spectral sequences of a bicomplex
that we will use.
\begin{Lemma}
\label{lem:4.2}
Let $(K,\dst',\dst'')$ be a bicomplex, $K=\sum_{p,q} K_{[p,q]}$,
and $\dst=\dst'+\dst''$ the total
differential of $K$.

The first spectral sequence of the bicomplex
$E'=\{({E'}^r,d^r)\},\,\, {E'}^r=\sum_{p,q} {E'}^r_{[p,q]}$  has the property:
\[
({E'}^0,\,d^0)\simeq (K,\,\dst''),\quad ({E'}^1,\,d^1)\simeq (H(K,\dst''),\,\dst'),
 \text{ so that }  \,\,
{E'}^2_{[p,q]}\simeq H_p(H_q(K,\dst''),\,\dst').
\]
For the second spectral sequence
$E''=\{({E''}^r,d^r)\},\,\, {E''}^r=\sum_{p,q} {E''}^r_{[p,q]}$
the roles of $\dst',\dst''$
are reversed:
\[
({E''}^0,\,d^0)\simeq (K,\,\dst'),\quad ({E''}^1,\,d^1)\simeq (H(K,\dst'),\,\dst''),
 \text{ so that }\,\,
  {E''}^2_{[p,q]}\simeq H_p(H_q(K,\dst'),\,\dst'').
\]
The spectral sequences are functors on the bicomplexes.

Any of the above spectral sequences $E$ converges to the homology of $K$ with
respect to the total differential $\dst$
whenever for every $n$ the set
\[
\{(p,q)\,|\,p+q=n,\,\,E^2_{[p,q]}\neq 0\}
\]
is finite.
\end{Lemma}
The above condition for the convegence could be relaxed but this form
is enough for our purposes. It is farely traditional although slightly
differs from the one in Theorem 6.1 of [ML, Chapter XI]. The
arguments of Proposition 3.2 of [ML, Chapter XI] are easily modified
to prove the convegence under our condition. We leave details to the reader. \\

%Instead of applying Lemma~\ref{lem:4.2} directly to
Now we decompose bicomplexes $G_{X}$ ( and their sub- or
quotient-bicomplexes $G_{X^o}$ )
into a sum of smaller bicomplexes.

%..!
Introduce notations:
\[
\La_i^{\pm}=\La^i\lsp{d_1^{\pm},d_3^{\pm},d_3^{\pm}}, \,\,
\] and ( $0\leq i,j\leq 3 ):$
\[
\begin{array}{cc}
\La_i^+\La_j^-[x]=\La_i^+\La_j^-\otimes_\CC \CC[x_1,x_2,x_3],&
\La_i^+\La_j^-[\dprt]=\La_i^+\La_j^-\otimes_\CC \CC[\dprt_1,\dprt_2,\dprt_3].
\end{array}
\]
Let
\begin{equation}
\label{eq:4.9}
\begin{array}{ll}
G_A (a,b)_{[p,q]}=\La^+_{a-p} \La^-_{b-q}  [x]     z_+^{p}    z_-^{q}\,, \,\,\,&
G_C (a,b)_{[p,q]}=\La^+_{a-p}\La^-_{b-q} [\dprt]  z_+^{p}    z_-^{q}, \\
 & \\
G_B (a,b)_{[p,q]}=\La^+_{a-p} \La^-_{b-q} [x] \,{\dprt}_+^{-p}{\dprt}_-^{-q}\,
,\,\,&
G_D (a,b)_{[p,q]}=\La^+_{a-p}\La^-_{b-q}[\dprt]\,\dprt_+^{-p}\dprt_-^{-q}.
\end{array}
\end{equation}
Then $G_X$ decomposes in a direct sum of subcomplexes:
\begin{equation}
\label{eq:4.7}
G_X=\oplus_{a,b} G_X (a,b)\,,
\text{ where }
%\left\{
%\begin{array}{lll}
%a\geq 0,b\geq 0 & \text{for} & X=A,\, A^o,\,C,\,C^o,\\
%a\leq 3,b\leq 3 & \text{for} & X=B,\,B^o,\,D,\, D^o.
%\end{array}
%\right.
%\end{equation}
%and at the same time
%\begin{equation}
%\label{eq:4.8}
%
     G_X(a,b)=\bigoplus_{p,q} G_X (a,b)_{[p,q]}\,.
%
%\text{ where }
%\left\{
%\right.
\end{equation}
We have the induced decomposition
$G_{X^o}=\oplus_{a,b} G_{X^o} (a,b)\,.\,$

Note that
the above equalities are isomorphisms of $s\ell (3)$-modules but not those
of $s\ell (2)$-modules, since
$G_X (a,b)=\{f\in V_X \,| \,(z_+\dprt_+)f=af,\, (z_-\dprt_-)f=bf\,\}$
with the obvious action of $z_\pm\dprt_\pm$ on $G_X$.

%
%\begin{array}{cll}
%
%a\geq  p \geq \max(0,a-3)\,,
% \,\, b\geq q \geq \max(0, b-3), & \text{ for } & X=A,\, A^o,\,C,\,C^o\\
%
%\min(a,0)\geq  p  \geq (a-3),\,
%  \min(b,0)\geq  q \geq (b-3), & \text{ for }  & X=B\,,B^o,\,D,\, D^o.
%\end{array}
%\]
%.!

%The decomposition (\ref{eq:4.7}) shows that
Thus,
in order to know $H^{m,n}(G_{X^o})$
it is enough to compute homology  $H^{m,n}(G_{X^o}(a,b))$ and this is
our further goal. We will use formulae (\ref{eq:4.9}), (\ref{eq:4.7})
in these computations.\\

Let us consider $s\ell (3)$-modules
$\La^i=\La^i\lsp{x_1,x_2,x_3}$ for $i \geq 0$, and
let  $\La^i=0$ for $i<0$. Of course
$\La^i=0$ for $i>3$ too.

\begin{Proposition}
\label{prop:4.3}
If $a,b > 3$, then (as $s\ell (3)$-modules)
%=A-case
\[
H^{m,n}(G_{A^o}(a,b),\tdn)=H^{m,n}(G_{A}(a,b),\tdn)\simeq
\left\{
\bAr{lll}
0,
& \text{ for }& m>0,
\\
\La^{a+b-n},
& \text{ for }& m=0;
\eAr
\right.
\]
%=C-case
\[
H^{m,n}(G_{C^o}(a,b),\tdn)=H^{m,n}(G_{C}(a,b),\tdn)\simeq
\left\{
\bAr{lll}
0,
& \text{ for }& m<0,
\\
\La^{a+b-n-3},
& \text{ for }& m=0.
\eAr
\right.
\]
%=B-case
If $a,b< 0$,  then (as $s\ell (3)$-modules)
\[
H^{m,n}(G_{B^o}(a,b),\tdn)=H^{m,n}(G_{B}(a,b),\tdn)\simeq
\left\{
\bAr{lll}
0,
& \text{ for }& m>0,
\\
\La^{a+b-n},
& \text{ for }& m=0,
\eAr
\right.
\]
%=D-case
\[
H^{m,n}(G_{D^o}(a,b),\tdn)=H^{m,n}(G_{D}(a,b),\tdn)\simeq
\left\{
\bAr{lll}
0,
& \text{ for }& m<0,
\\
\La^{a+b-n-3},
& \text{ for }& m=0.
\eAr
\right.
\]
\end{Proposition}
(Note that for $X=B$ or $D$ the above formulae show that
the homology
$H^{0,n}(G_{X^o}(a,b),\tdn)$ can be non-zero
only for negative
$n$, namely when \\
$ -3+a+b \leq n \leq a+b < 0 $ for $B$ and
$ -6+a+b \leq n \leq -3+a+b < 0 $ for $D$.) \\

\prf\/
First of all it follows from (\ref{eq:4.9}) that, under restrictions of
the proposition, $G_X=G_{X^o}$, because we are to care about the difference
only if $(G_X)_{[0,0]}\neq 0$ which is not the case here.

We use the spectral sequences
of Lemma~\ref{lem:4.2} for the evaluation  of $H(G_X)$ and
the first spectral sequence  happens
to be sufficient  for the proof.
As we compute  the ${E'}^2$-term,
we notice that we have got a one-row
spectral sequence that necessarily  degenerates
(i.e. all the higher differentials are zero), thus
${E'}^2\simeq{E'}^\infty\simeq H(G_X,\tdn)$ as $s\ell (3)$-modules.

Let us start with the $A$-case. By  Lemma~\ref{lem:4.2}
\[
{E'}^2_{[p,q]}(G_A(a,b)) \simeq H_p(H_q(G_A(a,b),\dst''),\,\dst'),
\]
thus we are to begin with considering
$G_A(a,b)$ as a complex with the differential
$\dst''=\De^-\dprt_-$.

We see from (\ref{eq:4.9})
that it splits into a sum of  subcomplexes
\begin{equation}
\label{eq:4.10}
\cdots
\leftarrow
\La^+_{a-p} \La^-_{b-q+1}  [x]     z_+^{p}    z_-^{q-1}\,
%\overset{\De^-\dprt_-}
{\longleftarrow}
\,\La^+_{a-p} \La^-_{b-q}  [x]     z_+^{p}    z_-^{q}
\leftarrow
\cdots
\end{equation}
Observing how the differential acts we conclude that
the complex (\ref{eq:4.10})
is isomorphic to the tensor product of the following complex
\begin{equation}
\label{eq:4.11}
0
\leftarrow
\La^-_{3}[x] z_-^{b-3}
\leftarrow
\cdots
\leftarrow
\La^-_{b-q+1}  [x] z_-^{q-1}\,
%\overset{\De^-\dprt_-}
\longleftarrow
\,\La^-_{b-q}  [x]    z_-^{q}
\leftarrow
\cdots
\leftarrow
\La^-_{0}[x]z_-^{b}
\leftarrow
0
\end{equation}
(note that  $b>3$)
with the vector space
$\La^+_{a-p} z_+^{p}$ which is not affected by the differential.
% implies that $\La^-_{b-q}$
%ranges from $\La^-_{0}$ for $q=b$ to $\La^-_{3}$ for $q=b-3$).

The complex (\ref{eq:4.11}) is nothing but a De Rham complex
with the ``grading variable'' $z_-$ added. This
implies that it has non-zero homologies only at its right end (with our
direction of arrows), and those are isomorphic to $\CC z_-^{b}$.

Therefore the complex (\ref{eq:4.10}) also has its non-zero
homologies only at one place and those are isomorphic to
$\La^+_{a-p} z_+^{p}z_-^{b}$.

This shows us that all non-zero
terms in $ {E'}^1_{[p,q]}(G_A(a,b)) $ are confined to one row $q=b$
and that $\dst'$ is zero on this row. Thus ${E'}^2={E'}^1$ and for
a one-row spectral sequence
${E'}^2=\ldots={E'}^\infty$, hence
we have arrived at the following answer
\begin{equation}
\label{eq:4.12}
{E'}^\infty_{[p,q]}(G_A(a,b))=
\left\{
\bAr{cll}
0 & \text{ for } & q\neq b, \\
\La^+_{a-p} z_+^{p}z_-^{b} &
\text{ for } & q = b.
\eAr
\right.
\end{equation}
%An one-row spectral sequence degenerates
At the same time
\begin{equation}
\label{eq:4.12'}
\sum_{m}H^{m,n}(G_A(a,b))  \simeq\sum_{p+q=n}{E'}^\infty_{[p,q]}(G_A(a,b))
\simeq {E'}^\infty_{[n-b,b]}(G_A(a,b))
\simeq \La^+_{a+b-n} z_+^{n-b}z_-^{b} \,.
\end{equation}
Now (\ref{eq:4.12'}) shows that $H^{m,n}(G_A(a,b))=0$ for $m\neq 0$ and
that it has the stated value for $m=0$.
This proves the proposition for $X=A$.

%>B
The proof for $X=B$ is quite similar, we
are to  deal  with complexes
\begin{equation}
\label{eq:4.10B}
\cdots
\leftarrow
\La^+_{a-p} \La^-_{b-q+1}[x]\dprt_+^{-p}\dprt_-^{1-q}\,
%\overset{\De^-\dprt_-}
{\longleftarrow}
\,\La^+_{a-p} \La^-_{b-q}[x]\dprt_+^{-p}\dprt_-^{-q}
\leftarrow
\cdots
\end{equation}
isomorphic to the tensor product of a De Rham complex
\begin{equation}
\label{eq:4.11B}
0
\leftarrow
\La^-_{3}[x]\dprt_-^{3-b}
\leftarrow
\cdots
\leftarrow
\La^-_{b-q+1}  [x]  \dprt_-^{1-q}\,
%\overset{\De^-\dprt_-}
{\longleftarrow}
\,\La^-_{b-q}  [x] \dprt_-^{-q}
\leftarrow
\cdots
\leftarrow
\La^-_{0}[x]\dprt_-^{-b}
\leftarrow
0\,
\end{equation}
($b<0$), and a vector space
$\La^+_{a-p}\dprt_+^{-p}$ which is not affected by the differential.

It gives
\begin{equation}
\label{eq:4.13}
{E'}^2_{[p,q]}(G_B(a,b))=
\left\{
\bAr{cll}
0 & \text{ for } & q\neq b, \\
\La^+_{a-p} \dprt_+^{-p}\dprt_-^{-b} &
 \text{ for } & q = b.
\eAr
\right.
\end{equation}
We have got the same configuration with one non-zero row, hence
${E'}^2=\ldots={E'}^\infty$. At the same time
\begin{equation}                                                                \label{eq:4.12B}
\sum_{m}H^{m,n}(G_B(a,b))  \simeq\sum_{p+q=n}{E'}^\infty_{[p,q]}(G_B(a,b))
\simeq {E'}^\infty_{[n-b,b]}(G_B(a,b))
\simeq \La^+_{a+b-n} \dprt_+^{b-n}\dprt_-^{-b} \,,
\end{equation}
and
this immediately implies the proposition for $X=B$.

%>C
Going to $X=C$ we represent $G_C(a,b)$ as the sum
of subcomplexes of the form
\begin{equation}
\label{eq:4.14}
\cdots
\leftarrow
\La^+_{a-p} \La^-_{b-q+1}  [\dprt]     z_+^{p}    z_-^{q-1}\,
%\overset{\De^-\dprt_-}
{\longleftarrow}
\,\La^+_{a-p} \La^-_{b-q}  [\dprt]     z_+^{p}    z_-^{q}
\leftarrow
\cdots
\end{equation}
which differs from (\ref{eq:4.10}) in that respect that
here we have modules over $\CC[\dprt]$ instead
of $\CC[x]$.
% and this means another action of the differential
%although it is given by the same formula.
Again
%we conclude first of all that
the complex (\ref{eq:4.14})
is isomorphic to the tensor product of
\begin{equation}
\label{eq:4.15}
0
\leftarrow
\La^-_3[\dprt]z_-^{b-3}
\leftarrow
\cdots
\leftarrow
\La^-_{b-q+1}  [\dprt]       z_-^{q-1}\,
%\overset{\De^-\dprt_-}
{\longleftarrow}
\,\La^-_{b-q}  [\dprt]    z_-^{q}
\leftarrow
\cdots
\leftarrow
\La^-_0[\dprt]z_-^{b}
\leftarrow
0\,
\end{equation}
($b>3$), and a vector space
$\La^+_{a-p} z_+^{p}$ which is not affected by the differential.

Now with its differential
the complex (\ref{eq:4.15}) is essentially a Koszul complex
(dual of the De Rham complex) and
this
implies that it has non-zero homologies only at its very left end
(with our direction of arrows),
and those are isomorphic to $\La^-_3 z_-^{b-3}$.
Thus the complex (\ref{eq:4.14}) also has its non-zero
homologies only at its left end and those are isomorphic to
$\La^+_{a-p}\La^-_3 z_+^{p}z_-^{b-3}$.

We have got a one-row spectral sequence again, but with the
row $q=b-3$
and again $\dst'$ is zero on this row.
Hence
\begin{equation}
\label{eq:4.16}
{E'}^2_{[p,q]}(G_C(a,b))=
\left\{
\bAr{cll}
0 & \text{ for } & q\neq b-3, \\
\La^+_{a-p}\La^-_3 z_+^{p}z_-^{b-3} &
\text{ for } & q = b-3.
\eAr
\right.
\end{equation}
Now ${E'}^2_{[p,q]}={E'}^{\infty}_{[p,q]}$,
\begin{equation}
\label{eq:4.12C}
\sum_{m}H^{m,n}(G_C(a,b))
%\simeq\sum_{p+q=n}{E'}^\infty_{[p,q]}(G_C(a,b))
\simeq {E'}^\infty_{[n-b+3,b-3]}(G_C(a,b))
\simeq \La^+_{a+b-n-3}\La^-_3 z_+^{n-b+3}z_-^{b-3} \,,
\end{equation}
and, taking into account the
isomorphism $\La^-_3\simeq \CC$, we conclude that we proved the
proposition for $X=C$.

%>D
For $X=D$ we arrive similarly at the complex
\begin{equation}
\label{eq:4.13D}
\cdots
\leftarrow
\La^+_{a-p} \La^-_{b-q+1}[\dprt]\dprt_+^{-p}\dprt_-^{1-q}\,
%\overset{\De^-\dprt_-}
{\longleftarrow}
\,\La^+_{a-p} \La^-_{b-q}[\dprt]\dprt_+^{-p}\dprt_-^{-q}
\leftarrow
\cdots\,,
\end{equation}
which is the tensor product of a Koszul complex
\begin{equation}
\label{eq:4.14D}
0
\leftarrow
\La^-_3[\dprt]\dprt_-^{3-b}
\leftarrow
\cdots
\leftarrow
\La^-_{b-q+1}  [\dprt]  \dprt_-^{1-q}\,
%\overset{\De^-\dprt_-}
{\longleftarrow}
\,\La^-_{b-q}  [\dprt] \dprt_-^{-q}
\leftarrow
\cdots
\leftarrow
\La^-_0[\dprt]\dprt_-^{-b}
\leftarrow
0\,
\end{equation}
($b<0$), and a vector space
$\La^+_{a-p}\dprt_+^{-p}$.
So we get a formula
\begin{equation}
\label{eq:4.17}
{E'}^2_{[p,q]}(G_D(a,b))=
\left\{
\bAr{cll}
0 & \text{ for } & q\neq b-3, \\
\La^+_{(a-p)}\La^-_3 \dprt_+^{-p}\dprt_-^{3-b} &
\text{ for } & q = b-3.
\eAr
\right.
\end{equation}
Again ${E'}^2_{[p,q]}={E'}^{\infty}_{[p,q]}$, and
\begin{equation}
\label{eq:4.12D}
\sum_{m}H^{m,n}(G_D(a,b))
%\simeq\sum_{p+q=n}{E'}^\infty_{[p,q]}(G_D(a,b))
\simeq {E'}^\infty_{[n-b+3,b-3]}(G_D(a,b))
\simeq \La^+_{a+b-n-3} \La^-_3\dprt_+^{b-n-3}\dprt_-^{3-b} \,.
\end{equation}
This is enough to prove the proposition for $X=D$. \epf

Looking back at the proof we notice that the roles of $a$ and $b$ are
not symmetric. The restrictions on $b$ are important in that respect
that the complexes (\ref{eq:4.11}), (\ref{eq:4.11B}), (\ref{eq:4.14}),
(\ref{eq:4.14D}) would not be cut short and remain the full length
De Rham or Koszul complexes.
But the restrictions on $a$
have not been used.
%for calculating
%$E^2_{[p,q]}$, formulae
%(\ref{eq:4.12}), (\ref{eq:4.13}), (\ref{eq:4.16}), (\ref{eq:4.17}),
%remain valid.

Therefore leaving the proof the same
we can relax the rectrictions on $a$, arriving at the following
corollary.
\bCo
\label{cor:4.4}
If
%$a\geq 0$,
$b >3$,  then
%=A-case
\[
H^{m,n}(G_{A^o}(a,b),\tdn)=H^{m,n}(G_{A}(a,b),\tdn)\simeq
\left\{
\bAr{lll}
\La^{a+b-n}\,\,\,& \text{ for }& m=0,\,n\geq b, \qquad
\\
0 & &\text{ otherwise },
\eAr
\right.
\]
%=C-case
\[
H^{m,n}(G_{C^o}(a,b),\tdn)=H^{m,n}(G_{C}(a,b),\tdn)\simeq
\left\{
\bAr{lll}
\La^{a+b-n-3} & \text{ for }& m=0,\,n\geq b-3.
\\
0 & &\text{ otherwise }.
\eAr
\right.
\]
%=B-case
If
%$a\leq 3$,
$b< 0$, then
\[
H^{m,n}(G_{B^o}(a,b),\tdn)=H^{m,n}(G_{B}(a,b),\tdn)\simeq
\left\{
\bAr{lll}
\La^{a+b-n}\,\,\,& \text{ for }& m=0,\, n\leq b,\qquad
\\
0 & &\text{ otherwise },
\eAr
\right.
\]
%=D-case
\[
H^{m,n}(G_{D^o}(a,b),\tdn)=H^{m,n}(G_{D}(a,b),\tdn)\simeq
\left\{
\bAr{lll}
\La^{a+b-n-3} & \text{ for }& m=0,\,n\leq b-3
\\
0 & &\text{ otherwise }.
\eAr
\right.
\]
\eCo

\bCo
\label{cor:4.5}
By interchanging $a$ and $b$ in Corollary~\ref{cor:4.4} we get
valid statements as well.
\eCo
This is because we can use the second spectral sequence of Lemma~\ref{lem:4.2}
in the proof instead of the first one. \\

We also have to remember that for
some values of
$a$ complexes $G_X(a,b)$ are entirely zero. Thus
we can assume that
$a\geq 0$ for $X=A, A^o,C,C^o$
and $a\leq 3$ for $X=C,C^o,D,D^o$, and
we are left with the cases when $0\leq a,b \leq 3$. Here
the answer becomes somewhat different and the proof demands more
elaborate arguments, but goes along the same lines of
computing the homology via spectral sequence.

\begin{Proposition}
\label{prop:4.6}
Let $ 0 \leq a\leq b \leq 3$ then
\[
%=A-case
\bAr{rl}
H^{m,n}(G_{A^o}(a,b),\tdn)\simeq &
\left\{
\bAr{lll}
%>A
\La^{a+b-n}\qquad\,\,& \text{ for }& m=0,\,n\geq b,
\\
\La^{1+a+b-n}\,\,\quad & \text{ for }& m=1,\,0\leq n\leq a, \qquad
\\
0 & &\text{ otherwise },
\eAr
\right. \\
 & \\
%=C-case
H^{m,n}(G_{C^o}(a,b),\tdn)\simeq &
\left\{
\bAr{lll}
\La^{a+b-n-3} \quad\, & \text{ for }& m=0,\,n\geq 0,
\\
0 & &\text{ otherwise },
\eAr
\right. \\
 & \\
%=B-case
H^{m,n}(G_{B^o}(a,b),\tdn)\simeq &
\left\{
\bAr{lll}
\La^{a+b-n}\qquad\,\,\,& \text{ for }& m=0,\, n\leq 0,\,\,\,\,\,\qquad
\\
0 & &\text{ otherwise }.
\eAr
\right.
%\\
% &
\eAr\]
%=D-case
Let $ 0 \leq b\leq a \leq 3$ then
\[
\bAr{rl}
H^{m,n}(G_{D^o}(a,b),\tdn)\simeq &
\left\{
\bAr{lll}
\La^{a+b-n-3} & \text{ for }& m=0,\,n\leq b-3
\\
\La^{-1+a+b-n-3} & \text{ for }& m=-1,\,a-3\leq n\leq 0.
\\
0 & &\text{ otherwise }.
\eAr
\right.
\eAr
\]
\end{Proposition}

\begin{proof}
We see immediately that the statement is true for the
following  complexes
\[
\bAr{ll}
G_{A^o}(0,0)=\CC+\lsp{x_1,x_2,x_3}\,, &
G_{C^o}(0,0)=0\,, \\
G_{B^o}(3,3)=0\,, &
G_{D^o}(3,3)= \CC+\lsp{\dprt_1,\dprt_2,\dprt_3} \,,
\eAr
\]
with trivial
differentials. We exclude these cases from consideration further on.

We already noticed that variables $z_\pm$ (resp. $\dprt_\pm$)
play only the role of grading variables, so
we can eliminate them as it is done below.

Let us define a bicomplex
\[
(\tilde{G}_A(a,b))_{[p,q]}=\left\{
\bAr{cc}
\La^+_{a-p}\La^-_{b-q}[x] & \text{ for } p\geq 0, q\geq 0,\\
0  & \text{ overwise},
\eAr\right.
\]
with the differentials $\dst'=\De^+ , \dst''=\De^- $. Comparing
with (\ref{eq:4.9}) we conclude that there exists an isomorphism
of bicomplexes $\al:{G_A(a,b)} \ar \tilde{G}_A(a,b)$.

Let
\[
(G_{B'}(a,b))_{[p,q]}=\left\{
\bAr{cc}
\La^+_{a+1}\La^-_{b+1}[x] & \text{ for } p= 0, q= 0,\\
0  & \text{ overwise}.
\eAr\right.
\]
Following (\ref{eq:AB0}) the bicomplex $G_{A^o}(a,b)$ is the kernel of the morphism
$\tdn_2:G_{A}(a,b)\ar G_{A'}(a,b)$ and the morphism can be included into
a commutative diagram
\[
\bAr{ccc}
G_{A}(a,b) & \arr & G_{A'}(a,b) \\
\downarrow & & \parallel \\
\tilde{G}_{A}(a,b) & \arr & G_{A'}(a,b) \\
\eAr
\]
where the morphism at the lower row is equal to $\De^+\De^-$. This
shows that $\al$ induces an isomorphism between $G_{A^o}(a,b)$
and the bicomplex
\[
\tilde{G}_{A^o}(a,b)=
\Ker (\De^+\De^-:\tilde{G}_{A}(a,b)  \arr  G_{A'}(a,b)\,)\,,
\]
thus it also induces an isomorphism of their homologies.\\

The bicomplex $\tilde{G}_{A}(a,b)$ can be represented
by the following diagram
(we allow ourselves to omit the zero components and morphisms
with a zero source or target):
\[
\bAr{ccccccc}
\La^+_{a}\La^-_{0}[x]
& \leftarrow \cdots\leftarrow
& \La^+_{k+1}\La^-_{0}[x]
&\leftarrow
& \La^+_{k}\La^-_{0}[x]
& \leftarrow \cdots\leftarrow
& \La^+_{0}\La^-_{0}[x]
\\
\downarrow &\cdots &\downarrow &       &\downarrow &\cdots &\downarrow \\
\vdots     &\cdots &\vdots     &\cdots &\vdots     &\cdots & \vdots \\
\La^+_{a}\La^-_{j}[x]
& \leftarrow \cdots\leftarrow
& \La^+_{k+1}\La^-_{j}[x]
&\leftarrow
& \La^+_{k}\La^-_{j}[x]
& \leftarrow \cdots\leftarrow
& \La^+_{0}\La^-_{j}[x]
\\
\downarrow &\cdots &\downarrow &       &\downarrow &\cdots &\downarrow \\
\vdots     &\cdots &\vdots     &\cdots &\vdots     &\cdots & \vdots \\
\La^+_{a}\La^-_{b}[x]
& \leftarrow \cdots\leftarrow
& \La^+_{k+1}\La^-_{b}[x]
&\leftarrow
& \La^+_{k}\La^-_{b}[x]
& \leftarrow \cdots\leftarrow
& \La^+_{0}\La^-_{b}[x]
\eAr
\]
where the row (resp. column) maps are the De Rham differentials $\De^+$
(resp. $\De^-$).

To represent $\tilde{G}_{A^o}(a,b)$ by a similar
diagram we need only to change it in the lower-left corner
and put there $\Ker(\De^+\De^-)$.

It follows that the ${E'}^1$-term of the spectral sequence
of the bicomplex $\tilde{G}_{A^o}(a,b)$ is represented by
the following diagram:
\[
\bAr{ccccccc}
\La^+_{a}
& \leftarrow \cdots\leftarrow
& \La^+_{k+1}
&\leftarrow
& \La^+_{k}
& \leftarrow \cdots\leftarrow
& \La^+_{0}
\\
0 &\leftarrow \cdots\leftarrow
 &0 & \leftarrow &0 &\leftarrow \cdots\leftarrow  &0 \\
\cdots &\cdots\cdots &\cdots & \cdots &\cdots &\cdots\cdots &\cdots \\
0 &\leftarrow \cdots\leftarrow
 &0 & \leftarrow &0 &\leftarrow \cdots\leftarrow  &0 \\
\frac{\Ker\De^+\De^-} {\IM \De^-}
& \leftarrow \cdots\leftarrow
& \frac{\La^+_{k+1}\La^-_{b}[x]} {\IM \De^-}
&\leftarrow
& \frac{\La^+_{k}\La^-_{b}[x]} {\IM \De^-}
& \leftarrow \cdots\leftarrow
&\frac{ \La^+_{0}\La^-_{b}[x]} {\IM \De^-}
\eAr
\]
with only two non-zero rows, those where $q=0$
and $q=b$. Here  $b\neq 0$  as $0\leq a\leq b$ and
we have excluded $a=b=0$.

The ${E'}^2$-term will have the similar ``two rows'' structure,
and this together with
$a\leq b$ imply that
for every differential $d^r_{[p,q]},\,r\geq 2$, either its source
or its target is zero. Thus all the differentials are trivial and
therefore ${E'}^2=\cdots={E'}^\infty$.

Let us compute ${E'}^2$.
First of all
the differential $d'$ is induced by $\De^+$ and evidently it is
trivial on the upper row, hence the row  descends to ${E'}^2$
unchanged.

The following lemma helps us to compute the terms in the lower row.
\bLe
\label{lem:4.7}
Let $R(a,b)$ be the complex
\[
\bAr{ccccccc}
\frac{ \La^+_{0}\La^-_{b}[x]} {\IM \De^-}
&\overset{\De^+}{\arr}
&\,\,\frac{ \La^+_{1}\La^-_{b}[x]} {\IM \De^-}\,\,
&\overset{\De^+}{\arr}
  \cdots \overset{\De^+}{\arr}
&\,\, \frac{\La^+_{k}\La^-_{b}[x]} {\IM \De^-}\,\,
&\overset{\De^+}{\arr}
 \cdots \overset{\De^+}{\arr}
&\frac{\Ker\De^+\De^-} {\IM \De^-}
\eAr
\]
and $ S(a,b)$ be the complex
\[
\bAr{ccccccc}
\mx{\footnotesize{$
\De^-( \La^+_{0}\La^-_{b}[x]) $}}
&\overset{\De^+}{\ar}
&\mx{\footnotesize{$
\De^-( \La^+_{1}\La^-_{b}[x])$}}
&\overset{\De^+}{\ar}
  \cdots \overset{\De^+}{\ar}
&\mx{\footnotesize{$
 \De^-(\La^+_{k}\La^-_{b}[x])$}}
&\overset{\De^+}{\ar}
 \cdots \overset{\De^+}{\ar}
&S(a,b)_b\,,
\eAr
\]
where the last term $ S(a,b)_b=
\Ker\left(
%\De^+:
\mx{\footnotesize{$\De^-(\La^+_{ a}\La^-_{b}[x])$}}\overset{\De^+}{\ar}
\mx{\footnotesize{$\De^-(\La^+_{a+1}\La^-_{b}[x])$}}\right)$.\\
Then
\alphaparenlist
  \begin{enumerate}
  \item %%a
$\De^-$ induces an isomorphism of complexes
$R(a,b)\ar S(a,b)$
for $b>0$,
  \item %%b
 homologies of $R(a,b)$ (and $S(a,b)$) for $b>0$ are isomorphic to:
\[
\bAr{ccccccc}
\La^{b+1}\,,\quad
&\La^{b+2}\,,\quad
&\ldots \quad,\,\,
&\La^{b+k+1}\,,\quad
&\ldots \quad,\,\,
&\La^{a+b+1}\,.\quad
\eAr
\]
\end{enumerate}
\eLe
In the notations of
Lemma~\ref{lem:4.7}
we write the terms of the lower
non-trivial row
of the ${E'}^2$-term of the spectral sequence
as follows:
\begin{equation}
\label{eq:lem4.7}
{E'}^2_{[p,0]}(\tilde{G}_{A^o}(a,b))=H_{a-p}(R(a,b))
\,
\end{equation}
(in the part (b) of the lemma these terms are explicitely calculated).

%cp
\pLe~\ref{lem:4.7}.\/
The statement (a) is
obvious.
Going to (b)
%Before evaluating the homology
%We leave it to the reader to check the isomorphism. \\
let us
first of all
notice  that the second complex has a nice property:
\[
H_i(S(a,b))=H_i(S(a+1,b)) \text{ for } 0\leq i \leq a\,.
\]
Thus it is sufficient to compute its homologies for large $a$,
and because of (\ref{eq:lem4.7}) it is enough
to compute ${E'}^2(\tilde{G}_{A^o}(a,b))$ when $a$ is large.

Let $a>3$ (and thus $a>b$). Evidently  ${E'}^2(G_{A^o}(a,b))$
has  two non-zero rows, for $q=0$ and $q=b$, and
therefore among differentials $d^r_{[p,q]}, \,r\geq 2$, all but
those for $r=b+1$, $\,q=0,\,b<p\leq a$ are equal to zero.

On the other hand total homologies of  $\tilde{G}_{A^o}(a,b)$
are isomorphic to total homologies of  ${G}_{A^o}(a,b)$
and their description is given in Corollary~\ref{cor:4.5}.
It follows from Corollary~\ref{cor:4.5} that
\[
\sum_{p+q=n}
{E'}^\infty_{[p,q]}(\tilde{G}_{A^o}(a,b)) =0 \text{ for } n< a\,,
\text{
and }\sum_{p+q=a}
{E'}^\infty_{[p,q]}(\tilde{G}_{A^o}(a,b)) =\La^+_b.
\]
All this forces us to conclude that the differential
$d^{(b+1)}_{[p,0]},\,b<p\leq a$, determines isomorphism
\begin{equation}
\label{eq:lem4.7'}
H_{a-p}(R(a,b))= {E'}^{b+1}_{[p,0]}(\tilde{G}_{A^o}(a,b))
\arr {E'}^{b+1}_{[p-b-1,b]}(\tilde{G}_{A^o}(a,b))
=\La^+_{a+b-p+1}\,.
\end{equation}
%\vspace*{-6ex}
This proves the lemma.
\epf

It is important to notice that the above morphism
$d^{(b+1)}_{[p,0]}$ is induced by $\tdn$ hence it diminishes the
degree of $x$'s by 1. Therefore
the space ${E'}^{b+1}_{[p,0]}(\tilde{G}_{A^o}(a,b))$
above is represented by elements of $x$-degree 1
because elements in $ {E'}^{b+1}_{[p-b-1,b]}(\tilde{G}_{A^o}(a,b)) $
have $x$-degree zero.

>From (\ref{eq:lem4.7}) and (\ref{eq:lem4.7'}) we get
\begin{equation}
\label{eq:lem4.7''}
{E''}^2_{[p,0]}(\tilde{G}_{A^o}(a,b))=H_{a-p}(R(a,b))
=\La^{a+b-p+1}\,(\text{in $x$-degree 1})
\,.
\end{equation}

To continue proving Proposition~\ref{prop:4.6}, we return to our previous
data $0\leq a\leq b \leq 3$ and conclude that if
 $0\leq n\leq a$ then
\[
\sum_{m}H^{m,n}(\tilde{G}_{A^o}(a,b))\simeq
{E'}^{\infty}_{[n,0]}(\tilde{G}_{A^o}(a,b))
\simeq {E'}^{2}_{[n,0]}(\tilde{G}_{A^o}(a,b)) \simeq
\La^{a+b-n+1}\,(\text{in $x$-degree 1})\,,
\]
hence $H({G}_{A^o}(a,b))^{1,n}\simeq \La^{a+b-n+1}$,
and $H({G}_{A^o}(a,b))^{m,n}=0$ for $m \neq 1$.
With this we have got the rest of the statement for $X=A^o$.\\

Similarly for $X=B$ we define a ``$\dprt_\pm$-free version'',
a bicomplex
\[
(\tilde{G}_B(a,b))_{[p,q]}=\left\{
\bAr{cc}
\La^+_{a-p}\La^-_{b-q}[x] & \text{ for } p\leq 0, q\leq 0,\\
0  & \text{ overwise},
\eAr\right.
\]
with the differentials $\dst'=\De^+ , \dst''=\De^- $,
and looking at
(\ref{eq:4.9}) we see that there is a natural isomorphism
of bicomplexes $\al:{G_B(a,b)} \ar \tilde{G}_B(a,b)$.

Let
\[
(G_{A'}(a,b))_{[p,q]}=\left\{
\bAr{cc}
\La^+_{a-1}\La^-_{b-1}[x] & \text{ for } p= 0, q= 0,\\
0  & \text{ overwise}.
\eAr\right.
\]
It is clear that $\tdn_2$ maps $G_{A'}(a,b)$ into $G_{B}(a,b)$ having
$G_{B^o}(a,b)$
as the cokernel,
and that $\al$ induces an isomorphism
between $G_{B^o}(a,b)$ and
the cokernel of the morphism
 $\De^+\De^-$ from $G_{A'}(a,b)$ into $\tilde{G}_{B}(a,b)$,
the latter cokernel we will denote by $\tilde{G}_{B^o}(a,b)$.

Notice that for $a=b=0$ we have $G_{A'}(0,0)=0$, so
$G_{B}(0,0)=G_{B^o}(0,0)$ and it is easy to check that the proof
of Proposition~\ref{prop:4.3} works and provides the result.
Let us suppose $b>0$.

Again
we represent $\tilde{G}_{B}(a,b)$ by the following diagram:
\[
\bAr{ccccccc}
\La^+_{3}\La^-_{b}[x]
& \leftarrow \cdots\leftarrow
& \La^+_{k+1}\La^-_{b}[x]
&\leftarrow
& \La^+_{k}\La^-_{b}[x]
& \leftarrow \cdots\leftarrow
& \La^+_{a}\La^-_{b}[x]
\\
\downarrow &\cdots &\downarrow &       &\downarrow &\cdots &\downarrow \\
\vdots     &\cdots &\vdots     &\cdots &\vdots     &\cdots & \vdots \\
\La^+_{3}\La^-_{j}[x]
& \leftarrow \cdots\leftarrow
& \La^+_{k+1}\La^-_{j}[x]
&\leftarrow
& \La^+_{k}\La^-_{j}[x]
& \leftarrow \cdots\leftarrow
& \La^+_{a}\La^-_{j}[x]
\\
\downarrow &\cdots &\downarrow &       &\downarrow &\cdots &\downarrow \\
\vdots     &\cdots &\vdots     &\cdots &\vdots     &\cdots & \vdots \\
\La^+_{3}\La^-_{3}[x]
& \leftarrow \cdots\leftarrow
& \La^+_{k+1}\La^-_{3}[x]
&\leftarrow
& \La^+_{k}\La^-_{3}[x]
& \leftarrow \cdots\leftarrow
& \La^+_{a}\La^-_{3}[x]
\eAr
\]
where the row (resp. column) maps are the De Rham differentials
$\De^+$ (resp. $\De^-$) and
the upper-right corner corresponds to $[p,q]=[0,0]$ so
the diagram is situatied in the third quarter.
For $\tilde{G}_{B^o}(a,b)$  we are to change  the upper-right corner
to $\Coker(\De^+\De^-)$.

%cB
To compute the ${E'}^1$-term of the spectral sequence we consider
the column complexes and evidently they have non-zero homology
only at the upper ($q=0$) row. Having in mind that $b>0$ one
easily checks that
the ${E'}^1$-term
of the spectral sequence
of the bicomplex $\tilde{G}_{B^o}(a,b)$ is represented by
the following diagram:
\[
\bAr{ccccccc}
\mx{\footnotesize{$
\De^-( \La^+_{3}\La^-_{b-1}[x]) $}}
& \leftarrow \cdots\leftarrow
&\mx{\footnotesize{$
\De^-( \La^+_{k+1}\La^-_{b-1}[x])$}}
&\leftarrow
&\mx{\footnotesize{$
 \De^-(\La^+_{k}\La^-_{b-1}[x])$}}
& \leftarrow \cdots\leftarrow
&\frac{\Ker \De^-}{\IM \De^+\De^-}
\,
\\
0 &\leftarrow \cdots\leftarrow
 &0 & \leftarrow &0 &\leftarrow \cdots\leftarrow  &0 \\
\vdots &\cdots &\vdots & \cdots &\vdots &\cdots &\vdots \\
0 &\leftarrow \cdots\leftarrow
 &0 & \leftarrow &0 &\leftarrow \cdots\leftarrow  &0
\eAr
\]
Now notice that
\[
\bAr{lclcl}
\frac{\Ker \De^-}{\IM \De^+\De^-}&\simeq&
\frac{\De^-(\La^+_{a}\La^-_{b-1}[x])}{\De^+\De^-(\La^+_{a-1}\La^-_{b-1}[x])}&\simeq&
\Coker\left(
%\De^+:
\mx{\footnotesize{$\De^-(\La^+_{ a}\La^-_{b-1}[x])$}}
\overset{\,\De^+}{\leftarrow}
\mx{\footnotesize{$\De^-(\La^+_{a-1}\La^-_{b-1}[x])$}}\right)
\eAr
\]

Now comparing the upper row with the complex $S(3,b-1)$ of Lemma~\ref{lem:4.7}
we have no difficulty  determining its homology. The upper row
of the diagram for ${E'}^2$ becomes
\[
\bAr{ccccccc}
\La^{a+b+3}\,,\quad
&\ldots\,\,\,, \quad
&\La^{k+b+1}\,,\quad
&\La^{k+b}\,,\quad
&\ldots \,\,\,,\quad
&\La^{a+b}\,.\quad
\eAr
\]
There are no more non-zero differentials, and we obtain the result
 in this case.\\

The reasoning for $X=D$ goes along the same lines and the situation is in some
sense dual to this in the case $X=A$. We will stress the main points leaving
it to the reader to fill in  details.

The diagram  representing $\tilde{G}_{D^o}(a,b)$ is the following
(we skip the evident definition of $\tilde{G}_{D^o}(a,b)$):
\[
\bAr{ccccccc}
\La^+_{3}\La^-_{b}[\dprt]
& \leftarrow \cdots\leftarrow
& \La^+_{k+1}\La^-_{b}[\dprt]
&\leftarrow
& \La^+_{k}\La^-_{b}[\dprt]
& \leftarrow \cdots\leftarrow
& \Coker(\De^+\De^-)
\\
\downarrow &\cdots &\downarrow &       &\downarrow &\cdots &\downarrow \\
\vdots     &\cdots &\vdots     &\cdots &\vdots     &\cdots & \vdots \\
\La^+_{3}\La^-_{j}[\dprt]
& \leftarrow \cdots\leftarrow
& \La^+_{k+1}\La^-_{j}[\dprt]
&\leftarrow
& \La^+_{k}\La^-_{j}[\dprt]
& \leftarrow \cdots\leftarrow
& \La^+_{a}\La^-_{j}[\dprt]
\\
\downarrow &\cdots &\downarrow &       &\downarrow &\cdots &\downarrow \\
\vdots     &\cdots &\vdots     &\cdots &\vdots     &\cdots & \vdots \\
\La^+_{3}\La^-_{3}[\dprt]
& \leftarrow \cdots\leftarrow
& \La^+_{k+1}\La^-_{3}[\dprt]
&\leftarrow
& \La^+_{k}\La^-_{3}[\dprt]
& \leftarrow \cdots\leftarrow
& \La^+_{a}\La^-_{3}[\dprt]
\eAr
\]
where the row (resp. column) maps $\De^+$ (resp. $\De^-$)
are Koszul differentials,
the upper-right corner corresponds to $[p,q]=[0,0]$,
the diagram is situatied in the third quarter and its other
corners are $[0,b-3]$, $[a-3,0]$, $[a-3,b-3]$.

Calculating homologies of the vertical complexes we get
(provided $b>0$) the following diagram:
\[
\bAr{ccccccc}
\mx{\footnotesize{$
\De^-( \La^+_{3}\La^-_{b-1}[\dprt]) $}}
& \leftarrow \cdots\leftarrow
&\mx{\footnotesize{$
\De^-( \La^+_{k+1}\La^-_{b-1}[\dprt])$}}
&\leftarrow
&\mx{\footnotesize{$
 \De^-(\La^+_{k}\La^-_{b-1}[\dprt])$}}
& \leftarrow \cdots\leftarrow
&\frac{\Ker \De^-}{\IM \De^+\De^-}
\,,
\\
0 &\leftarrow \cdots\leftarrow
 &0 & \leftarrow &0 &\leftarrow \cdots\leftarrow  &0 \\
\vdots &\cdots &\vdots & \cdots &\vdots &\cdots &\vdots \\
0 &\leftarrow \cdots\leftarrow
 &0 & \leftarrow &0 &\leftarrow \cdots\leftarrow  &0 \\
\La^+_{3}\La^-_{3}
& \leftarrow \cdots\leftarrow
& \La^+_{k+1}\La^-_{3}
&\leftarrow
& \La^+_{k}\La^-_{3}
& \leftarrow \cdots\leftarrow
& \La^+_{a}\La^-_{3}
\eAr
\]
Clearly the term at the upper-right corner is isomorpic to
\[
\bAr{ccccc}
\frac{\Ker \De^-}{\IM \De^+\De^-}&\simeq&
\frac{\IM \De^-}{\IM \De^+\De^-}&\simeq&
\Coker\left(
\mx{\footnotesize{$\De^-(\La^+_{ a}\La^-_{b-1}[\dprt])$}}
\overset{\,\De^+}{\leftarrow}
\mx{\footnotesize{$\De^-(\La^+_{a-1}\La^-_{b-1}[\dprt])$}}\right)
\eAr
\]
and we evaluate the homologies of the complex in the upper row by
the following lemma (which is analogous to Lemma~\ref{lem:4.7}).
\bLe
\label{lem:4.8}
Let $b>0$ and $T(a,b)$ be the complex
\[
\bAr{ccccccc}
\mx{\footnotesize{$
\De^-( \La^+_{3}\La^-_{b-1}[\dprt]) $}}
&\overset{\,\De^+}{\leftarrow}
&\mx{\footnotesize{$
\De^-( \La^+_{2}\La^-_{b-1}[\dprt])$}}
&\overset{\,\De^+}{\leftarrow}
  \cdots \overset{\,\De^+}{\leftarrow}
&\mx{\footnotesize{$
 \De^-(\La^+_{k}\La^-_{b-1}[\dprt])$}}
&\overset{\,\De^+}{\leftarrow}
 \cdots \overset{\,\De^+}{\leftarrow}
&T(a,b)_{a}\,,
\eAr
\]
where $ T(a,b)_{a}=
\Coker\left(
%\De^+:
\mx{\footnotesize{$\De^-(\La^+_{a+1}\La^-_{b-1}[\dprt])$}}
\overset{\,\De^+}{\leftarrow}
\mx{\footnotesize{$\De^-(\La^+_{a}\La^-_{b-1}[\dprt])$}}\right)$.\\
Then
%\alphaparenlist
%  \begin{enumerate}
%  \item %%a
 homologies of $T(a,b)$  are isomorphic to:
\[
\bAr{cccccc}
\La^{b-1}\,,\quad
&\La^{b-2}\,,\quad
&\ldots\,, \quad
&\La^{b-k}\,,\quad
&\ldots \,,\quad
&\La^{b-1+a-3}\,,\quad
\eAr
\]
and are represented by the elements linear in $\dprt_i,\, i=1,2,3$.
%\end{enumerate}
\eLe
We leave it to the reader to prove
the lemma by the same trick of changing $a$ and
comparing the first and the second spectral sequences of the bicomplex
in question. \\

The condition $0\leq b\leq a$ implies that
no futher differentials of the spectral sequence
are non-zero. Thus ${E'}^2={E'}^\infty$
and this gives us the result in the $D$-case.

For $C$-case
%we come again to the first quarter and
the
diagram representing ${E'}^1(\tilde{G}_{C^o}(a,b))$ situated
again in the first quarter and it is
\[
\bAr{ccccccc}
0 &\leftarrow \cdots\leftarrow
 &0 & \leftarrow &0 &\leftarrow \cdots\leftarrow  &0 \\
\vdots &\cdots &\vdots & \cdots &\vdots &\cdots &\vdots \\
0 &\leftarrow \cdots\leftarrow
 &0 & \leftarrow &0 &\leftarrow \cdots\leftarrow  &0 \\
\frac{\Ker (\De^+\De^-)}{\IM \De^-}
& \leftarrow \cdots\leftarrow
&\mx{\footnotesize{$
\De^-( \La^+_{k+1}\La^-_{b}[\dprt])$}}
&\leftarrow
&\mx{\footnotesize{$
 \De^-(\La^+_{k}\La^-_{b}[\dprt])$}}
& \leftarrow \cdots\leftarrow
&\mx{\footnotesize{$
\De^-( \La^+_{0}\La^-_{b}[\dprt]) $}}
\,,
\\
\eAr
\]
Clearly the term at the lower-left corner is isomorpic to
\[
\bAr{ccc}
\frac{\Ker (\De^+\De^-)}{\IM \De^-}&\simeq&
\Ker\left(
\mx{\footnotesize{$\De^-(\La^+_{ a}\La^-_{b-1}[\dprt])$}}
\overset{\,\De^+}{\leftarrow}
\mx{\footnotesize{$\De^-(\La^+_{a-1}\La^-_{b-1}[\dprt])$}}\right)
\eAr
\]
and the values of the horisontal homologies could be found
from those of $T(0,b)$ given by Lemma~\ref{lem:4.8}.
This closes the last case and completes the proof of Propositon~\ref
{prop:4.6}.
\end{proof}
\bCo
\label{cor:4.9}
By interchanging $a$, $b$ in Proposition~\ref{prop:4.6} we get
a valid statement as well.
\eCo
We are to change rows into columns and use the other spectral
sequence of a bicomplex.\\

Given $n \in \ZZ, \,y \in \CC$, let
$P(n,y)$ be an irreducible $s\ell (2) \oplus g\ell(1)$-module
with  highest weight $(n,y)$ when $n\geq 0$ and $P(n,y)=0$ when
$n<0$.
\bTh
\label{th:4.10}
There are the following isomorphisms of $\gind 0$-modules
(sums below run over $0\leq i\leq 3$):
\alphaparenlist
  \begin{enumerate}
  \item %%a
\[
H^{m,n}(G_{A^o})=\left\{
\bAr{ll}
\mx{$\sum$}\, \La^i\otimes P(n-i,\,-\tfrac{1}{3}i-n)&
\text{for } m=0, \,n \geq 0, \\
& \\
\mx{$\sum$}\, \La^i\otimes P(i-n-1,\,-\tfrac{1}{3}i-n+1)&
\text{for } m=1 , \,0 \leq n \leq 3, \,\,\quad\\
& \\
\qquad \qquad
0 & otherwise \,.
\eAr\right.
\]

  \item %%b
\[
 H^{m,n}(G_{B^o})=\left\{
\bAr{ll}
\mx{$\sum$}\, \La^i\otimes P(-n+i,\,-\tfrac{1}{3}i-n+2)&
\quad \text{for } m=0 , \,n \leq 0,\qquad \,\,\,\,\,\,\\
& \\
\qquad \qquad
0 & \quad otherwise\,.
\eAr\right.
\]

  \item %%c
\[
 H^{m,n}(G_{C^o})=\left\{
\bAr{ll}
\mx{$\sum$}\, \La^i\otimes P(n+3 -i,\,-\tfrac{1}{3}i-n-3)&
\,\,\,\text{for } m= 0, \,n \geq 0, \quad \qquad \\
& \\
\qquad \qquad
0 & \,\,\, otherwise\,.
\eAr\right.
\]

  \item %%d
\[
H^{m,n}(G_{D^o})=\left\{
\bAr{ll}
\mx{$\sum$}\, \La^i\otimes P(-n-3+i,\,-\tfrac{1}{3}i-n-1)&
\text{for } m=0 ,\, n \leq 0, \\
& \\
\mx{$\sum$} \,\La^i\otimes P(n -i+2,\,-\tfrac{1}{3}i-n-2)&
\text{for } m=-1 , \,-2\leq n \leq 0, \\
& \\
\qquad \qquad
0 & otherwise\,.
\eAr\right.
\]
   \end{enumerate}
\eTh

\begin{proof}
To prove the statements we use the decomposition (\ref{eq:4.7}) and
then collect the information about homologies for various values
of $a,b$
from Propositions~\ref{prop:4.3},~\ref{prop:4.6}
and Corollaries~\ref{cor:4.4},~\ref{cor:4.5},~\ref{cor:4.9}, having
in mind that $h_3$ acts on $G_{X^o}(a,b)$ as multiplication by $a-b$.

The action of $Y$ has to be computed too.
It is easy to compute it for $E^2$
by choosing a representative for a homology class.
Then it descends to $E^\infty$ and
because of the convergence of the spectral sequences we immediately
determine the action on homologies. We leave the
details to the reader.
\end{proof}
%eth4.10

%bf2

\vspace{1ex}

%\vspace*{1cm}
\begin{figure}[htbp]
  \begin{center}
    \leavevmode
  \setlength{\unitlength}{0.25in}
%%%%%%%%%%%%%%%%%%%%%%%%%%%%%%%%%%%%%%%%%%%%%%%   corner labels
%\begin{picture}(20,19)
\begin{picture}(21,19)

\put(18.7,17.5){A}
\put(18.7,-1){B}

\put(1,17.5){C}
\put(1,-1){D}

%%%%%%%%%%%%%%%%%%%%%%%%%%%%%%%%%%%%%%%%%%%%%%%   vertical lines
\put(8.5,17.5){\line(0,-1){7} }%5 C-vertical line
\put(8.4,18){r}

\put(11.5,17.5){\line(0,-1){8.5} }%6 A-vertical line
\put(11.4,18){r}

\put(10,-.5){\line(0,1){9.5} }% D-vertical line
\put(9.9,-1){r}

\put(13,-.5){\line(0,1){8} }% B-vertical line
\put(12.9,-1){r}

%%%%%%%%%%%%%%%%%%%%%%%%%%%%%%%%%%%%%%%%%%%%%%%   horizontal lines
\put(13,7.5){\line(1,0){5} }%5 B-horizontal line
\put(18.5,7.4){p}

\put(1.8,9){\line(1,0){8} }%6 full horizontal line
\put(11.5,9){\line(1,0){6.5} }%6 full horizontal line
\put(1,8.9){q}
\put(18.5,8.9){p}

\put(1.8,10.5){\line(1,0){6.6} }%7 C-horizontal line
\put(1,10.4){q}

%%%%%%%%%%%%%%%%%%%%%%%%%%%%%%%%%%%%    dots set up in vertical rows
\thicklines
\multiput(2.5,0)(0,1.5){12}{\circle{.25} }%1
\multiput(4,0)(0,1.5){12}{\circle{.25} }%2
\multiput(5.5,0)(0,1.5){12}{\circle{.25} }%3
\multiput(7,0)(0,1.5){12}{\circle{.25} }%4

\multiput(8.5,0)(0,1.5){4}{\circle{.25} }%5
\multiput(8.5,9)(0,1.5){6}{\circle*{.25} }%5
%\multiput(8.3,5.8)(0,1.5){2}{$\blacksquare$ }%5
\multiput(8.5,6)(0,1.5){2}{\circle*{.25} }

\multiput(10,0)(0,1.5){5}{\circle*{.25} }%6

%\multiput(9.8,7.3)(0,1.5){2}{$\blacksquare$ }%6
\multiput(10,7.5)(0,1.5){2}{\circle*{.25} }

\multiput(11.5,12)(0,1.5){4}{\circle*{.25} }%7

%\multiput(11.3,8.8)(0,1.5){2}{$\blacksquare$ }%7
\multiput(11.5,9)(0,1.5){2}{\circle*{.25} }

\multiput(13,0)(0,1.5){7}{\circle*{.25} }%8
\multiput(13,13.5)(0,1.5){3}{\circle{.25} }

%\multiput(12.8,10.3)(0,1.5){2}{$\blacksquare $}%8
\multiput(13,10.5)(0,1.5){2}{\circle*{.25}}

\multiput(14.5,0)(0,1.5){12}{\circle{.25} }%9
\multiput(16,0)(0,1.5){12}{\circle{.25} }%10
\multiput(17.5,0)(0,1.5){12}{\circle{.25} }%11
\thinlines

%%%%%%%%%%%%%%%%%%%%%%%%%%%%%%%%%%%%%%%%%%   endings
\thinlines
\drawline(1.95,-.55)(2.4,-.1)%1v-ends
\drawline(1.95, .95)(2.4,1.4)%1v-ends
\drawline(1.95,2.45)(2.4,2.9)%1v-ends
\drawline(1.95,3.95)(2.4,4.4)%1v-ends
\drawline(1.95,5.45)(2.4,5.9)%1v-ends
\drawline(1.95,6.95)(2.4,7.4)%1v-ends
\drawline(1.95,8.45)(2.4,8.9)%1v-ends

\drawline(1.95,11.45)(2.4,11.9)%1v-ends
\drawline(1.95,12.95)(2.4,13.4)%1v-ends
\drawline(1.95,14.45)(2.4,14.9)%1v-ends
\drawline(1.95,15.95)(2.4,16.4)%1v-ends

\drawline(3.45,-.55)(3.9,-.1)%1h-ends
\drawline(4.95,-.55)(5.4,-.1)%1h-ends
\drawline(6.45,-.55)(6.9,-.1)%1h-ends
\drawline(7.95,-.55)(8.4,-.1)%1h-ends
\drawline(9.45,-.55)(9.9,-.1)%1h-ends

\drawline(13.95,-.55)(14.4,-.1)%1h-ends
\drawline(15.45,-.55)(15.9,-.1)%1h-ends
\drawline(16.95,-.55)(17.4,-.1)%1h-ends

\multiput(18.2,.7)(0,1.5){12}{\vector(-1,-1){.55}}%v-ends
\multiput(3.2,17.2)(1.5,0){4}{\vector(-1,-1){.55}}%h-ends
\multiput(12.2,17.2)(1.5,0){5}{\vector(-1,-1){.55}}%h-ends

%%%%%%%%%%%%%%%%%%%%%%%%%%%%%%%%%%%%%%%%%%   dashed lines

\multiput(3.85,16.35)(1.5,0){4}{\vector(-1,-1){1.2}}%%1
\multiput(12.85,16.35)(1.5,0){4}{\vector(-1,-1){1.2}}%%1

\multiput(3.85,14.85)(1.5,0){4}{\vector(-1,-1){1.2}}%%2
\multiput(12.85,14.85)(1.5,0){4}{\vector(-1,-1){1.2}}%%2

\multiput(3.85,13.35)(1.5,0){4}{\vector(-1,-1){1.2}}%%3
\multiput(12.85,13.35)(1.5,0){4}{\vector(-1,-1){1.2}}%%3

\multiput(3.85,11.85)(1.5,0){4}{\vector(-1,-1){1.2}}%%4
\multiput(12.85,11.85)(1.5,0){4}{\vector(-1,-1){1.2}}%%4

\dashline[-10]{.1}(3.85,10.35)(2.35,8.85)%%5
       \put(7.33,9.37){\vector(-1,-1){.25}}
\dashline[-10]{.1}(5.35,10.35)(3.85,8.85)%%5
       \put(5.8,9.37){\vector(-1,-1){.25}}
\dashline[-10]{.1}(6.85,10.35)(5.35,8.85)%%5
       \put(4.3,9.37){\vector(-1,-1){.25}}
\dashline[-10]{.1}(8.35,10.35)(6.85,8.85)%%5
       \put(2.9,9.35){\vector(-1,-1){.25}}

\multiput(12.85,10.35)(1.5,0){4}{\vector(-1,-1){1.2}}%%5

\multiput(3.85,8.85)(1.5,0){5}{\vector(-1,-1){1.2}}%%6
\dashline[-10]{.1}(14.35,8.85)(12.85,7.35)%%6
\dashline[-10]{.1}(15.85,8.85)(14.35,7.35)%%6
\dashline[-10]{.1}(17.35,8.85)(15.85,7.35)
\dashline[-10]{.1}(2.25,10.25)(2,10)

%%%%%%%%%%%%%%%%%%%%%%%%%%%%%%%%arrow heads for dashed lines 7
      \put(13.4,7.9){\vector(-1,-1){.25}}
      \put(14.9,7.9){\vector(-1,-1){.25}}
      \put(16.4,7.9){\vector(-1,-1){.25}}

\multiput(3.85,7.35)(1.5,0){5}{\vector(-1,-1){1.2}}%%7
\multiput(14.35,7.35)(1.5,0){3}{\vector(-1,-1){1.2}}%%7

\multiput(3.85,5.85)(1.5,0){5}{\vector(-1,-1){1.2}}%%8
\multiput(14.35,5.85)(1.5,0){3}{\vector(-1,-1){1.2}}%%8

\multiput(3.85,4.35)(1.5,0){5}{\vector(-1,-1){1.2}}%%9
\multiput(14.35,4.35)(1.5,0){3}{\vector(-1,-1){1.2}}%%9

\multiput(3.85,2.85)(1.5,0){5}{\vector(-1,-1){1.2}}%%10
\multiput(14.35,2.85)(1.5,0){3}{\vector(-1,-1){1.2}}%%10

\multiput(3.85,1.35)(1.5,0){5}{\vector(-1,-1){1.2}}%%11
\multiput(14.35,1.35)(1.5,0){3}{\vector(-1,-1){1.2}}%%??5

%\put(10,1.5){\vector(-1,-1){1.35}}%%5
%\multiput(14.5,1.5)(1.5,0){2}{\vector(-1,-1){1.35}}%%11
%\put(16,1.5){\vector(-1,-1){1.35}}%%5
%\put(17.5,1.5){\vector(-1,-1){1.35}}%%5
%\put(17.5,1.5){\vector(-1,-1){1.35}}%%5

\end{picture}
\vspace{3ex}
    \caption{}
    \label{fig:2}
  \end{center}
\end{figure}

We see that Theorem~\ref{th:4.10}
provides us with the information about the homologies of the
complexes $\Gr M_{X^o}$ (and hence $M_X$ for $X=AB, CD$).
In the following  Figure~\ref{fig:2} this information
is presented graphically.
The white circles %in Figure~\ref{fig:2}
mark
those places
$(m,\!n)$ where the homologies
$H^{m,n}(\Gr M_{X^o})$ are zero,
and black nodes mark
those positions where they may be non-zero.

%ef2

\begin{proof}[ Proof of Proposition~\ref{prop:3.2}.]
If either $m\geq2$ or $m=1,\,n\neq 0,1,2$, then,
by Theorem~\ref{th:4.10}, $H^{m,n}(G_X)=0$ for $X=A^o,B^o$.
Equations (\ref{eq:H-AB}) show that $H^{m,n}(G_{AB})=0$ under
the same conditions on $m,n$, and hence $H^{m,n}(M_{AB})=0$
by Proposition~\ref{prop:3.1}. In the same way
using Theorem~\ref{th:4.10} and equations (\ref{eq:H-CD})
we prove the second part of the proposition.
\end{proof}

For a linear map $\tdn: M'\ar M''$ we shall
 use the notations:
\bEq \label{eq:4.26}
\bAr{lcl}
\Ker(\tdn|M')    &=& \Ker(\tdn:M'\ar M'')\,,\\
\IM(\tdn|M'')    &=& \IM(\tdn:M'\ar M'')=\IM(\tdn|_{M'})\,,\\
\Coker(\tdn|M'') &=& \Coker(\tdn:M'\ar M'')\,.
\eAr
\eEq
\bPr
\label{prop:4.11}
Morphism $\tdn_3$ induces the following isomorphisms:

\vspace*{-4ex}
\[
\bAr{ll}
 & \\
\Coker(\tdn|M^{0,n}_A)\arr \Ker(\tdn|M^{0,n-3}_C)\,,&
n\geq 4,\\ & \\
\Coker(\tdn|M^{0,3}_A)\arr \Ker(\tdn_2|M^{0,0}_C)\,,& \\ & \\
\Coker(\tdn|M^{0,n}_B)\arr \Ker(\tdn|M^{0,n-3}_D)\,,&
n< 0,\\ & \\
\Coker(\tdn_2|M^{0,0}_B)\arr \Ker(\tdn|M^{0,-3}_D)\,.&
\eAr
\]
\ePr
\begin{proof}
To prove that a morphism of filtered modules with a differential
induces an isomorphism on homologies it is sufficient
to show that it induces an isomorphism of the initial terms
of their spectral sequences and the initial terms are
of the form $H(\Gr M_X)$.

Because of (\ref{eq:3.4}) and (\ref{eq:AB0}), (\ref{eq:CD0})
it is enough to
establish that $\tdn_3$ induces isomorphisms:
\[
H^{0,n}(G_{A^o})\ar H^{0,n-3}(G_{C^o}) \text{ for } n\geq 3,
\quad
H^{0,n}(G_{B^o})\ar H^{0,n-3}(G_{D^o})  \text{ for }  n\leq 0.
\]
Theorem~\ref{th:4.10} shows that the corresponding
homologies are indeed isomorphic $\gind 0$-modules, hence
we should check  only
that $\tdn_3$ maps each of the highest weight vectors in
$ H^{0,n}(G_{A^o})$ (resp.  $H^{0,n}(G_{B^o})$ ) to a non-zero
element of $H^{0,n-3}(G_{C^o})$ (resp. $H^{0,n-3}(G_{D^o})$~).
The
representatives for highest weight vectors are
\[
z_+^{n},\,\,
d_1^+z_+^{n-1} z_-,\,\,
d_1^+d_2^+z_+^{n-2} z_-^{2},\,\,
d_1^+d_2^+d_3^+z_+^{n-3}, z_-^{3}
\]
for $H^{0,n}(G_{A^o})$, and
\[
\partial^{-n}_-,\,\,
d_1^+\partial^{-n-1}_-\partial_+,\,\,
d_1^+d_2^+\partial^{-n-2}_-\partial^{2}_+, \,\,
d_1^+d_2^+d_3^+\partial^{-n-3}_-\partial^{3}_+
\]
 for $H^{0,n}(G_{B^o})$. Now it is immediate to see that the images
are non-zero.
\end{proof}
\bPr
\label{prop:4.12}
Morphisms $\tdn'_4$, $\tdn''_4$
induce  isomorphisms:

\vspace*{-4ex}
\[
\bAr{ll}
 & \\
\tdn'_4:& \Coker(\tdn|M^{0,2}_A)\arr \Ker(\tdn|M^{-1,0}_D)\,,\\
& \\
\tdn''_4:& \Coker(\tdn|M^{1,0}_A)\arr \Ker(\tdn|M^{0,-2}_D)\,.
\eAr
\]
\ePr
\begin{proof}
Again it is sufficient to establish that the maps induce
isomorphisms of the initial terms of the spectral sequences
that are isomorphic $\gind 0$-modules because of
Theorem~\ref{th:4.10}. Thus  again we have to check the images of
the highest vectors of $H^{0,2}(G_{A^o})$ and $H^{1,0}(G_{A^o})$.

For $H^{0,2}(G_{A^o})$  the representatives of the highest
weight vectors are
\[
z_+^{2},\,\,
d_1^+z_+^{1} z_-,\,\,
d_1^+d_2^+ z_-^{2},\,\,
\]
and their images
\[
d^+_{123}\De^{\!-},\,
d^+_1 d^-_1 d^+_{23}\De^{\!-},\,
d^+_{12}(d^-_{12}d^+_3+d^-_1d^+_2d^-_3)\De^{\!-},\,
\]
are indeed non-zero in $H^{0,-2}(G_{D^o})$.

Similarly for $H^{1,0}(G_{A^o})$  the representatives  are
\[
x_1,\,\, d^+_1x_2-d^+_2x_1,\,\,
d^+_1d^+_2x_3+d^+_2d^+_3x_1+d^+_3d^+_2x_1.
\]
We use (\ref{eq:2.12n}) and calculate in
$G_D=\La(\gind{-1})\otimes V_D$. We leave to the reader to
make the calculations and see that the results are indeed
non-zero.
\end{proof}

%s5
\section{ Homology of $M_X$ and secondary singular vectors.}
\label{sec:5}

We return to the spectral sequence for $M_X$,
decribed in Section~\ref{sec:3}.
We are particulary interested in eight cases when $X=A$ or $D$,
which have been left unfinished in the previous section.
The description of the initial term $E^0=H(\Gr M_X)$
of the spectral sequences in these cases follows from Theorem~\ref{th:4.10}
and was presented in Figure~2.
In this section we
compute the subsequent terms of these spectral sequences
and hence the homology.\\

\bTh
\label{th:5.1}
There are the following isomorphisms of $E(3,6)$-modules:
\romanparenlist
\begin{enumerate}
\item{\rm} %i
$\,H^{0,0}(M_{A})= \CC$, $\,\,\,H^{1,2}(M_{A})=0$,
\vspace{1ex}
\item{\rm} %ii
$\,H^{0,1}(M_{A})\simeq H^{1,1}(M_{A})\simeq I(0,0;1;-1)$,
\vspace{1ex}
\item{\rm} %iii
$\,H^{0,0}(M_{D})= 0$, $\,\,\,H^{-1,-2}(M_{D})=\CC$,
\vspace{1ex}
\item{\rm} %iv
$\,H^{0,-1}(M_{D})\simeq H^{-1,-1}(M_{D})\simeq I(0,0;1;-1)$.
\end{enumerate}
\eTh

In the following we denote by $\tdn$ any of the morphisms
$\tdn,\tdn_1,\ldots,\tdn_6$ (see Figure~\ref{fig:1}) assuming that
it is clear from the context which map is considered.

%\vspace*{1cm}
\begin{figure}[htbp]
  \begin{center}
    \leavevmode
  \setlength{\unitlength}{0.25in}
%%%%%%%%%%%%%%%%%%%%%%%%%%%%%%%%%%%%%%%%%%%%%%%   corner labels
%\begin{picture}(20,19)
\begin{picture}(21,19)

\put(18.7,17.5){A}
\put(18.7,-1){B}

\put(1,17.5){C}
\put(1,-1){D}

%%%%%%%%%%%%%%%%%%%%%%%%%%%%%%%%%%%%%%%%%%%%%%%   vertical lines
\put(8.5,17.5){\line(0,-1){7} }%5 C-vertical line
\put(8.4,18){r}

\put(11.5,17.5){\line(0,-1){8.5} }%6 A-vertical line
\put(11.4,18){r}

\put(10,-.5){\line(0,1){9.5} }% D-vertical line
\put(9.9,-1){r}

\put(13,-.5){\line(0,1){8} }% B-vertical line
\put(12.9,-1){r}

%%%%%%%%%%%%%%%%%%%%%%%%%%%%%%%%%%%%%%%%%%%%%%%   horizontal lines
\put(13,7.5){\line(1,0){5} }%5 B-horizontal line
\put(18.5,7.4){p}

\put(1.8,9){\line(1,0){8.2} }%6 full horizontal line
\put(11.5,9){\line(1,0){6.5} }%6 full horizontal line
\put(1,8.9){q}
\put(18.5,8.9){p}

\put(1.8,10.5){\line(1,0){6.8} }%7 C-horizontal line
\put(1,10.4){q}

%%%%%%%%%%%%%%%%%%%%%%%%%%%%%%%%%%%%    dots set up in vertical rows
\thicklines
\multiput(2.5,0)(0,1.5){12}{\circle{.25} }%1
\multiput(4,0)(0,1.5){12}{\circle{.25} }%2
\multiput(5.5,0)(0,1.5){12}{\circle{.25} }%3
\multiput(7,0)(0,1.5){12}{\circle{.25} }%4

\multiput(8.5,0)(0,1.5){4}{\circle{.25} }%5
\multiput(8.5,9)(0,1.5){6}{\circle{.25} }%5
\multiput(8.3,5.9)(0,1.5){1}{$\bigstar$ }%5
\multiput(8.4,7.4)(0,1.5){1}{$\spadesuit$ }%5

\multiput(10,0)(0,1.5){6}{\circle{.25} }%6
%\multiput(9.8,7.3)(0,1.5){}{$\b $ }%6

\multiput(11.5,10.5)(0,1.5){5}{\circle{.25} }%7
%\multiput(11.3,10.3)(0,1.5){1}{\circle{.25} }%7
%\multiput(11.3,8.8)(0,1.5){2}{$\blacksquare$ }%7

\multiput(13,0)(0,1.5){7}{\circle{.25} }%8
\multiput(13,12)(0,1.5){4}{\circle{.25} }%8
\multiput(12.8,10.4)(0,1.5){1}{$\blacklozenge $}%8
%\multiput(12.8,11.8)(0,1.5){1}{\circle{.25}}%8

\multiput(14.5,0)(0,1.5){12}{\circle{.25} }%9
\multiput(16,0)(0,1.5){12}{\circle{.25} }%10
\multiput(17.5,0)(0,1.5){12}{\circle{.25} }%11
\thinlines

%%%%%%%%%%%%%%%%%%%%%%%%%%%%%%%%%%%%%%%%%%   endings
\thinlines
\drawline(1.95,-.55)(2.4,-.1)%1v-ends
\drawline(1.95, .95)(2.4,1.4)%1v-ends
\drawline(1.95,2.45)(2.4,2.9)%1v-ends
\drawline(1.95,3.95)(2.4,4.4)%1v-ends
\drawline(1.95,5.45)(2.4,5.9)%1v-ends
\drawline(1.95,6.95)(2.4,7.4)%1v-ends
\drawline(1.95,8.45)(2.4,8.9)%1v-ends

\drawline(1.95,11.45)(2.4,11.9)%1v-ends
\drawline(1.95,12.95)(2.4,13.4)%1v-ends
\drawline(1.95,14.45)(2.4,14.9)%1v-ends
\drawline(1.95,15.95)(2.4,16.4)%1v-ends

\drawline(3.45,-.55)(3.9,-.1)%1h-ends
\drawline(4.95,-.55)(5.4,-.1)%1h-ends
\drawline(6.45,-.55)(6.9,-.1)%1h-ends
\drawline(7.95,-.55)(8.4,-.1)%1h-ends
\drawline(9.45,-.55)(9.9,-.1)%1h-ends

\drawline(13.95,-.55)(14.4,-.1)%1h-ends
\drawline(15.45,-.55)(15.9,-.1)%1h-ends
\drawline(16.95,-.55)(17.4,-.1)%1h-ends

\multiput(18.2,.7)(0,1.5){12}{\vector(-1,-1){.55}}%v-ends
\multiput(3.2,17.2)(1.5,0){5}{\vector(-1,-1){.55}}%h-ends
\multiput(12.2,17.2)(1.5,0){5}{\vector(-1,-1){.55}}%h-ends

%%%%%%%%%%%%%%%%%%%%%%%%%%%%%%%%%%%%%%%%%%   dashed lines

\multiput(3.85,16.35)(1.5,0){4}{\vector(-1,-1){1.2}}%%1
\multiput(12.85,16.35)(1.5,0){4}{\vector(-1,-1){1.2}}%%1

\multiput(3.85,14.85)(1.5,0){4}{\vector(-1,-1){1.2}}%%2
\multiput(12.85,14.85)(1.5,0){4}{\vector(-1,-1){1.2}}%%2

\multiput(3.85,13.35)(1.5,0){4}{\vector(-1,-1){1.2}}%%3
\multiput(12.85,13.35)(1.5,0){4}{\vector(-1,-1){1.2}}%%3

\multiput(3.85,11.85)(1.5,0){4}{\vector(-1,-1){1.2}}%%4
\multiput(12.85,11.85)(1.5,0){4}{\vector(-1,-1){1.2}}%%4

\dashline[-10]{.1}(3.85,10.35)(2.35,8.85)%%5
       \put(7.33,9.37){\vector(-1,-1){.25}}
\dashline[-10]{.1}(5.35,10.35)(3.85,8.85)%%5
       \put(5.8,9.37){\vector(-1,-1){.25}}
\dashline[-10]{.1}(6.85,10.35)(5.35,8.85)%%5
       \put(4.3,9.37){\vector(-1,-1){.25}}
\dashline[-10]{.1}(8.35,10.35)(6.85,8.85)%%5
       \put(2.9,9.35){\vector(-1,-1){.25}}

\multiput(14.35,10.35)(1.5,0){3}{\vector(-1,-1){1.2}}%%5

\multiput(3.85,8.85)(1.5,0){4}{\vector(-1,-1){1.2}}%%6
\dashline[-10]{.1}(14.35,8.85)(12.85,7.35)%%6
\dashline[-10]{.1}(15.85,8.85)(14.35,7.35)%%6
\dashline[-10]{.1}(17.35,8.85)(15.85,7.35)
\dashline[-10]{.1}(2.25,10.25)(2,10)

%%%%%%%%%%%%%%%%%%%%%%%%%%%%%%%%arrow heads for dashed lines 7
      \put(13.4,7.9){\vector(-1,-1){.25}}
      \put(14.9,7.9){\vector(-1,-1){.25}}
      \put(16.4,7.9){\vector(-1,-1){.25}}

%@@

\dashline[-10]{.2}(10.3,16.8)(8.5,15)
       \put(8.8,15.35){\vector(-1,-1){.25}}

%\multiput(4,16.5)(1.5,0){4}{\vector(-1,-1){1.35}}%%1
\dashline[-10]{.2}(11.5,16.5)(8.5,13.5)
       \put(8.8,13.85){\vector(-1,-1){.25}}
%\multiput(13,16.5)(1.5,0){4}{\vector(-1,-1){1.35}}%%1

%\multiput(4,15)(1.5,0){4}{\vector(-1,-1){1.35}}%%2
\dashline[-10]{.2}(11.5,15)(8.5,12)%%2
       \put(8.8,12.35){\vector(-1,-1){.25}}
%\multiput(13,15)(1.5,0){4}{\vector(-1,-1){1.35}}%%2

%\multiput(4,13.5)(1.5,0){4}{\vector(-1,-1){1.35}}%%3
\dashline[-10]{.2}(11.5,13.5)(8.5,10.5)%%3
       \put(8.8,10.85){\vector(-1,-1){.25}}
%\multiput(13,13.5)(1.5,0){4}{\vector(-1,-1){1.35}}%%3

%\multiput(4,12)(1.5,0){4}{\vector(-1,-1){1.35}}%%4

\thicklines
\put(11.5,12){\vector(-1,-1){2.81}}
\put(13,9){\vector(-1,-1){2.81}}

%*!*?
\put(11.49,10.49){\vector(-1,-2){1.44}}
\put(11.51,10.51){\line(-1,-2){1.41}}
\put(11.47,10.47){\line(-1,-2){1.41}}

\dashline[-5]{.3}(12.8,10.5)(9.1,7.9)%%last line
      \put(9.1,7.9){\vector(-3,-2){.3}}
\thinlines

%\multiput(4,7.5)(1.5,0){5}{\vector(-1,-1){1.35}}%%7
\dashline[-10]{.2}(13,7.5)(10,4.5)%%7
      \put(10.4,4.9){\vector(-1,-1){.25}}
%\multiput(14.5,7.5)(1.5,0){3}{\vector(-1,-1){1.35}}%%7

%\multiput(4,6)(1.5,0){5}{\vector(-1,-1){1.35}}%%8
\dashline[-10]{.2}(13,6)(10,3)%%8
      \put(10.4,3.4){\vector(-1,-1){.25}}
%\multiput(14.5,6)(1.5,0){3}{\vector(-1,-1){1.35}}%%8

%\multiput(4,4.5)(1.5,0){5}{\vector(-1,-1){1.35}}%%9
\dashline[-10]{.2}(13,4.5)(10,1.5)%%9
      \put(10.4,1.9){\vector(-1,-1){.25}}
%\multiput(14.5,4.5)(1.5,0){3}{\vector(-1,-1){1.35}}%%9

%\multiput(4,3)(1.5,0){5}{\vector(-1,-1){1.35}}%%10
\dashline[-10]{.2}(13,3)(10,0)%%10
     \put(10.4,.4){\vector(-1,-1){.25}}
%\multiput(14.5,3)(1.5,0){3}{\vector(-1,-1){1.35}}%%10

%\multiput(4,1.5)(1.5,0){3}{\vector(-1,-1){1.35}}%%11

%\put(8.5,1.5){\vector(-1,-1){1.35}}%%5
%\put(10,1.5){\vector(-1,-1){1.35}}%%5

\dashline[-10]{.2}(13,1.5)(11,-.5)%%  end
\dashline[-10]{.2}(13,0)(12.5,-.5)%%  end

%%###

\multiput(3.85,7.35)(1.5,0){5}{\vector(-1,-1){1.2}}%%7
\multiput(14.35,7.35)(1.5,0){3}{\vector(-1,-1){1.2}}%%7
\multiput(3.85,5.85)(1.5,0){5}{\vector(-1,-1){1.2}}%%8
\multiput(14.35,5.85)(1.5,0){3}{\vector(-1,-1){1.2}}%%8
\multiput(3.85,4.35)(1.5,0){5}{\vector(-1,-1){1.2}}%%9
\multiput(14.35,4.35)(1.5,0){3}{\vector(-1,-1){1.2}}%%9
\multiput(3.85,2.85)(1.5,0){5}{\vector(-1,-1){1.2}}%%10
\multiput(14.35,2.85)(1.5,0){3}{\vector(-1,-1){1.2}}%%10
\multiput(3.85,1.35)(1.5,0){5}{\vector(-1,-1){1.2}}%%11
\multiput(14.35,1.35)(1.5,0){3}{\vector(-1,-1){1.2}}%%??5

%\put(10,1.5){\vector(-1,-1){1.35}}%%5
%\multiput(14.5,1.5)(1.5,0){2}{\vector(-1,-1){1.35}}%%11
%\put(16,1.5){\vector(-1,-1){1.35}}%%5
%\put(17.5,1.5){\vector(-1,-1){1.35}}%%5
%\put(17.5,1.5){\vector(-1,-1){1.35}}%%5

\end{picture}
\vspace{3ex}
    \caption{}
    \label{fig:3}
  \end{center}
\end{figure}

>From now on we combine the differential $E(3,6)$-modules $M_{AB}$  and $M_{CD}$
in one differential $E(3,6)$-module $\MM$ represented by
%the following
Figure~\ref{fig:3}. The $E(3,6)$-module $\MM$ equals to the direct sum
\[
\MM= \sum_{(m,n)\neq (0,0)} M_{A}^{m,n}\oplus
M_{B}\oplus M_{C} \oplus
\sum_{(m,n)\neq (0,0)} M_{D}^{m,n},
\]
and the maps are the same as in Figure~\ref{fig:1} except
for the map  $\tilde\tdn:  M_A^{1,1} \arr  M_D^{-1,-1}$ which
is defined to be equal to the composition:
\begin{displaymath}
  M_A^{1,1}
\overset{\triangledown}{\longrightarrow}
  M_A^{0,0}\simeq M_D^{0,0}
\overset{\triangledown}{\longrightarrow}
  M_D^{-1,-1} \, .
\end{displaymath}

%efig3

White nodes in Figure~\ref{fig:3} mark the positions where the
kernel of the outcoming map equals to the image of the incoming one
(i.e. the corresponding homology of $\MM$ is zero).
The black marks denote the places with  non-zero homology. Namely,
the star refers to the trivial module $\CC$, the diamond to
$I(0,0;1;-1)$
and the spade to the module $P=I(0,0;1;-1)\oplus \CC$.

The above description of the homologies
follows
from Theorem~\ref{th:5.1},
Propositions~\ref{prop:3.2}, \ref{prop:4.11}, \ref{prop:4.12},
and \ref{prop:5.25}.
Saying that $P$ is a direct sum we have taken into account that
an extension
\[
0\arr \CC \arr P \arr I(0,0;1;-1) \arr 0\,
\]
is necessary a direct sum, because
all eigenvalues of $Y$ on $I(0,0;1;-1)$ are strictly negative,
but $Y$ acts on $\CC$ with eigenvalue zero.

\bCo
\label{cor:Zcpls}
The complex $\MM$ decomposes in a direct sum of
$\ZZ$-graded complexes, that are infinite in both directions
and consist of generalized Verma modules $M^{m,n}_X$.
Homologies of these
complexes are all zero except those that appear at terms
$M^{1,1}_A \simeq M(1,0;1;-\tfrac{2}{3})$,
$M^{-1,-1}_D\simeq M(0,1;1;\tfrac{2}{3})$ and
$M^{-1,-2}_D\simeq M(0,1;2;\tfrac{5}{3})$.
\eCo

We use below
the notations (\ref{eq:4.26}).
\bCo
\label{cor:5.2}
For the $E(3,6)$-modules $M(p,q;r;y_X)$ corresponding to
the white nodes of Figure~\ref{fig:3} (and for $M(0,0;0;0)$)
$\Ker \tdn=\IM \tdn$
is a unique non-trivial submodule, it is
isomorphic to the irreducible quotient of the previous module of the complex,
and the quotient by it is isomorphic to $I(p,q;r;y_X)$.
Equivalently,
$\IM(\tdn|M^{m,n}_X)$
        is a unique non-trivial
                submodule of  $M^{m,n}_X$
except when
$M^{m,n}_X=M^{1,1}_A,\,M^{-1,-1}_D,\,M^{-1,-2}_D$.

Thus, all degenerate generalized Verma modules $M(p,q;r;y_X)$,
except for $M(1,0;1;y_A)$, $M(0,1;1;y_D)$ and $M(0,1;2;y_D)$,
have a unique non-trivial submodule.

\eCo
\bCo
\label{cor:eXsubm}
The only nontrivial submodules of the excluded above modules
$M(1,0;1;y_A)$,
$M(0,1;2;y_D)$
and
$M(0,1;1;y_D)$
are the following
\romanparenlist
\bN
\item %i
 $\IM\tdn\,\subset\,\Ker \tdn$ for $M(1,0;1;y_A)$,
with subquotients
respectively isomorphic to \\
 $I(2,0;2;y_A)$, $I(0,0;1;y_A)$, $I(1,0;1;y_A)$,
\item %ii
 $\IM\tdn\,\subset\,\Ker \tdn$ for $M(0,1;2;y_D)$,
with subquotients
  $I(0,0;1;y_D)$, $\CC$, $I(0,1;2;y_D)$,
\item %iii
 $S_i,\,i=0,1,2,3$ for $M(0,1;1;y_D)$, where
\[
S_0=\IM\tilde\tdn,\quad\,S_1=\IM\tdn,\,\quad S_2=U(L)q_+ +S_0,\quad
\,S_3=\Ker \tdn\,,
\]
and $q_+$ is defined before Lemma~\ref{lem:5.23}. One has:
$S_0=S_1\cap S_2$, $S_3=S_1+S_2$,
\[
S_0\simeq I(1,0;1;y_A),\quad S_3/S_0\simeq \CC \oplus I(0,0;1;y_A),\quad
M(0,1;1;y_D)/
S_3\simeq I(0,1;1;y_D)\,.
\]
 \eN
\eCo
The results of Corollary~\ref{cor:eXsubm} will be established while
proving Theorem~\ref{th:5.1}.
To prove Corollary~\ref{cor:5.2} we need the following lemma.
%
%%% new version
%
\bLe
\label{lem:5.3}
 Let
$M''\overset{\,\tdn''}{\leftarrow} M'
\overset{\,\tdn'}{\leftarrow} M$
be an exact sequence of highest weight modules over $L$
with highest weight vectors $m''$, $m'$ and $m$ respectively.
Suppose that $\tdn'(m)$ is a unique (up to a constant factor) non-trivial
singular vector of $M'$. Then
\[\IM\tdn' = \Ker\tdn''\]
is an irreducible $L$-submodule of $M'$.

Suppose also that $\tdn''(m')$ is a unique (up to a constant factor)
non-trivial singular vector of $M''$.
Then
\[
\IM\tdn''\simeq \Coker\tdn'=M'/\IM\tdn'
\]
is irreducible too, thus $\,\IM\tdn'$ is a unique non-trivial
submodule in $M'$.
\eLe
\begin{proof}
Due to exactness, the module $\Ker\tdn''$ is a quotient of the
highest weight module $M'$. Hence, having by the conditions
only trivial singular vectors, it is irreducible.

Furthermore,
%since $\tdn''$ is a non-zero map and
by exactness
we have the isomorphism of $\Coker\tdn'$ and $\IM\tdn''$.
Due to uniqueness
of a non-trivial singular vector in $M''$,
we conclude, as above, that $\IM\tdn''$ is irreducible.
\end{proof}

\begin{proof}
[Proof of Corollary~\ref{cor:5.2}.] We use
Corollary~\ref{cor:Zcpls}
to provide
exact sequences and Corollary~\ref{cor:2.11} to ensure
the uniqueness of non-trivial singular vectors.

We also get the fact that $\IM(\tdn|M^{0,0}_A)$
is irreducible applying the first part of Lemma~\ref{lem:5.3} to
the sequence
$ \CC \leftarrow M^{0,0}_A \leftarrow M^{1,1}_A $.
Then $M^{0,0}_A/\IM(\tdn|M^{0,0}_A)=
H^{0,0}(M_{A})\simeq \CC$ by
Theorem~\ref{th:5.1}, so we get the statement
 for $M^{0,0}_A=M^{0,0}_D$.
\end{proof}
\vspace{1ex}

To prove
 Theorem~\ref{th:5.1}
we are to compute the differentials in the spectral sequence
$\{E^i(M_X),\,\tdn^{(i)}\}$ for the homology $H(M_X)$, $X=A,D$,
and for this we need to consider representatives of
various cycles.

\bDe\label{def:5.4}
We say that an element $s\in M^{m,n}_X$ represents the highest
weight vector of $H^{m,n}(G_X)$ if
\bN
\item
$s$ is a non-zero $\gind 0$-highest weight vector (in $M^{m,n}_X$),
\item
the image $[s]$ of $s$ in $\Gr M_X$ is a cycle,
\item
$[s]$ belongs to $G_X \subset \Gr M_X$.
\eN
It follows that the homology class of $[s]$ in $H^{m,n}(G_X)$ is
a $\gind 0$-highest weight vector.\\
We say that $s_0,\ldots,s_k\in M^{m,n}_X $
represent a basis of
$H^{m,n}(G_X)$ if
\bN
\item[(1$'$)]
the span of ${s_0,\ldots,s_k}$ is a $\gind 0$-submodule in $M^{m,n}_X$,
\item[(2$'$)]
all $[s_i]$ are cycles and belong to  $G_X \subset \Gr M_X$,
\item[(3$'$)]
$\{\,[s_i]\,\}$ give us a basis for $H^{m,n}(G_X)$.
\eN
\eDe

\bDe\label{def:5.5}
We say that $s\in M^{m,n}_X$ represents a singular vector of $H^{m,n}(M_X)$
if
\bN
\item
$s$ is a non-zero $\gind 0$-highest weight vector,
\item
$s$ is a cycle in $M_X$,
\item
$e_0s\in \IM\tdn$, $e'_0s\in \IM\tdn$.
\eN
\eDe
\newcommand{\shc}{\equiv}
Whenever $c_1$, $c_2$ are cycles, the notation $c_1 \shc c_2$ means
$c_1$ and  $c_2$ {\it belong to the same homology class}.

Let us notice that
due to Remark~\ref{rem:W} the terms $E^i(M_{X})^{m,n}$ are
$W$-modules and clearly they are
$\gind 0$-modules as well. Thus we may consider
the action
on these terms
of the following subalgebra $\Ww$ of $E(3,6)$
\begin{equation}
\label{eq:5.1}
\Ww=W+\gind 0\simeq W\oplus s\ell(2)\,.
\end{equation}
Given a $\gind 0$-module $V$ with the trivial
action of $L_1^\Ww =L_1^W$
we get
 an isomorphism:
\begin{equation}
\label{eq:5.2}
\Sym (\fg_{-2}) \otimes_\CC V\simeq
\Ind^\Ww_{L_0^\Ww}( V)\,.
\end{equation}
We conclude that
(\ref{eq:Wisom}), (\ref{eq:3.4})
lead us to the following
isomorphisms of $\Ww$-modules:
\begin{equation}
\label{eq:5.3}
 \Gr M_X \simeq
\Ind^\Ww_{L_0^\Ww}(\Lambda (\fg_{-1}) \otimes V_X)\,,
\qquad
H^{m,n}(\Gr M_X) \simeq
\Ind^\Ww_{L_0^\Ww}(H^{m,n}(G_X))\,.
\end{equation}
Therefore we can  speak of $\Ww$-singular vectors in $\Gr M_X$ or
in $H^{m,n}(\Gr M_X)$ (see [KR1] for general properties
of singular vectors).

\begin{remark}
\label{rem:5.6}
Formulae (\ref{eq:1.3}) and
(\ref{eq:Wsub}) show that the central
element $\sum x_i\partial_i \in g\ell(3)\subset W_3$
corresponds to $\tfrac{3}{2}Y\in g\ell(3) \subset W \subset \Ww$.
\end{remark}
\vspace*{1ex}

Because of Remark~\ref{rem:5.6} we notice that
  the following $g\ell(3)\oplus s\ell(2)$-modules
are isomorphic to each other
\begin{equation}
\label{eq:5.4}
V=
F(0,1;0;-\tfrac{2}{3})
\simeq \La^2\otimes P(0,-\tfrac{2}{3})\simeq \gind {-2}
\simeq (\CC^3)^*\otimes \CC
\simeq V_D^{-1,0}=\lsp{\partial_1,\partial_2,\partial_3}
\,.
\end{equation}

Let $T^k = \Ind^\Ww_{L_0^\Ww}(\La^k V)$
 with $V$ defined by
the isomorphisms (\ref{eq:5.4}) above.
It is clear that with respect to its $W_3$-module structure
the module $T^k$ is the dual
to the differential forms module $\Om^k_3$. Hence
dualizing the De Rham complex $\Om^\bullet_3$
we get an exact sequence of $\Ww$-modules:
\begin{equation}
\label{eq:5.5}
0
\leftarrow
\CC \overset{\,\si_0}{\longleftarrow} T^0
\overset{\,\si_1}{\longleftarrow} T^1
\overset{\,\si_2}{\longleftarrow} T^2
\overset{\,\si_3}{\longleftarrow} T^3 \leftarrow 0\,.
\end{equation}

Denote $R=\Ker(\si_1)=\Coker (\si_3)$
and $Q=\Ind^\Ww_{L_0^\Ww}\,(\La^0\otimes P(1,-1))=
\Ind^\Ww_{L_0^\Ww}\,V_A^{0,1} $.
\bLe
\label{lem:5.7} There are the following isomorphisms of $\Ww$-modules:
\arabicparenlist
\begin{enumerate}
\item%{\rm :} %%1
$\,
E^0(M_{A})^{0,0}\simeq T^0
\,$, $\,\,\,\,\qquad\,
E^0(M_{A})^{1,1}\simeq T^1\oplus Q
$,
\vspace{1.5ex}
\item%{\rm :} %%2
$\,
E^0(M_{A})^{0,1}\simeq T^2\oplus Q
$, $\,\,\,\,\,\,
E^0(M_{A})^{1,2}\simeq T^3
$,
\vspace{1.5ex}
\item%{\rm :} %%3
$\,
E^0(M_{D})^{-1,-2}\simeq T^0
$,$\,\,\,\qquad
E^0(M_{D})^{0,-1}\simeq T^1\oplus Q
$,
\vspace{1.5ex}
\item%{\rm :} %%4
$\,
E^0(M_{D})^{-1,-1}\simeq T^2\oplus Q
\,$, $
\,E^0(M_{D})^{0,0}\,\simeq T^3
$,
\vspace{1.5ex}
\end{enumerate}
\eLe
Statements (1) and (2)
%of Lemma~\ref{lem:5.7}
(resp. statements (3), (4) ) follow from Theorem~\ref{th:4.10}
as soon as we take into account equations (\ref{eq:5.3})
and accomodate to
the new notations.

\bPr
\label{prop:5.8}
 There are the following isomorphisms of $\Ww$-modules:
\arabicparenlist
\begin{enumerate}
%\vspace{1.5ex}
\item%{\rm :} %%1
$\,
E^1(M_{A})^{0,0}\simeq \CC
\,$, $\,\,\,\qquad\quad\,\,
E^1(M_{A})^{1,1}\simeq R \oplus Q
$,
\vspace{1.5ex}
\item%{\rm :} %%2
$\,
E^1(M_{A})^{0,1}\simeq R\oplus Q
\,$, $\,\,\,\,\quad\,\,
E^1(M_{A})^{1,2}\simeq 0
$,
\vspace{1.5ex}
\item%{\rm :} %%3
the spectral sequence $\{E_i(M_{A})\}$
degenerates at $E^1$,
that is \\
$\,E^1(M_{A})^{m,n}\simeq E^\infty(M_{A})^{m,n}$.
\end{enumerate}
\ePr
To prove the proposition we shall calculate the differentials
$\tdn^{(i)}$
starting with $\tdn^{(0)}$. We need first to find
the representatives of the bases of $H^{m,n}(G_{A})\subset E^0(M_{A})$
for $(m,n)=(1,2),(0,1),(1,1),(0,0)$.
This is done in the following lemmae.
\vspace{1ex}

Let $\ze_i= d^-_iz_+ -d^+_iz_- $.
\bLe
\label{lem:5.9} The element
$
s=\tfrac{1}{3}(\al_-z_+^2 - \al_0z_+z_-  + \al_+ z_-^2)\in
M^{1,2}_{A}
$,
where
\[
\al_\pm=d^\pm_1d^\pm_2x_3+d^\pm_2d^\pm_3x_1+d^\pm_3d^\pm_1x_2\,,
\quad \al_0= f_3 \al_+ \,(\,= e_3 \al_-),
\]
represents the unique highest weight vector of $H^{1,2}(G_{A})$,
%(and also represents a basis of $H^{1,2}(G_{A})$).
and $\,\Ind^\Ww_{L_0^\Ww}\lsp{\,s\, }\simeq T^3$.
\eLe
\begin{proof}
Clearly $e_3\al_-=\al_0$, $\,e_3\al_0=2\al_+$, $\,e_3\al_+=0$, hence
\[
e_3 s=\tfrac{1}{3}(\al_0z_+^2
- \al_0z_+^2
- 2\al_+z_+z_-
+ 2\al_+ z_+z_-)=0\,.
\]
Similarly $f_3s=0$
and $s\ell(3)s=0$. Also $Y\!\cdot s=(-2)s$ and
\bEq \label{eq:5.6}
\tdn s=\dhind 1\ze_1+
        \dhind 2\ze_2+
        \dhind 3\ze_3\,.
\eEq
We see that  $\tdn s \in F_2 M_{A}$.
On the other hand  $s \in F_2 M_{A}$, therefore
$\tdn [s]=0$ by the definition of the
differential of $\Gr M_{A}$. That means $[s]$ is a cycle and it
evidently generates a one-dimensional subspace
$\CC\,s\,= H^{1,2}(G_{A})\simeq \La^0\otimes P(0,-2)$
in $H^{1,2}(\Gr M_{A})$.
\end{proof}
It follows from the lemma that
the image of $[s]$ in $H^{1,2}(\Gr M_A)=E^0(M_{A})^{1,2}$
is the trivial $\Ww$-sigular vector generating
$ E^0(M_{A})^{1,2}\simeq T^3$.
\bLe
\label{lem:5.10}
Vectors $z_+,\,\ze_1\in M^{0,1}_{A}$ represent the highest
weight vectors of $H^{0,1}(G_{A})$,
and
 %%%$z_+, z_-,\,\ze_1,\ze_2,\ze_3$
    $\{z_\pm,\,\ze_i\}$
represent the basis of $H^{0,1}(G_{A})$.
Also
$\Ind^\Ww_{L_0^\Ww}\lsp{z_+,z_- }\simeq Q$ and
$\,\Ind^\Ww_{L_0^\Ww}\lsp{\,\ze_i\,,i=1,2,3 }\simeq T^2$.
\eLe
%$\Ind^\Ww_{L_0^\Ww}\lsp{ }\simeq
This is easy to check, and
we leave it to the reader.
\bRe
\label{rem:5.11} Considering $E^0(M_{A})^{0,1}=H^{0,1}(M_{A})$
as a $\Ww$-module, we conclude that $[z_+]$ is a $\Ww$-singular vector that
generates a submodule $Q$
and $[\ze_1]$
is a $\Ww$-singular vector that
generates a $\Ww$-submodule isomorphic to $T^2$.
Of course $z_+$ also represents an
$\lag$-singular vector of $H^{0,1}(M_{A})$.
\eRe
%

%%% newcommands on Cyclic sums --

\newcommand{\Cedd}[3]{\mbox{\tt C}_e ( d^{#1}_\cdot d^{#2}_\cdot d^{#3}_\cdot )}

\newcommand{\Codd}[3]{\mbox{\tt C}_o ( d^{#1}_\cdot d^{#2}_\cdot d^{#3}_\cdot )}

\newcommand{\Cedx}[2]{\mbox{\tt C}_e ( d^{#1}_\cdot d^{#2}_\cdot x_\cdot )}

\newcommand{\Codx}[2]{\mbox{\tt C}_o ( d^{#1}_\cdot d^{#2}_\cdot x_\cdot )}

%%% %% %% %% %% %%%

Let
$\mx{\tt C}_{e} (u_\cdot v_\cdot w_\cdot)
=\sum_{even} u_i v_j w_k$
and
$\mx{\tt C}_{o} (u_{\cdot} v_{\cdot} w_{\cdot})
=\sum_{odd} u_i v_j w_k$ be the sums over all even and
odd permutations respectively, where we keep the
order of the letters but permute the indices,
for example,
\bEaz
\Cedd+-+&=&d^+_1d^-_2d^+_3  +d^+_2d^-_3d^+_1  +d^+_3d^-_1d^+_2\,,\\
\Cedx+-&=&d^+_1d^-_2x_3 +  d^+_2d^-_3x_1 +  d^+_3d^-_1x_2 \,.
\eEaz
By the definition of $\tdn$ one has: $\tdn \Cedx+- z_\pm = \Cedd+-\pm$ and
$\tdn \Codx+- z_\pm = \Codd+-\pm$.

 Also
\[\Cedd++- +\Codd++-=0\quad \text{ and }\quad\Cedx++ +\Codx++=0 \]
because $[d^+_i,d^+_j]=0$,
and similarly for the pair of minuses.

The relation $[d^+_i,d^-_j] +[d^+_j,d^-_i]  =0\,$ implies
\bEa
 \Cedx+- \,+\,\Codx-+ \,+\,\,
       \Codx+- \,\,+\,\,\Cedx-+ &=&0 \\ \nonumber
 \Cedd+-\pm +\Codd-+\pm +\Codd+-\pm +\Cedd-+\pm &=&0\,, \nonumber
\eEa
hence
\bEq \label{eq:5.8}
\Codd+-+ +\Cedd+-+ =0\,.
\eEq
One easily checks that
\bEa
\Cedd++-&=&-\,\Codd+-+ -\,(\dhind 1d^+_1 +\dhind 2d^+_2 + \dhind 3d^+_3 )\,,
               \label{eq:5.9}\\
\Cedd-++&=&-\,\Codd+-+ +\,(\dhind 1d^+_1 +\dhind 2d^+_2 + \dhind 3d^+_3 )\,,
               \label{eq:5.10}
\eEa
hence
\bEq
\label{eq:5.11}
\Cedd-++ +\Cedd++-=-2 \Codd+-+ =2\Cedd+-+ \,.
\eEq
\bLe
\label{lem:5.12}
The following elements  $ t_\pm,\ta_1,\ta_2,\ta_3\in M^{1,1}_{A}$
represent a basis of $H^{1,1}(G_{A})$:
\vspace*{-.5ex}
\bEaz
t_+&=&\quad
\Cedx++\,z_- - \Cedx+-z_+ + \Codx+-z_+ - \Codx-+z_+
\,,\\
t_-&=&
 -\Cedx--\,z_+ + \Cedx-+z_- - \Codx-+z_- + \Codx+-z_-
\,,\\
\ta_1&=&
d^+_2x_3z_- + d^-_3x_2z_+\,,\\
\ta_2&=&
d^+_3x_1z_- + d^-_1x_3z_+\,,\\
\ta_3&=&
d^+_1x_2z_- + d^-_2x_1z_+\,.\\
\eEaz

%\vspace*{-3ex}
%\hspace{-1em}
Here we have isomorphisms
 $\quad\Ind^\Ww_{L_0^\Ww}\lsp{t_+,t_-}\simeq Q$,
$\quad\Ind^\Ww_{L_0^\Ww}\lsp{\,\ta_i\,,i=1,2,3 }\simeq T^1$.
\eLe

This is also not difficult to prove. Theorem~\ref{th:4.10} shows
us the weight subspaces where the representatives should be and
we only have to compute the cycles and boundaries in these subspaces.
We leave it to the reader to fill in the details.
\bRe
\label{rem:5.13}
The elements $[t_+],\,[\ta_3]$
of $E^0(M_{A})^{1,1}=H^{1,1}(M_{A})$
are  $\Ww$-singular vectors, the first one
generates the $\Ww$-submodule
isomorphic to $Q$, the second
generates the submodule isomorphic to $T^2$.
At the same time $t_+$ represents a singular vector of
$H^{1,1}(M_{A})$, which is not a singular but a secondary
singular vector  of $M^{1,1}_A\simeq M(1,0;1;y_A)$.
\eRe
\begin{proof}
[Proof of Proposition~\ref{prop:5.8}]
We are ready now for the calculating of the differential \\
$\tdn^{(0)}: E^0(M_{A})^{1,2}\ar E^0(M_{A})^{0,1}$.
>From (\ref{eq:5.6}) it follows that
\bEq
\label{eq:3-2}
\tdn^{(0)}
[s]=\dhind 1[\ze_1] + \dhind 2[\ze_2] + \dhind 3[\ze_3] \,.
\eEq
Using Lemmae~\ref{lem:5.9}, \ref{lem:5.10}
and comparing with (\ref{eq:5.5})
we see immediately that
\[
\Ker (\tdn^{(0)}|E^0(M_{AB})^{1,2})=0\quad \text{ and }  \quad
\Coker(\tdn^{(0)}|E^0(M_{AB})^{0,1})=Q\oplus R\,.
\]
This proves (2).

For the differential
$\tdn^{(0)}: E^0(M_{A})^{1,1}\ar E^0(M_{A})^{0,0}$
we need to know $\tdn t_\pm,\, \tdn \ta_i$. Clearly
\bEq
\label{eq:namt}
\tdn t_+=
\Cedd++- - \Cedd+-+ + \Codd+-+ - \Codd-++=0
\eEq
because of (\ref{eq:5.8}) and (\ref{eq:5.11}). Now
$\tdn t_-=\tdn(f_3 t_+)= f_3(\tdn t_+)=0$. And
$\tdn \ta_3 = [d^+_1,d^-_2]= -\dhind 3$. Similarly
$\tdn \ta_i = - \dhind i$ for the other values of $i$ .

Keeping in mind that $E^0(M_{A})^{0,0}=H^{0,0}(M_{A})=T^0$ and
Lemma~\ref{lem:5.12} we conclude that
\[
\Ker (\tdn^{(0)}|E^0(M_{A})^{1,1})=Q\oplus R \quad \text{ and }  \quad
\Coker(\tdn^{(0)}|E^0(M_{A})^{0,0})= \CC\,,
\]
which gives us (1).

It is clear that
$\tdn^{(i)}: E^i(M_{A})^{1,2}\ar E^i(M_{A})^{0,1}$
is zero for any $i\geq 1$. Also we see that
if any of the differentials
$\tdn^{(i)}: E^i(M_{A})^{1,1}\ar E^i(M_{A})^{0,0}$
happens to be not zero then $E^\infty(M_{A})^{0,0}=0$ but this is
impossible because  $\IM(\tdn|M^{0,0}_{A})\neq M^{0,0}_{A}$. This proves (3).
\end{proof}
\bRe
\label{rem:5.14} Proposition~\ref{prop:5.8} shows that
 $\Ww$-modules
$E^\infty(M_{A})^{0,1}$ and $E^\infty(M_{A})^{1,1}$
are isomorphic.
We would like to have such an isomorphism
$\mx{$\ph: H^{0,1}(M_{A})\ar H^{1,1}(M_{A})$}$
that would
be not only an isomorphim of $\Ww$-modules but also
a morphism (hence isomorphism) of $\lag$-modules.

Let us notice that the elements
\bEa
\tht_1&=&\dhind 3 \ta_2  - \dhind 2 \ta_3 \,, \nonumber\\
\tht_2&=&\dhind 1 \ta_3  - \dhind 3 \ta_1 \,, \label{eq:NbasT1}\\
\tht_3&=&\dhind 2 \ta_1  - \dhind 1 \ta_2 \,  \nonumber
\eEa
represent a basis for the subspace of trivial %$\Ww$-
singular vectors
in the $\Ww$-submodule
\[
\Ker\tdn^{(0)}\,\cap\, (\,U(\Ww)\lsp{[\ta_i]}\,)\,\,\subset \,
E^0(M_{A})^{1,1}\,,
\]
which is
isomorphic to $R$.
Also the elements $\ze_i$ represent the basis of the subspace of trivial
%$\Ww$-
singular vectors of a
$\Ww$-submodule
$\Coker(\tdn^{(0)}|E^0(M_{A})^{0,1})$,
which is isomorphic to $R$ as well.
A similar pair of bases of singular vectors
for
$\Ww$-modules which are isomorphic to $Q$
 is $[t_\pm]$ and $[z_\pm]$.

It holds in $H^{0,1}(M_{A})$ that $d^-_1[z_+] -d^+_1[z_-] =[\ze_1]$.
On the other hand it holds ``in $M^{1,1}_{A}$ modulo boundaries''
that
\bEq
\label{eq:RbasT1}
d^-_1[t_+] - d^+_1[t_-] \shc 4 (\dhind 3[\ta_2] - \dhind 2[\ta_3])
\shc 4 [\tht_1]
\eEq
(which could be obtained by
straightforward but unfortunately hard computation).
Thus if we define
a $\Ww$-isomorphism
$\ph: H^{0,1}(M_{A})\ar H^{1,1}(M_{A})$
by conditions
\bEq
\label{eq:1-0}
\bAr{lcl}
\ph([z_\pm])&=&\tfrac{1}{4}[t_\pm]\,,\\
\ph([\ze_i])&=&[\tht_i]\,,
\eAr
\eEq
we get a map which is $L_{-}$-linear.
As it also maps a singular vector
$[z_+]$ to a
singular vector $\tfrac{1}{4}[t_+]$, so it is a $\lag$-homomorphism
by the arguments of Remark~\ref{rem:2.4}(b).
Clearly
%$\ph$ is  an $\Ww$-isomorphism, so
we got the needed isomorphism of the $E(3,6)$-modules.
\eRe

%CDcase

\bPr %% E^1 for CD
\label{prop:5.15}
 There are the following isomorphisms of $\Ww$-modules:
\arabicparenlist
\begin{enumerate}
\item%{\rm :} %%1
$\,
E^1(M_{D})^{-1,-2}\simeq \CC
\,$, $\,\,\,\,\qquad\,\,\!
E^1(M_{D})^{0,-1}\simeq R\oplus Q
$,
\vspace{1.5ex}
\item%{\rm :} %%2
$\,
E^1(M_{D})^{-1,-1}\simeq \,R \oplus Q
\,$, $\,\,\,\,
E^1(M_{D})^{0,0}\simeq 0
$.
\vspace{1.5ex}
\item%{\rm :} %%3
The spectral sequence $\{E_i(M_{D})\}$
degenerates at $E^1$,
that is \\
$\,E^1(M_{D})^{m,n}\simeq E^\infty(M_{D})^{m,n}$.
\end{enumerate}
\ePr
In order to prove this proposition, we shall use representatives for the
bases of $H^{m,n}(G_{D})$ and
calculate the differentials $\tdn^{(i)}$ but first we have to introduce
some notations and prove several lemmas.\\

%%%extra-new-command
\newcommand{\hDe}{\widehat{\De}}

Consider the associative algebra
$U(L_-)\otimes_\CC V_D \simeq M_D$.
Allowing ourselves a slight abuse of notations we denote
$\dhind i := \dpind i\otimes 1$,
$d^\pm_i := d^\pm_i\otimes 1$,
and $\dpind i := 1 \otimes \dpind i$.
Furthermore
we consider
$\De^\pm=d^\pm_1\dpind 1 + d^\pm_2\dpind 2  + d^\pm_3\dpind 3$
and define
\[
\hDe^\pm=
\dhind 1d^\pm_1 + \dhind 2d^\pm_2  + \dhind 3d^\pm_3\,.
\]

The following two lemmae are straightforward to check.(For odd elements $x,y$
from $M_D$ we mean $[x,y]=xy+yx$ as usual.)
\bLe
\label{lem:5.16}
$\/\qquad\quad \,[\,d^\pm_i,\,\De^\pm]\,=\,0\,$,
\bEaz
\, [\,d^+_1,\,\De^-]&=&-[\,d^-_1,\,\De^+]\,=\,\dhind2\dpind3-\dhind3\dpind2\,,\\
\, [\,d^+_2,\,\De^-]&=&-[\,d^-_2,\,\De^+]\,=\,\dhind3\dpind1-\dhind1\dpind3\,,\\
\, [\,d^+_3,\,\De^-]&=&-[\,d^-_3,\,\De^+]\,=\,\dhind1\dpind2-\dhind2\dpind1\,.
\eEaz
\eLe
\bLe
\label{lem:5.17}
$[\,d^\pm_i,\,\hDe^\ep]=0$ for $\ep=\,\scriptstyle{-}$ or $\scriptstyle{+}\,$.
\eLe
It follows  from these two lemmae that, whetever $\ep$, one has
\bEq
\label{eq:5.12}
\,[\,\De^\pm,\,\hDe^\ep]=0\,,\quad
\,[\,\hDe^\pm,\,\hDe^\ep]=0\,,\quad
\,[\,\De^\pm,\,\De^\ep]=0\,.
\eEq

Let us supplement the notations introduced in (\ref{eq:2.12}), (\ref{eq:2.13})
defining $d:=d^-_1d^-_2d^-_3$. Clearly
\bEaz
f_3\,a=b, & &f_3\,b=2c,\quad\,f_3\,c=3d,\quad\,f_3 \,d=0\,, \\
e_3\,d=c, & &e_3\,c=2b,\quad\,e_3\,b=3a,\quad\,e_3\,a=0\,.
\eEaz
It follows from Lemma~\ref{lem:5.17} that
\[
[\,a,\hDe^\pm]=0,\quad[\,b,\hDe^\pm]=0,\quad
[\,c,\hDe^\pm]=0,\quad[\,d,\hDe^\pm]=0\,.
\]
\bLe
\label{lem:5.18}\hspace{-2em}
\arabicparenlist
\bN
\item %1
$\/\hspace{.4em}
[\,a,\De^-]\,=\,\hDe^+\De^+\,=\,-\De^+\hDe^+$,\\
\vspace{.5ex}
$\,[\,b,\De^-]\,=\,\hDe^+\De^-  + \hDe^-\De^+=\hDe^+\De^- - \De^+\hDe^-$,\\
\vspace{.5ex}
$\,[\,c,\De^-]\,=\,\hDe^-\De^-\,=\,-\De^-\hDe^-$,\\
\vspace{.5ex}
$\/\quad d\,\De^-\,=\,0,\qquad\De^-d\,=\,0$,
\vspace{.5ex}
\item %2
$\/\quad\,a\,\De^+\,=\,0,\qquad\De^+a\,=\,0$,\\
\vspace{.5ex}
$\,[\,b,\De^+]\,=\,\De^+\hDe^+\,=\,-\hDe^+\De^+$,\\
\vspace{.5ex}
$\,[\,c,\De^+]\,=\,\De^+\hDe^-  + \De^-\hDe^+=\De^+\hDe^- - \hDe^+\De^-$,\\
\vspace{.5ex}
$\,[\,d,\De^+]\,=\,\De^-\hDe^-\,=\,-\hDe^-\De^-$,
\vspace{.5ex}
\item %3
$0\,=\,\De^+a\,=\,\De^-a+\De^+b\,=\,\De^-b+\De^+c
\,=\,\De^-c+\De^+d\,=\,\De^-d$,\\
\vspace{.5ex}
$0\,=\,a\,\De^+\!=\,a\,\De^-\!+\,b\,\De^+\!=\,b\,\De^-\!+\,c\,\De^+\!
=\,c\,\De^-\!+\,d\,\De^+\!=\,d\,\De^-$, \\
and exactly the same for $\hDe^\pm$
(with a hat).
\eN
\eLe
\begin{proof} We can check that
\bEq
\label{eq:5.13}
\, [\,a\,,\,d^-_i]=\hDe^+d^+_i \,,\qquad\, [\,d\,,\,d^+_i]=-\hDe^-d^-_i \,,
\eEq
and $\,[\,a\,,\De^-]\,=\,\hDe^+\De^+\,$,
$\,[\,d\,,\De^+]\,=\,\De^-\hDe^-\,$, follows. Now we
apply $f_3$ and get the
rest in (1), or apply $e_3$ and get the rest in (2).

Similarly  $\De^+a\,=\,a\De^+\!=0\,$, $\,\hDe^+a\,=\,a\hDe^+\!=0\,$,
then we apply $f_3$ and get (3).
\end{proof}

\begin{Lemma}
  \label{lem:5.19}
Consider the
element
\[
  \xi = a \Delta^- \partial^2_+ +
  b \Delta^- \partial_+ \partial_- + c \Delta^- \partial^2_- \in
M (0,1;2;y_D).
\]
It has the following properties:
\alphaparenlist
\begin{enumerate}
\item %%a
  $\fg_0 \cdot \xi =0$,
\item %%b
  $e'_0 \cdot \xi =0$,
\item %%c
  $e_0 \cdot \xi \in Im \triangledown$, but $e_0 \cdot \xi
  \neq 0$, and
  $\triangledown \xi =0$,
\item %%d
  $\xi$ represents a basis for $H^{-1,-2}(G_{CD})$.
\end{enumerate}
Here $\,\,\Ind^\Ww_{L_0^\Ww}\lsp{\,\xi\,}\simeq T^0$.
\end{Lemma}

%%%
\begin{proof}
The proof of (a) immediate. We get (b)
by a straightforward calculation.
To get (c) it is enough to notice that
\[
e_0 \xi =
-2(d^-_2d^+_3+d^+_3d^-_3)
\tdn\partial_+ - 4d^-_2d^-_3\tdn\partial_-
=\tdn\left(-2(d^-_2d^+_3+d^+_3d^-_3)
\partial_+ - 4d^-_2d^-_3\partial_-\right)\,,
\]
and that
\[
\tdn\xi=
a \Delta^-\De^+ \partial^3_+ +
(a \Delta^-\De^-  +   b \Delta^-\De^+) \partial^2_+ \partial_-
+ ( b \Delta^- \De^- + c \Delta^- \De^+)\partial_+ \partial^2_-
+ c \Delta^- \partial^3_-=0\,,
\]
as it follows form (\ref{eq:5.12}) and Lemma~\ref{lem:5.18}(3).

For (d) we are to prove that $\xi \not\in \IM \triangledown$
so $[\xi]\neq 0$ in $H^{-1,-2}(G_{D^o})\simeq \CC$.
If $\xi \in \IM \tdn$ then $H^{-1,-2}(M_{D})=0,$ and
this would imply the existence of a non-zero differential
\[
\Sym(\gind{-2})\otimes H^{0,-1}(G_{D})\arr
\Sym(\gind{-2})\otimes H^{-1,-2}(G_{D})
\]
in the spectral sequence for $H(M_{D})$ with its image containing
$[\xi]$. This is impossible because $Y$-eigenvalue of $\xi$
is $0$, but
all eigenvalues
of $Y$ on $ \Sym(\gind{-2})\otimes H^{0,-1}(G_{D})$
are negative
as it follows from the description of the homologies in
Theorem~\ref{th:4.10}.
\end{proof}
\bCo
\label{cor:5.20}
$\xi$ is a secondary singular
vector of the $E(3,6)$-module $M(0,1;2;y_D)$.
\eCo

\bLe
\label{lem:5.21}\hspace{-2em}
\arabicparenlist
\bN
\item %1
      $ac=ca=a\hDe^-=-\hDe^-a$, $\,\,ab=ba=0$,
\item %2
      $db=bd=-d\hDe^+=\hDe^+d$, $\,\,dc=cd=0$,
\item %3
      $ad+da=b\hDe^-=\hDe^+c$,  $\,\,2ad+bc=da$.
\eN
\eLe
\begin{proof}
It is not difficult to check that $ac=a\hDe^-$ and $ca=-\hDe^-a$.
Then
\[
 ab=e_3\,ac=e_3\,a\hDe^-=a\hDe^+=0\,,
\]
and we get (1).  Now let
\[
\la=2ad+c\hDe^+=2ad-b\hDe^-\,.
\]
Then $e_3\la=2ac+2b\hDe^+=0$ because of (1) and Lemma~\ref{lem:5.18}(3).
As the $s\ell(2)$-weight of $\la$ is zero, so $f_3\la=0$, which gives us
$2bd-2c\hDe^-=0$. The latter means $bd=c\hDe^-=-d\hDe^+$.

At the same time $\la=-2da+b\hDe^-$
because $\la$ is invariant under
the Weyl reflection in $SL(2)$ and this implies the first part of (3).
We now get the rest of (2) from $e_3\la=0$ and the rest of (3) applying
$f_3$ to $ac=a\hDe^-$.
\end{proof}
Let
\vspace{-3ex}
\bEaz
q_+&=&\tfrac{1}{3}(-2a\De^-\dpind+  - \, b\De^-\dpind- )\,,
\qquad\qquad\qquad\/\\
q_-&=&\tfrac{1}{3}(\,\,c\De^+\dpind+  + 2d\De^+\dpind- )\,,
\qquad\qquad\qquad\/
\eEaz

 $\qquad\ka_i=d^-_iq_+ - d^+_iq_-$.
\bLe
\label{lem:5.23}\hspace{-2em}
\arabicparenlist
\bN
 \item
$e_3\,q_+=0$, $e'_0\,q_+=0$, $\tdn q_+=0$,
$e_0\,q_+=\tdn(-2(d^+_2d^-_3 + d^-_2d^+_3))\neq 0$,
 \item
vectors $\,q_+,\,\ka_1$ represent
the highest weight vectors of $H^{-1,-1}(G_D)$,\\
and $q_+,q_-,\,\ka_1,\ka_2,\ka_3$ represent a basis of $H^{-1,-1}(G_D)$,
\item
$q_+$ represents the (highest weight) singular vector of $H^{-1,-1}(M_D)$,
\item
$q_+$ is the secondary (highest weight) singular vector in $M(0,1;1;y_D)$.
\eN
We have isomorphisms
$\,\,\Ind^\Ww_{L_0^\Ww}\lsp{q_+,q_-}\simeq Q$,
$\quad\Ind^\Ww_{L_0^\Ww}\lsp{\,\ka_i\,,i=1,2,3}\simeq T^2 $.
\eLe
We leave it to the reader to check the calculations. The rest follows.

\bLe
\label{lem:5.22}
The element $\la=2ad+c\hDe^+=-2da+b\hDe^- \in M^{0,0}_{D}$ represents
a basis for $H^{0,0}(G_D)$, the singular vector of $H^{0,0}(M_D)$,
$\,\,\Ind^\Ww_{L_0^\Ww}\lsp{\,\la\,}\simeq T^3$,
and
\[
\tdn\la= \hDe^-q_+ - \hDe^+q_-=\dhind1 \ka_1 +\dhind 2\ka_2 +\dhind 3\ka_3
\,.
\]
\eLe
\begin{proof}
Let us start with checking the last statement:
\bEaz
\tdn\la&=&2ad\De^+\dpind+    + c\hDe^+\De^+\dpind+ + c\hDe^+\De^-\dpind-\\
       &=&2a\De^-\hDe^-\dpind+  - b\hDe^-\De^+\dpind+
+ c\hDe^+\De^-\dpind-\\
       &=&a\De^-\hDe^-\dpind+    + c\hDe^+\De^-\dpind-\\
       &=&a\De^-\hDe^-\dpind+    + d\hDe^+\De^+\dpind-\,.
\eEaz
On the other hand
\bEaz
\hDe^-q_+ - \hDe^+q_-&=&\tfrac{1}{3}\left(
\hDe^-(-2a\De^-\dpind+  - \, b\De^-\dpind- )-
\hDe^+( c\De^+\dpind+  + 2d\De^+\dpind- )\right)\\
&=&\tfrac{1}{3}\left(
(-2\hDe^-a\De^- + \hDe^+c\De^+)\dpind+
+
(-\hDe^-b\De^- + 2d\hDe^+\De^+)\dpind-\right)\\
&=&\tfrac{1}{3}\left(
3a\hDe^-\De^- \dpind+
+
3d\hDe^+\De^+\dpind-\right)\,.
\eEaz
Hence we get the equality. It implies that
 $\tdn[\la]=0$ for $[\la]\in \Gr M_D$,
so $[\la]$ is a cycle. The rest is immediate.
\end{proof}

Let
$\quad\,r_+\,=\,a\,(d\,\dpind- + c\,\dpind+)\,$,
$\quad r_-\,=\,d\,(b\,\dpind- + a\,\dpind+)\,,\quad$ and
\bEaz
%r_+&=&a\,(d\,\dpind- + c\,\dpind+)\,,\\
%r_-&=&d\,(b\,\dpind- + a\,\dpind+)\,,\\
\ro_1&=&\tfrac{1}{2}\,( d^+_2d^+_3\dpind-
                   +  d^-_2d^-_3\dpind+ )\,,\\
\ro_2&=&\tfrac{1}{2}\,( d^+_3d^+_1\dpind-
                   +  d^-_3d^-_3\dpind+ )\,,\\
\ro_3&=&\tfrac{1}{2}\,( d^+_1d^+_2\dpind-
                   +  d^-_1d^-_2\dpind+ )\,.
\eEaz

\bLe
\label{lem:5.24}
Elements $r_\pm,\ro_i \in M^{0,-1}_D$ have the properties:
\arabicparenlist
\bN
\item
The vector $r_+$ coincides (up to a sign) with
the non-trivial
singular vector of $M(0,0;1;y_D)$ defined in (\ref{eq:2.10nn}):
$r_+=-w_3$,
\item
vectors $r_+,\,\ro_1$ represent
the highest weight vectors of $H^{0,-1}(G_D)$,\\
and $r_+,r_-,\,\ro_1,\ro_2,\ro_3$ represent the basis of
$H^{0,-1}(G_D)$,
\item
$d^-_1r_+ - d^+_1r_-=\dhind 3\ro_2 - \dhind 2\ro_3$,
\item
$\tdn \ro_3 \shc -\tfrac{2}{3}\xi$ modulo boundaries in $M^{-1,-2}_D$.
\eN
Here $\quad\Ind^\Ww_{L_0^\Ww}\lsp{r_+,r_-}\simeq Q\,\, $ and
$\quad\Ind^\Ww_{L_0^\Ww}\lsp{\,\ro_i\,,i=1,2,3}\simeq T^1$.
\eLe
Equation (\ref{eq:2.10nn}) written in our notations is
$w_3= (da+\hDe^-b)\dpind- + \hDe^-a\dpind+$. (1) now follows from
Lemma~\ref{lem:5.21}.
The calculations needed to check the rest of
the statements are quite straightforward,
we leave them to the reader.
\begin{proof}
[Proof of Proposition~\ref{prop:5.15}.]
As we compare equations (\ref{eq:5.6}) with (\ref{eq:3-2})
and equations  (\ref{eq:1-0}) with  Lemma~\ref{lem:5.24}(4),
we notice that the differential $\tdn^{(0)}$ for $E^0(M_D)$
is given
by the similar formulae as $\tdn^{(0)}$ for $E^0(M_A)$. Thus
it is possible to repeat
the arguments of the proof of Proposition~\ref{prop:5.8}
and conclude that the values
of homologies are as stated.
\end{proof}
\bPr
\label{prop:5.25}\hspace{-2em}
\arabicparenlist
\bN
\item
If $f\in M^{0,1}_A$ and $f=f'z_+ + f''z_-$ where
$f',f''\in U(L_-)$, then\\
$\qquad
\tdn_6 f=-(f'r_+ + f''r_-)\,,
$
\item
 $\tdn_6\cdot\tdn=0$,
\item
the morphism
 $\Coker(\tdn|M^{0,1}_A)\ar \Ker(\tdn|M^{0,-1}_D)$
 defined by $\tdn_6$  is an isomorphism.
\eN
\ePr
\begin{proof}
Evidently the morphism given by the formula in (1) maps $z_+$
to $w_3$, and commutes with $\gind 0$ and $L_-$, hence it is
a morphism of $\lag$-modules, thus equal to $\tdn_6$.
To show that   $\tdn_6\cdot\tdn=0$ it is enough to check that
$\tdn_6$ annihilates vector $d^+_iz_+$,
%$d^+_iz_- + d^-_iz_+$ and $d^-_iz_-$
that generates $\IM \tdn $ over $U(L)$.
For it we have
\[
\tdn_6\, d^+_iz_+\,=\,d^+_ir_+\,=\,d^+_i\,a(d\,\dpind- + c\,\dpind+)\,=0
\]
because $d^+_ia=0$.
The last statement follows.
\end{proof}
\bCo
\label{cor:5.26}
$H^{0,1}(M_A)\simeq \Ker(\tdn_6|M^{0,-1}_D)$
is an irreducible module,
isomorphic to $I(0,0;1;-1)$.
\eCo
\begin{proof}
The submodule $\Ker(\tdn_6|M^{0,-1}_D)$ is generated by the singular
vector $w_3$ and, by Theorem~\ref{th:2.10},
is has no other singular vectors,
hence (by Proposition 1.3h from [KR1]) it is irreducible.
As it is isomorphic to $H^{0,1}(M_A)$, it has to
be a quotient of $M(0,0;1;-1)$, thus  $I(0,0;1;-1)$.
\end{proof}

\begin{proof}
[Proof of Theorem~\ref{th:5.1}.]
The statement (i) of the theorem follows from Proposition~\ref{prop:5.8}
and (iii) follows from Proposition~\ref{prop:5.15}.
>From Remark~\ref{rem:5.14} we know
that $H^{0,1}(M_A)\simeq  H^{1,1}(M_A)$,
and Corollary~\ref{cor:5.26} shows
that $H^{0,1}(M_A)\simeq  I(0,0;1;-1)$.

Again as we compare
equations (\ref{eq:RbasT1}) with  Lemma~\ref{lem:5.24}(3),
we see that
the arguments of Remark~\ref{rem:5.14} can be applied to
 $M^{-1,-1}_D$, $M^{0,-1}_D$ as well.
This
shows us that
$H^{0,-1}(M_D)\simeq  H^{-1,-1}(M_D)\simeq H^{0,1}(M_A)$.
That is all what we need.
\end{proof}

%s6

\section{The size of $I (p,q;r;y)$}
\label{sec:6}

We define the character of a $E(3,6)$-module $V$ by the usual
formula:
\begin{displaymath}
  \ch V = tr_V  t^{-3Y} \, .
\end{displaymath}
If $V$ is a degenerate highest weight module, then $\ch V$ is a Laurent
series in $t$, whose coefficients are dimensions of eigenspaces
of the operator $-3Y$.  Since, as a $\fg_0$-module
\begin{equation}
  \label{eq:6.1}
  M(p,q;r;y) = \Sym (\fg_{-2}) \otimes \Lambda (\fg_{-1})
        \otimes F (p,q;r;y)
\end{equation}
and the eigenvalues of $-3Y$ on $\fg_{-1}$ and $\fg_{-2}$ are $1$
and $2$ respectively, we obtain:
\begin{equation}
  \label{eq:6.2}
  \ch\, M (p,q;r;y) = t^{-3y} \dim F (p,q;r;y) R(t) ,
\end{equation}
where  $R(t) = (1+t)^6 / (1-t^2)^3$.

Next, let us compute the characters of the modules $I
(p,0;r;y_A)$.  By Corollary~\ref{cor:Zcpls}, we have an exact sequence
\begin{displaymath}
  0 \leftarrow I (p,0;r; \tfrac{2}{3}p-r) \leftarrow M
  (p,0;r;\tfrac{2}{3}p-r) \leftarrow M (p+1,0;r+1;
  \tfrac{2}{3} p-r-\tfrac{1}{3}) \leftarrow \cdots ,
\end{displaymath}
unless $p=r=0$ or $1$.  Hence, using (\ref{eq:6.2}), we obtain:
\begin{equation}
    \ch\, I (p,0;r;\tfrac{2}{3}p-r)
  = t^{3r-2p} R(t) \sum^{\infty}_{j=0} (-1)^j
  \frac{(j+p+2)(j+p+1)(j+r+1)}{2}t^j
\end{equation}
unless $p=r=0$ or $1$.  In order to sum this series for  $|t|<1$,
we use the identity
\begin{equation}
  \label{eq:6.3}
  \sum^{\infty}_{j=0} (-1)^j \binom{j+m}{m} t^j
  = \frac{1}{(1+t)^{m+1} } \, .
\end{equation}
We get (unless $p=r=0$ or $1$):
\begin{equation}
  \label{eq:6.4}
  \ch\, I (p,0;r;\tfrac{2}{3}p-r) =t^{3r-2p}
  R(t) \left ( \frac{3}{(1+t)^4} + \frac{2p+r-2}{(1+t)^3}
    + \frac{p^2+2pr-p}{2(1+t)^2}
    + \frac{p^2r + pr}{2(1+t)}
\right) \, .
\end{equation}
In particular, for $r>0$ we have:
\begin{equation}
  \label{eq:6.5}
  \ch\, I (0,0;r;-r) = t^{3r}R(t)
  \frac{(r+1)+(r-2)t}{(1+t)^4} \, .
\end{equation}

Next, we compute the characters of the modules $I (0,q;r;y_D)$.
By Corollary~\ref{cor:Zcpls}, we have an exact sequence
\begin{displaymath}
  0 \to I (0,q;r;r-\tfrac{2}{3}q) \to
  M(0,q+1;r+1;r-\tfrac{2}{3}q-\tfrac{1}{3}) \to \cdots
\end{displaymath}
unless $(q,r)=(0,0,)$, $(1,1)$, $(0,1)$ and $(1,2)$, in the last
%%% two cases %?!
      case
the exactness being broken by a $1$-dimensional space.
Hence, except for these four cases, we have:
\begin{equation}
  \label{eq:6.6}
  \ch I (0,q;r;r-\tfrac{2}{3}q) = t^{4q-6r}
  \ch I (q+1,0;r+1;r-\tfrac{2}{3}q-\tfrac{1}{3}) \, .
\end{equation}

One computes similarly the characters of $B$ and $C$~type, but
they are more cumbersome and we omit them.

\begin{Definition}
  Given a $\Sym(\fg_{-2})$-module $V$, we define
  \begin{displaymath}
     \hbox{size } V = \frac{1}{4} \hbox{ rank }_{\Sym(\fg_{-2})} V   \, .
  \end{displaymath}

\end{Definition}

It is clear that size $V$ can be expressed via $\ch V$ as follows:
  \begin{equation}
\label{eq:6.7}
     \hbox{size } V = \frac{1}{4} \lim_{t \to -1}
     (1-t^2)^3 \ch V \, .
  \end{equation}
In particular, we obtain from (\ref{eq:6.2}):
\begin{equation}
  \label{eq:6.9}
  \hbox{size } M (p,q;r;y) = 16 \dim F (p,q;r;y) \, .
\end{equation}
One computes immediately the sizes of modules $I (p,0;r;y_A)$ and
 $I (p,0;r;y_D)$ using (\ref{eq:6.4}), (\ref{eq:6.6}) and
 (\ref{eq:6.7}).  The sizes of $I(p,0;r;y_B)$ and  $I(p,0;r;y_C)$
 are then computed by adding a finite number of terms using exact
 sequences from Figure~3; it follows that in both cases the size
 is a polynomial of degree $\leq 2$ in $p$ and of degree $\leq 1$
 in $r$.  The final result of the calculations is given by the
 following.

\Alphaparenlist
 \begin{Theorem}\label{th:6.2}
   \begin{enumerate}
   \item %%A
     If $(p,r) \neq (0,0)$ or $(1,1)$, then
     \begin{eqnarray*}
       \hbox{size } I(p,0;r;\tfrac{2}{3}p-r)
       &=& 2r (2p^2+4p+1) + (2p^2+2p-1) \, ;\\
       \hbox{size } I (1,0;1;-\tfrac{1}{3}) &=& 16 \, .
     \end{eqnarray*}

\item %%B
  $\hbox{size } I (p,0;r;\tfrac{2}{3}p+r+2)
  = 2r(2p^2 +4p+1) + (6p^2 + 14p+5)$.

\item %%C
 $ \hbox{size } I (0,q;r;-\tfrac{2}{3}q-r-2)
  = 2r (2q^2 +8q+7)+(2q^2+10q+11)$.

\item %%D
  If $(q,r) \neq (0,0)$ or $(1,1)$, then
  \begin{eqnarray*}
    \hbox{size } I (0,q;r;r-\tfrac{2}{3}q) &=&
    2r(2q^2+6q+7) + (6q^2+22q+17) \, ;\\
\hbox{size } I (0,1;1;\tfrac{1}{3}) &=& 74 \, .
  \end{eqnarray*}

   \end{enumerate}
 \end{Theorem}

\begin{remark}

It follows from (\ref{eq:6.1}) and Figure~3  that the sizes of
the even and odd parts of all modules $I(p,q;r;y)$ are equal.
\end{remark}

%s7
\section{The secondary $\ZZ$-grading of $E(5,10)$ as a
  $E(3,6)$-module.}
\label{sec:7}

Recall that in Section~\ref{sec:1} we defined the secondary
$\ZZ$-grading
\begin{displaymath}
  E(5,10) = \prod_{j \geq -1} U^j
\end{displaymath}
by letting
\begin{displaymath}
  \deg x_i =0=\deg \partial_i \hbox{ for } i=1,2,3 \, ; \,
  \deg x_j =1=-\deg \partial_j \hbox{ for } j=4,5;
  \deg d=-\tfrac{1}{2} \, ,
\end{displaymath}
so that $U^0$ is isomorphic to $E (3,6)$, and each $U^j$ is a
$E(3,6)$-module.  Each of the subspaces $U^j$ carries a
$\ZZ$-grading by finite-dimensional subspaces induced by the
consistent grading of $E(5,10)$ defined in Section~\ref{sec:1}:
\begin{displaymath}
  E(5,10)^j = \prod_{i \in \ZZ} \quad U^j_i \, .
\end{displaymath}
Each of the $E(3,6)$-modules $U^j$ is a linearly compact space,
hence the dual modules are discrete spaces:
\begin{equation}
  \label{eq:7.1}
  U^{j*} = \oplus_{j \in \ZZ} U^{j*}_i \, .
\end{equation}
Note that all $U^{j*}_i$ (and $U^j_i$) are $\fg_0$-modules.

Since  $E (5,10)^j_i =0$ for $i<-2$, it follows that all
$E(3,6)$-modules $U^{j*}$ are $L_0$-locally finite, hence they
are objects from the category $\Pw (L,L_0)$ considered in \cite{KR1}.

\begin{Theorem}
  \label{th:7.1}
\alphaparenlist
\begin{enumerate}
\item %%a
  The $E(3,6)$-module $U^{j*}$ has size $2j+3$,
  $j=-1,0,1,\ldots$.

\item %%b
  One has the following isomorphisms of $E(3,6)$-modules:
  \begin{eqnarray*}
    U^{-1*} &\simeq& I (0,0;1;-1), \,
    U^{0*} \simeq I (1,0;0;\tfrac{2}{3}) \, , \\
    U^{j*} &\simeq& I (0,0;j-1;j+1) \hbox{ for } j \geq 1 \, .
  \end{eqnarray*}

\end{enumerate}
\end{Theorem}

\begin{proof}
  It is straightforward to check~(a) using the construction of $U^j$.

It follows from (\ref{eq:7.1}) that $U^{j*}$ has a maximal
submodule $U^{j*}_{(0)}$ such that $U^{j*}/U^{j*}_{(0)}$ is an
irreducible module from $\Pw (L,L_0)$.  It is easy to see that
the lowest non-zero space in (\ref{eq:7.1}) is isomorphic to the
$\fg_0$-modules $F(0,0;1;-1)$, $F(1,0;0;\tfrac{2}{3})$ and
$F(0,0;j-1;j+1)$ for $j=-1$, $j=0$ and $j \geq 1$, respectively.
Hence the $E(3,6)$-module $U^{j*}/U^{j*}_{(1)}$ is isomorphic to
$I(0,0;1;-1)$, $I(1,0;0;\tfrac{2}{3})$ and $(I(0,0;j-1;j+1))$,
respectively.  It follows from Theorem~\ref{th:6.2} that these
$E(3,6)$-modules have the same size as $U^{j*}$ (given by (a)).
Since $U^{j*}$ as $\Sym(\fg_{-2})$-modules have no torsion, it
follows that $U^{j*}_{(1)}=0$, hence~(b).
\end{proof}

We conclude this section by an explicit description of the
%$E(3,6)$-
module $U^{-1*} = I(0,0;1;-1)$, which is a (non-trivial)
$E(3,6)$-
module of minimal possible size ($=1$).

Denote by $\Omega^k_3$ the space of differential $k$-forms over $\CC
[[x_1,x_2,x_3]]$ and by $\Omega^k_{3 c\ell}$ the subspace of closed
forms.  The Lie algebra $W_3$ of all formal vector fields acts on
$\Omega^k_3$ via Lie derivative, leaving $\Omega^k_{3 c\ell}$
invariant.  For $\lambda \in \CC$, define a $\lambda$-twisted
$W_3$-module structure on $\Omega^k_3$, which we denote by
$(\Omega^k_3)^{\lambda}$, by letting
\begin{displaymath}
  D\omega =L_D \omega + \lambda (\DIV D) \omega \, , \quad
  D \in W_3, \,\omega \in \Omega^k_3 \, .
\end{displaymath}
Note, that we have a canonical isomorphism of $W_3$-modules:
\begin{equation}
  \label{eq:7.2}
  (\Omega^3_3)^{\lambda} \simeq (\Omega^0_3)^{1+\lambda} \, .
\end{equation}

Recall the following explicit construction of $E(3,6)$ \cite{CK}:
\begin{equation}
  \label{eq:7.3}
  E(3,6)_{\bar{0}} = W_3 + \Omega^0_3 \otimes s\ell (2) \, ,\quad
  E(3,6)_{\bar{1}} = (\Omega^1_3)^{-\frac{1}{2}}\otimes \CC^2
  \, ,
\end{equation}
with the obvious bracket on $E(3,6)_{\bar{0}}$ and between
$E(3,6)_{\bar{0}}$ and $E(3,6)_{\bar{1}}$, and the following
bracket on $E(3,6)_{\bar{1}}$:
\begin{displaymath}
  [\omega \otimes v,\omega' \otimes v'] =
  (\omega \wedge \omega') \otimes (v \otimes v')
  + (d \omega \wedge \omega' + \omega \wedge d \omega')
  \otimes v \cdot v' \, ,
\end{displaymath}
where $v \cdot v' \in S^2\CC^2 \simeq s\ell (2)$,
$v \wedge v' \in\Lambda^2\CC^2\simeq\CC$.
We have identified here $\Omega^0_3$ with
$(\Omega^3_3)^{-1}$ (see~(\ref{eq:7.2})), and also use the
identification
\begin{equation}
  \label{eq:7.4}
  W_3 \simeq (\Omega^2_3)^{-1}  \, .
\end{equation}
(We fix a volume form in $\CC^3$ when we define the twisted
action.)

\begin{Proposition}
  \label{prop:7.1}
The $E(3,6)$-module $I=I(0,0;1;-1)$ is constructed explicitly as
follows:
\begin{displaymath}
  I_{\bar{0}} = (\Omega^0_3)^{\frac{1}{2}} \otimes \CC^2,
  \quad I_{\bar{1}} = \Omega^2_{3c\ell} \, .
\end{displaymath}
$E(3,6)_{\bar{0}}$ acts in the obvious way ( $\Omega^0_3 \otimes
s\ell (2) $ acts trivially on $I_{\bar{1}}$), and
$E(3,6)_{\bar{1}}$ acts as follows ( $\omega_i \in \Omega^i_3,\,\, u,v
\in \CC^2$):
\begin{eqnarray*}
  (\omega_1 \otimes u) (\omega_0 \otimes v) &=&
  \frac{}{} d (\omega_0\omega_1) \otimes (u \wedge v)\,  , \\
  (\omega_1 \otimes u) \,\omega_2 &=& (\omega_1 \wedge \omega_2) \otimes u
  \, ,
\end{eqnarray*}
where we use identifications (\ref{eq:7.3}),
$\,u \wedge v \in\Lambda^2\CC^2\simeq\CC$ and
$\omega_1 \wedge \omega_2
\in (\Omega^3_3)^{-\frac{1}{2}} \simeq (\Omega^0_3)^{\frac{1}{2}}$.
\end{Proposition}

\begin{proof}
It is straightforward from definitions.
\end{proof}

%s8
\section{A relation to the Standard Model.}
\label{sec:8}

Recall that the $\fg_0$-module $\fg_{-1}$ is isomorphic to $F
(1,0;1;-\tfrac{1}{3})$.  The action of the compact form $\fk
=su(3) + su (2) + i \RR Y$ of $\fg_0$ on $\fg_{-1}$
exponentiates to a faithful representation of the compact group
\begin{displaymath}
  K= (S U (3) \times SU (2) \times U(1))/C \, ,
\end{displaymath}
where $C$ is a central subgroup of order $6$.  Recall that the group
$K$ is the group of symmetries of the Standard Model.

It is
straightforward to check the following.

\begin{Lemma}
  \label{lem:8.1}
The $\fg_0$-module $F(p,q;r;y)$ exponentiates to $K$ iff the
following two conditions hold:
\vspace*{-1ex}
\begin{eqnarray}
  \label{eq:8.1}
  y \in \tfrac{1}{3} \ZZ \, , \\
\label{eq:8.2}
   2 (p-q) + 3r-3y \in 6 \ZZ \, .
\end{eqnarray}

\end{Lemma}

Since the action of $\fg_0$ on $\fg_{-1}$ and hence on $\fg_{-2}$
exponentiates to $K$ (this, in fact, is true for $\fg_1$ and hence
for all $\fg_j$ as well), we obtain from the isomorphism
(\ref{eq:1.10})
that a
$E(3,6)$-module $M(p,q;r;y)$ restricted to $\fg_0$ exponentiates
to $K$ iff (\ref{eq:8.1}) and (\ref{eq:8.2}) hold.  In
particular, we obtain the following corollary of Lemma~\ref{lem:8.1}
 and Theorem~\ref{th:1.2}.

 \begin{Corollary}
   \label{cor:8.1}
All degenerate $E(3,6)$-modules $I(p,q;r;y_X)$ exponentiate to $K$.
 \end{Corollary}

 \begin{Definition}
   \label{def:8.3}
   A $K$-module $F (p,q;r;y)$ is called a fundamental particle
   multiplet if the following two properties hold:
   \begin{eqnarray}
     \label{eq:8.3}
     \hbox{when restricted to $SU(3)$, only the 1-dimensional,}\\
       \nonumber\hbox{the two fundamental and
                       the adjoint representations occur,}
   \end{eqnarray}
%
%   \begin{eqnarray}
%     \label{eq:8.3}
%     (p,q) &=& (1,0), (0,1) \hbox{ or } (0,0);
%     r=0 \hbox{ or }1 \, ; \\
%%
  \begin{eqnarray}
\label{eq:8.4}
     \tfrac{1}{2}|y+h|  \leq  1, \hbox{ where } h
     \hbox{ is any eigenvalue of }
     H= \left(
       \begin{array}{cr}
         1 & 0 \\ 0 & -1
       \end{array}  \right) \in isu (2)\, .
   \end{eqnarray}

 \end{Definition}

Condition~(\ref{eq:8.4}) means that modulus of charges (given
by Gell-Mann-Nishijima formula) of all particles in the multiplet
do not exceed $1$.

It is immediate to see that all $K$-modules $F(p,q;r;y)$ for
which (\ref{eq:8.3}) and (\ref{eq:8.4}) hold are listed in the
left hand side of Table 1.  The right half contains all the
fundamental particles of the Standard Model: the upper part consists
of three generations of quarks and the middle part of three generations of
leptons (these are all fundamental fermions from which the matter
is built), and the lower part consists of the fundamental bosons
(which mediate the strong and electroweak interactions).

\begin{table}[htbp]
  \begin{center}
    \caption{}
    \label{tab:1} %%%{Table 1.}
\begin{tabular}{c c | ccc }
multiplets & charges  && particles\\
\hline \\[-1ex]
$(01,1,1/3)$ & $2/3,-1/3$ & $\binom{u_L}{d_L}$ & $\binom{c_L}{s_L}$
    & $\binom{t_L}{b_L}$\\[1ex]
$(10,1,-1/3)$ & $-2/3,1/3$ & $\binom{\tilde{u}_R}{\tilde{d}_R}$ &
     $\binom{\tilde{c}_R}{\tilde{s}_R}$ & $\binom{\tilde{t}_R}{\tilde{b}_R}$\\[1ex]
$(10,0,-4/3)$ & $-2/3$ & $\tilde{u}_L$ & $\tilde{c}_L$
    & $\tilde{t}_R$\\[1ex]
$(01,0,4/3)$ & $2/3$ & $u_R$ & $c_R$ & $ t_R$\\[1ex]
$(01,0,-2/3)$ & $-1/3$ & $d_R$ & $s_R$ &$b_R$\\[1ex]
$(10,0,2/3)$ & $1/3$ & $\tilde{d}_L$ & $\tilde{s}_L$ & $
   \tilde{b}_L$\\[-1ex]
%
%\rule{.25in}{1pt} \quad \rule{.25in}{1pt} \quad\rule{.25in}{1pt}
%\quad\rule{.25in}{1pt} \\
\setlength{\unitlength}{0.1in}
\begin{picture}(10,3)(0,-1)
  \multiput(0,0)(2,0){6}{\line(1,0){1.5}}
\end{picture}
&
\setlength{\unitlength}{0.1in}
\begin{picture}(10,3)(0,-1)
  \multiput(0,0)(2,0){6}{\line(1,0){1.5}}
\end{picture}
&
\setlength{\unitlength}{0.1in}
\begin{picture}(5,3)(1,-1)
  \multiput(0,0)(2,0){6}{\line(1,0){1.5}}
\end{picture}
&
\setlength{\unitlength}{0.1in}
\begin{picture}(5,3)(0,-1)
  \multiput(0,0)(2,0){5}{\line(1,0){1.5}}
\end{picture}
&
\setlength{\unitlength}{0.1in}
\begin{picture}(5,3)(0,-1)
  \multiput(0,0)(2,0){4}{\line(1,0){1.5}}
\end{picture}
\\
$(00,1,-1)$ & $0,-1$ & $\binom{\nu_L}{e_L}$
   & $\binom{\nu_{\mu L}}{\mu_L}$ & $\binom{\nu_{\tau
       L}}{\tau_L}$\\[1ex]
$(00,1,1)$ & $0,1$ & $\binom{\tilde{\nu}_R}{\tilde{e}_R}$
  & $\binom{\tilde{\nu}_{\mu R}}{\tilde{\mu}_R}$
  & $\binom{\tilde{\nu}_{\tau R}}{\tilde{\tau}_R}$\\[1ex]
$(00,0,2)$ & $1$ & $\tilde{e}_L$ & $\tilde{\mu}_L$
   & $\tilde{\tau}_L$\\[1ex]
$(00,0,-2)$ & $-1$ & $e_R$ & $\mu_R$ & $\tau_R$\\[1ex]
\hline \\[-1ex]
$(11,0,0)$ & $0$ & gluons\\[1ex]
$(00,2,0)$ & $1,-1,0$ & $W^+,W^-,Z$ & (gauge bosons)\\[1ex]
$(00,0,0)$ & $0$ & $\gamma$ & (photon)\\[1ex]
$(11,0, \pm 2)$ & $\pm 1$ & --

\end{tabular}
  \end{center}
\end{table}

One can show that the direct sum of degenerate $E(3,6)$-modules
\begin{displaymath}
  I (0,0;1;-1) \oplus I (1,0;0;\tfrac{2}{3}) \oplus
  I(0,0;0;2) \oplus I(0,0;0;-2)
\end{displaymath}
contains all the fundamental particle multiplets once, except for
$(01,1,\tfrac{1}{3}) $ which is contained twice.

%sAp
%\textbf
\section{APPENDIX:\quad A spectral sequence for a filtered
  module \\ with a differential
that does not preserve the filtration.}
\label{sec:app}

Usually a spectral sequence is constructed for a differential
filtered module but we relax the conditions
and suppose that there is a module with a differential
and a filtration where
{ the differential does not preserve the filtration}
but only the condition (\ref{eq:A1}) below is true.
We show that the construction still works with minor alterations.

Let $A$ be a module with a filtration (we follow more or less the
notations of [M, Ch.~XI, Section~3]):
\begin{displaymath}
  \cdots \subset F_{p-1} A \subset F_pA \subset F_{p+1} A \subset
  \cdots \qquad (p \in \ZZ)
\end{displaymath}
and with a differential $d:A \to A$, $d^2=0$, such that
\begin{equation}
  \label{eq:A1}
  d (F_p A) \subset F_{p-s+1}A
\end{equation}
for some fixed $s$ and every $p \in \ZZ$.  The usual case of
differential module corresponds to $s=1$, but for our main
application $s=0$.

We claim that there is a spectral sequence $E=\{ E^r, d^r \}_{r
\in \ZZ}$, which, as usual, is a sequence of $\ZZ$-graded modules
$E^r=\sum_{p \in \ZZ} E^r_p$, each with a differential
\begin{displaymath}
  d^r : E^r_p \to E^r_{p-r}
\end{displaymath}
and with isomorphisms
\begin{equation}
  \label{eq:A2}
  H (E^r,d^r) \simeq E^{r+1} \, ,
\end{equation}
and for this spectral sequence there are natural isomorphisms:
\begin{equation}
  \label{eq:A3}
  E^r_p \simeq F_p A/F_{p-1} A \hbox{ for } r \leq s-1 \quad
\hbox{ and }
  d^r=0 \hbox{ if } r < s-1 \, ,
  \,\, d^{s-1}=\gr\, d \, ,
\end{equation}
hence
\begin{equation}
  \label{eq:A4}
  E^s_p \simeq H (F_pA /F_{p-1}A) \, .
\end{equation}
In other words, $E^s$ is isomorphic to the homology of the
module $\Gr A$ with respect to the induced differential
\begin{equation}
  \label{eq:A5}
  \gr\, d : F_pA/F_{p-1} A \to
F_{p-s+1} A/F_{p-s}A \, .
\end{equation}

Let us mention that if $A$ is a graded filtered module with a
grading of some kind (bigrading, etc.) then the $E^r$ inherit
similar gradings.\\

To construct the spectral sequence we keep on with the usual
construction of the spectral sequence for a differential module.

Introduce submodules
\begin{eqnarray}
  \label{eq:A6}
  Z^r_p &=& \{ a| \,\,a \in F_p A \, , \, da \in F_{p-r} A \} \, ,\\
\noalign{\nonumber{\hbox{define subquotients}}}\\
\vspace{-1ex}
\label{eq:A7}
E^r_p &=& (Z^r_p + F_{p-1}A)/ (dZ^{r-1}_{p+r-1} +F_{p-1}A) \, ,
\end{eqnarray}
and differentials $d^r:E^r_p \to E^r_{p-r}$ as the homomorphisms
induced on the subquotients by the differential $d$ of $A$.

In the case $r \leq s-1$ one has:  $Z^r_p = F_p A$, and clearly
\begin{equation}
  \label{eq:A8}
  dZ^r_{p+r-1} = d (F_{p+r-1}A) \subset F_{p+r-s} A \subset
  F_{p-1}A \, .
\end{equation}
Therefore
\begin{displaymath}
  E^r_p = F_pA/F_{p-1}A \,\,\hbox{ for } r \leq s-1 \, .
\end{displaymath}
On the other hand (\ref{eq:A8}) shows that
\begin{displaymath}
  dZ^r_{p+r} \subset F_{p+1+r-s}A \,\,\hbox{ if } r<s-1 \, .
\end{displaymath}
Hence
\begin{displaymath}
  dZ^r_{p+r} \subset F_{p-1 } A \,\,\hbox{ and so  }\, d^r \equiv 0
     \hbox{ for } r<s-1 \, .
\end{displaymath}
Thus $E^{s-1}$ coincides with $\Gr A$ and $d^{s-1}$ coincides
with $\gr\, d$ and (\ref{eq:A2}) for $r \leq s-1$ implies (\ref{eq:A3})
and~(\ref{eq:A4}).\\

To prove (\ref{eq:A2}) we notice first that $Z^r_p \cap
F_{p-1}A=Z^{r-1}_{p-1}$, hence (\ref{eq:A7}) implies
\begin{equation}
  \label{eq:A9}
  E^r_p \simeq Z^r_p / (dZ^{r-1}_{p+r-1} + Z^{r-1}_{p-1})\, .
\end{equation}
Here we use the standard module isomorphism:
\begin{displaymath}
  (U+W)/(V+W) \simeq U/(V+U \cap W)
\end{displaymath}
for submodules $U,V,W$ of a module $A$ such that $U\supset V$.

We write in the same way
\begin{displaymath}
  E^r_{p-r} \simeq Z^r_{p-r} / (dZ^{r-1}_{p-1} +Z^{r-1}_{p-r-1})
  \, .
\end{displaymath}
But
\begin{displaymath}
  \{ x|\,\,x \in Z^r_p , \,\, dx \in Z^{r-1}_{p-r-1} \}
= \{ x|\,\,x \in Z^r_p , \,\, dx \in F_{p-r-1} A\} = Z^{r+1}_p \, .
\end{displaymath}
This forces us to conclude that $\ker (d^r\!:E^r_p \to E^r_{p-r})$
coincides with the image in $E^r_p$ of $Z^{r+1}_p \subset
Z^r_p$.

Thus, taking into account that $Z^{r+1}_p \supset
dZ^{r-1}_{p+r-1}$ and that $Z^{r+1}_p \cap Z^{r-1}_{p-1} =
Z^r_{p-1}$, we use (\ref{eq:A9}) to establish
the following isomorphism:
\begin{equation}
\label{eq:A10}
  \ker (d^r\!:E^r_p \to E^r_{p-r}) \simeq Z^{r+1}_p/
  (dZ^{r-1}_{p+r-1} + Z^r_{p-1}) \, .
\end{equation}
Now, because of (\ref{eq:A9}),
\begin{displaymath}
  E^r_{p+r} \simeq Z^r_{p+r} /(dZ^{r-1}_{p+2r-1} +
     Z^{r-1}_{p+r-1}) \, ,
\end{displaymath}
hence, as $d Z^r_{p+r} \subset Z^r_p$, we see that
\begin{eqnarray*}
  \Im (d^r:E^r_{p+r} \to E^r_p) &\simeq &
  (dZ^r_{p+r} + Z^r_{p-1})/ (dZ^{r-1}_{p+r-1} + Z^r_{p-1})\,.
\end{eqnarray*}
This, together with (\ref{eq:A10})
\begin{eqnarray*}
H(E^r_p) & \simeq & Z^{r+1}_p /(dZ^r_{p+r}+Z^r_{p-1})
 \simeq E^{r+1}_p \, .
\end{eqnarray*}
So (\ref{eq:A3}) is established.\\

Quite similar to the filtered differential module situation ([M,
Ch.~XI, Prop.~3.2]) we get the convergence of the
spectral sequence under additional conditions on the filtration.

\bPr
  If $\cup_p F_pA=A$ and for some $N, \,\, F_{-N} A=0$,
  then the spectral sequence converges.
\ePr

The latter means that
  for every $p$ and large enough $r$ we get a commutative diagram
  of natural morphisms
  \begin{eqnarray*}
    \begin{array}[]{clclcl}
E^r_p & \to & E^{r+1}_p & \to \cdots \\
& \searrow
&&\searrow \downarrow\\
&&&\!\!\!\!\!\!\!\! F_p (H(A))/F_{p-1}(H(A))=\Gr_p (H(A))
    \end{array}
  \end{eqnarray*}
that identifies $\displaystyle{\Gr_p (H(A)) \simeq
\lim_{\substack{\longrightarrow\\r}}\,  E^r_p}$, or
$\Gr (H(A)) \simeq E^{\infty}_p$.

The morphisms
\begin{eqnarray*}
    \begin{array}[]{clclcl}
E^r_p & \to & E^{r+1}_p & \to \cdots
    \end{array}
 \end{eqnarray*}
are defined
because $Z^r_{-N}=0$ so $E^r_{-N}=0$, therefore
$\ker (d^r\!:E^r_p \to E^r_{p-r}) =E^r_p$ for given $p$ and
$r$ large
enough.

Moreover $E^r_p = F_p A/(dZ^{r-1}_{p+r-1}+F_{p-1}A)$ and
$E^r_p \to E^{r+1}_{p}$ are surjective for large~$r$.
Also for
large $r$, $Z^r_p = (\ker d) \cap F_p A$ and then (\ref{eq:A7}) shows
that the inductive limit $E^{\infty}_p$ of the system $\{ E^r_p
\to E^{r+1}_{p} \to \cdots \}$ is
\begin{displaymath}
  E^{\infty}_p =\lim_{\substack{\longrightarrow\\r}} \, E^r_p
    = ((\ker d) \cap F_pA + F_{p-1}A)/
    (\cup_r dZ^{r-1}_{p+r-1} + F_{p-1}A) \, .
\end{displaymath}
But $\cup_r dZ^{r-1}_{p+r-1} = (dA) \cap F_pA$, hence
\begin{displaymath}
  E^{\infty}_p \simeq ((\ker d) \cap F_pA +F_{p-1}A)
  /(dA \cap F_pA + F_{p-1}A) = \Gr_p (H(A)) \, ,
\end{displaymath}
as stated above.

%%%%%%%%%%%%%%%---><---%%%%%%%%%%%%%%

\vspace{6ex}

\textbf{Authors' addresses:}
\begin{list}{}{}

\item  Department of Mathematics, MIT,
Cambridge MA 02139,
USA\\
email:~~kac@math.mit.edu

\vspace{1ex}

\item   Department~of~Mathematics, NTNU, Gl\o shaugen,
N-7491 Trondheim,
Norway \\
email:~~rudakov@math.ntnu.no

\end{list}

\end{document}